\documentclass[notitlepage,prd,showpacs,preprintnumbers,nofootinbib,superscriptaddress, twocolumn]{revtex4-2}
\usepackage{amssymb,amsmath,placeins,epsfig,array,bm}
\usepackage[usenames,dvipsnames]{color}
\usepackage[normalem]{ulem}
\usepackage[breaklinks,colorlinks,urlcolor=blue,citecolor=blue,linkcolor=blue]{hyperref}
\usepackage{float}
\usepackage{hyperref}
\usepackage{fontawesome}

\def\gsim{\mathrel{\raise.3ex\hbox{$>$\kern-.75em\lower1ex\hbox{$\sim$}}}}

\newcommand{\ie}{{\it i.e.~}}  \newcommand{\eg}{{\it e.g.~}}

\newcommand{\beq}{\begin{equation}} \newcommand{\eeq}{\end{equation}}
\newcommand{\bea}{\begin{eqnarray}} \newcommand{\eea}{\end{eqnarray}}

\newcommand{\Eq}[1]{Eq.~(\ref{#1})}  
  
\newcommand{\Fig}[1]{Fig.~\ref{#1}} 

\newcommand{\App}[1]{Appendix~\ref{#1}}

\newcommand{\githubmaster}{\href{https://github.com/SamWitte/Anteater_DataShare}{\faGithub}}
\definecolor{dullblue}{rgb}{0,0.298,0.49}
\definecolor{darkred}{rgb}{0.545,0,0}
\definecolor{blue2}{cmyk}{1, 0.1, 0.1, 0}
\hypersetup{colorlinks,linkcolor={darkred},citecolor={dullblue},urlcolor={dullblue}}

\definecolor{blue2}{cmyk}{1, 0.1, 0.1, 0}

\begin{document}

\title{Axion-Photon Conversion in Neutron Star Magnetospheres: \\   \normalfont{The Role of the Plasma in the Goldreich-Julian Model}}

\author{Samuel J.~Witte}
\email{s.j.witte@uva.nl}
\affiliation{GRAPPA Institute, Institute for Theoretical Physics Amsterdam and Delta Institute for Theoretical Physics, University of Amsterdam,
Science Park 904, 1098 XH Amsterdam, The Netherlands}
\author{Dion Noordhuis}
\affiliation{GRAPPA Institute, Institute for Theoretical Physics Amsterdam and Delta Institute for Theoretical Physics, University of Amsterdam,
Science Park 904, 1098 XH Amsterdam, The Netherlands}
\author{Thomas D. P. Edwards}
\affiliation{The Oskar Klein Centre for Cosmoparticle Physics, AlbaNova University Center, Roslagstullsbacken 21, SE-10691 Stockholm, Sweden}
\author{Christoph Weniger}
\email{c.weniger@uva.nl}
\affiliation{GRAPPA Institute, Institute for Theoretical Physics Amsterdam and Delta Institute for Theoretical Physics, University of Amsterdam,
Science Park 904, 1098 XH Amsterdam, The Netherlands}
\begin{abstract}
The most promising indirect search for the existence of axion dark matter uses radio telescopes to look for narrow spectral lines generated from the resonant conversion of axions in the magnetospheres of neutron stars. Unfortunately, a large list of theoretical uncertainties has prevented this search strategy from being fully accepted as robust. In this work we attempt to address major outstanding questions related to the role and impact of the plasma, including: $(i)$ does refraction and reflection of radio photons in the magnetosphere induce strong inhomogeneities in the flux, $(ii)$ can refraction induce premature axion-photon de-phasing, $(iii)$ to what extent do photon-plasma interactions induce a broadening of the spectral line, $(iv)$ does the flux have a strong time dependence, and $(v)$ can radio photons sourced by axions be absorbed by the plasma. We present an end-to-end analysis pipeline based on ray-tracing that exploits a state-of-the-art auto-differentiation algorithm to propagate photons from the conversion surface to asymptotically large distances. Adopting a charge symmetric Goldreich-Julian model for the magnetosphere, we show that for reasonable parameters one should expect a strong anisotropy of the signal, refraction induced axion-photon de-phasing, significant line-broadening, a variable time-dependence of the flux, and, for large enough magnetic fields, anisotropic absorption.  Our simulation code is flexible enough to serve as the basis for follow-up studies with a large range of magnetosphere models. 
\end{abstract}
\maketitle

\section{Introduction}
The QCD axion has emerged in recent years as one of the most compelling candidates to explain dark matter; this is predominantly because the axion is a fundamental ingredient in the most favored solution to the strong CP problem~\cite{Peccei:1977hh, Peccei:1977ur,Weinberg:1977ma, Wilczek:1977pj}, which attempts to explain why CP appears to be very accurately conserved in QCD. Recent experimental advancements have allowed laboratory searches to probe vast swathes of the axion parameter space~\cite{Sikivie:1983ip, Sikivie:1985yu,Irastorza:2018dyq, Sikivie:2020zpn}, however their progress is severely constrained by our current ignorance of the axion mass.\footnote{This is because many searches exploit resonance effects which require tuning experimental setups to specific axion masses~\cite{Hagmann:1998cb,Asztalos:2001tf,Asztalos:2009yp,Silva-Feaver:2016qhh,Du:2018uak,Zhong:2018rsr,TheMADMAXWorkingGroup:2016hpc,Majorovits:2017ppy,Brun:2019lyf,Ouellet:2018beu,Salemi:2021gck}.} Since the relic axion abundance is inherently linked to the axion mass, one might hope to infer the mass directly from measurements of the cold dark matter energy density; unfortunately, difficulties in modeling tpological defects self-consistently from formation to decay~\cite{Davis:1986xc,Hagmann:2000ja,Wantz:2009it,Hiramatsu:2012gg,Kawasaki:2014sqa,Klaer:2017ond,Gorghetto:2018myk,Buschmann:2019icd,Gorghetto:2020qws,Buschmann:2021sdq} (and an ignorance of the unknown scale at which the PQ symmetry was broken) lead to a broad range of estimates roughly spanning the range $10^{-6} \lesssim m_a \lesssim 10^{-3}$ eV \cite{Preskill:1982cy,Dine:1982ah,Abbott:1982af}. 

One promising search avenue which may shed light on the axion mass, and thus dramatically simplify terrestrial axion experiments, is to look for the decay and/or conversion of non-relativistic axion dark matter into photons in astrophysical environments~\cite{Raffelt:1987im,Sigl:2017sew,Caputo:2018ljp,Caputo:2018vmy,Ghosh:2020hgd,Pshirkov:2007st,Huang:2018lxq,Hook:2018iia,Safdi:2018oeu,Battye:2019aco,Leroy:2019ghm,Foster:2020pgt,Buckley:2020fmh,Edwards:2020afl,Edwards:2019tzf,Darling:2020uyo,Nurmi:2021xds}. These processes yield nearly monochromatic spectral features, with the characteristic frequency of the line providing direct information about the axion mass. 
Axion-photon conversion can be efficient in the presence of an external magnetic field, and becomes resonantly enhanced when the background plasma has an oscillation frequency $\omega_p \sim \sqrt{4\pi\alpha n_e/ m_e} $\footnote{More generally, the plasma mass in a cold dilute plasma is given by $\omega_p = \sqrt{\sum_i 4\pi\alpha n_i / m_i}$, where $n_i$ and $m_i$ are the number density and mass of species $i$. } roughly equivalent to the mass of the axion~\cite{Raffelt:1987im}. Neutron stars have recently been proposed as prime targets to search for radio signals from axion-photon conversion due to their plasma rich magnetospheres and enormous magnetic fields~\cite{Pshirkov:2007st,Huang:2018lxq,Hook:2018iia,Safdi:2018oeu,Battye:2019aco,Leroy:2019ghm,Foster:2020pgt,Buckley:2020fmh,Edwards:2020afl,Edwards:2019tzf,Darling:2020uyo,Darling:2020plz}. Although potentially capable of probing QCD axion dark matter, such observations are currently impeded by large theoretical uncertainties which prevent ongoing searches from reliably estimating the expected signal. 

The aforementioned uncertainties can be largely broken down into two categories: $(i)$ those related to the astrophysical description and characterization of magnetospheres, and $(ii)$ those related to the production and propagation of the radio photons generated from axion-photon conversion. The former of these concerns reflects an overarching ignorance of the complicated phenomena that occur in extremely energetic strong-field environments. There is promise, however, that these uncertainties may be largely mitigated in the near future, as up-coming observations from telescopes such as SKA~\cite{skatele,kramer2006periodically,konar2016neutron,tauris2014understanding,Keane:2014vja,Karastergiou:2014cka,antoniadis2015multi} and advances in the simulations of magnetospheres push our understanding to new limits (see \eg~\cite{philippov2015ab,petri2016theory,cerutti2017electrodynamics,chen2017particle,brambilla2018electron,carrasco2020magnetosphere,chen2020filling} for recent developments in magnetosphere modeling).  The latter uncertainty, encompassing open questions such as: how axions and photons mix in an inhomogeneous three-dimensional plasma; how photons propagate through the magnetosphere; if photon-plasma interactions induce a broadening of the spectral line; and if these photons can be absorbed, is more readily addressable given the current knowledge of the community. 

In this paper, we perform an end-to-end calculation of the radio signal in the  Goldreich-Julian (GJ) magnetosphere model. That is to say, our formalism self-consistently follows individual photons from their point of genesis on the conversion surface to asymptotically large distances where they can be treated as free streaming radio waves. We do this using a state-of-the-art auto-differentiation ray-tracing algorithm, which allows us to carefully track axion-photon de-phasing, photon reflections and refraction, energy exchanges with the ambient plasma, and photon absorption. In addition, we introduce a Monte Carlo (MC) sampling algorithm to efficiently draw photon trajectories from generic phase space distributions. We emphasize that this work contains a complete pipeline capable of computing the expected radio signature arising from axion-photon conversion in any magnetosphere model with arbitrary axion phase space distributions, and thus represents a major step towards ensuring the reliability of indirect axion searches.

\section{Photon Propagation}\label{sec:Ray}
Previous attempts to compute the radio flux arising from axion-photon conversion in magnetospheres have all but ignored the complicated electrodynamics detailing how photons sourced near the neutron star surface propagate to the light cylinder\footnote{The light cylinder is defined by the radial distance at which the co-rotating plasma would be required to move at the speed of light, \ie $R_{LC} \sim 1 / \omega_{NS}$, where $\omega_{NS}$ is the rotational frequency of the neutron star. Outside of this radius the magnetosphere model breaks down. }~\cite{Huang:2018lxq,Hook:2018iia,Safdi:2018oeu,Battye:2019aco,Leroy:2019ghm}.  A subset of the present authors made a first attempt in Ref.~\cite{Leroy:2019ghm} to improve upon these assumptions by following the trajectories of individual photons, with the goal of discerning the time-variation and viewing angle sensitivity of radio observations~\cite{Leroy:2019ghm}. This work, however, adopted the simplifying assumption of a free space dispersion relation (\ie photon trajectories are unaffected by the plasma), implying that their technique could not track photon refraction, reflection, absorption, and energy losses/gains due to the time variability of the plasma, quantities which are fundamental for understanding radio telescope sensitivity. We begin here by discussing how to move past the aforementioned assumptions, and properly account for the effect of the plasma on the propagation of radio photons.

The linear response of a plasma to the presence of electromagnetic modes is fully characterized by the dielectric tensor ${{\epsilon_{ij}}}$. Once this quantity is known, the dispersion relations governing the behavior of electromagnetic modes in the plasma are obtained by solving Maxwell's equations in Fourier space; in practice, this amounts to solving for the roots of the determinant of  $|| \, n^2 \, \delta_{ij} - n_i \, n_j - \epsilon_{ij}||$, where $n_i=k_i/\omega$ is the refractive index, and $\vec{k}$ and $\omega$ are respectively the wavenumber and frequency of the Fourier mode of interest.  For a cold and neutral plasma, this determinant yields a $10^{\rm th}$ order polynomial, highlighting the fact that the plasma response can in general be quite complex and diverse~\cite{swanson2012plasma}. In the large magnetization and low frequency limit (a good approximation for radio photons in neutron star magnetospheres), this can be reduced to a $6^{\rm th}$ order polynomial, allowing for the identification of three unique modes (each with a $\pm$ solution). These are often called the magnetosonic-t (sometimes also called the magnetoacoustic-t or X mode), the Alfv\'{e}n\footnote{This mode is synonymous with the magnetosonic-t waves in the limit of purely parallel propagation. }, and Langmuir-O modes. In the infinite magnetic field limit, the first of these has the trivial dispersion relation $\omega^2 = k^2$ -- notice that semi-relativistic axions cannot excite this mode, and thus it is not of interest. The Alfv\'{e}n mode is a subluminous (\ie $n  = k / \omega > 1$) density perturbation that is unavoidably damped at large radii. The Langmuir-O mode on the other hand is a superluminous wave with both longitudinal and transverse components, and naturally evolves into a transverse ordinary `O' mode\footnote{Ordinary modes are those described by the conventional cold isotropic plasma dispersion relation, \ie $\omega^2 = k^2 + \omega_p^2$.} as the wave propagates away from the neutron star~\cite{gedalin1998long}; this is the only mode that axions can excite, and thus it is this mode that we must track in our analysis. We derive in Appendix~\ref{sec:photonDisp} the generalized dispersion relation of the Langmuir-O mode for long wavelength radiation traveling through a highly magnetized plasma, and show an illustration of the phase diagram for each mode at various angles of propagation. In the limit that the plasma is non-relativistic (we defer the study of a boosted plasma to Appendix~\ref{sec:photonDisp}), the dispersion relation of the Langmuir-O mode is given by 
\begin{equation}\label{eq:nr_dspR}
\omega^2 = \frac{1}{2}\left(k^2 + \omega_p^2 + \sqrt{k^4 + \omega_p^4 + 2k^2\omega_p^2 (1- 2\cos^2\tilde{\theta})} \right) \, ,
\end{equation}
where $\tilde{\theta}$ is the angle between the magnetic field and $\vec{k}$ (for reference, we point out that \Eq{eq:nr_dspR} is consistent with one of the roots of the dispersion relation obtained in Eq. 55 of~\cite{gedalin1998long}, in limit of a non-relativistic plasma). Notice that because this is a superluminal mode, \ie the phase velocity $v_p \geq 1$, we need not be worried about Landau damping~\cite{swanson2012plasma}.

\begin{figure}
	\includegraphics[width=0.49\textwidth, trim={14cm 25cm 12cm 25cm},clip]{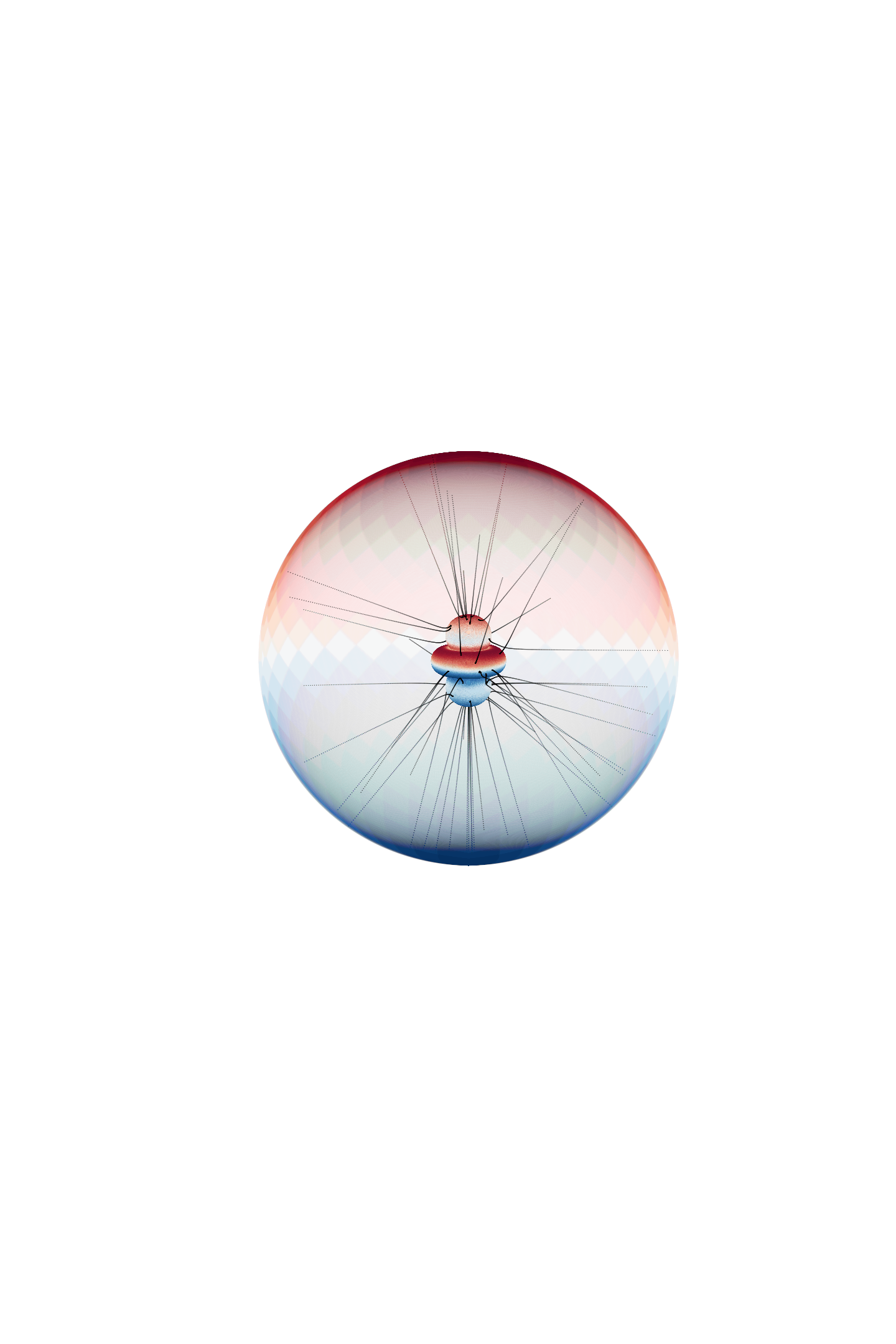}
	\caption{\label{fig:SphericalMap} Illustrative example of the ray-tracing procedure implemented in this manuscript. Photons (black lines) are sourced on the conversion surface in the magnetosphere and are propagated to distances $d = R_{LC}$ (represented by large sphere). The color coding on the conversion surface denotes the sky location where photons are most likely to end up (with red/blue corresponding to north/south pole). This plot, along with all others (including an additional animation illustrating the time evolution of the ray tracing algorithm) can be reproduced with \githubmaster.	}
\end{figure}

In a three-dimensional inhomogeneous plasma, one can derive a set of equations (under a generalized WKB theory\footnote{The WKB method (or approximation) is a procedure for finding approximate solutions to linear differential equations with space-time dependent coefficients (see \eg~\cite{bender2013advanced})}) that  governs the evolution of a wave packet as it propagates through a medium. These are called the ray-tracing equations, and for a pre-defined dispersion relation $\omega(\vec{x}, \vec{k}, t)$ are given by~\cite{swanson2012plasma}
\begin{gather}
\frac{d \vec{x}}{dt} = \nabla_k \, \omega(\vec{x}, \vec{k}, t) \, , \label{eq:raytrace1} \\
\frac{d \vec{k}}{dt} = -\nabla_x \, \omega(\vec{x}, \vec{k}, t) \, , \label{eq:raytrace2}\\
\frac{d\omega}{dt} = \partial_t \omega(\vec{x}, \vec{k}, t) \, \label{eq:raytrace3} \, ,
\end{gather}
The first two equations allow for photon trajectories to be carefully tracked from deep in the magnetosphere to distances $d \sim R_{LC}$, while the  final of these equations defines the differential energy injected to (or imparted from) the plasma per unit path length. Integrating Eq.~(\ref{eq:raytrace3}) along the trajectory yields the net energy change of the photon from production to detection.
%
%

At distances far from the surface of the neutron star, the strong-field assumption used in the derivation of the dielectric tensor breaks down; this is expected to occur when the cyclotron frequency of electrons is comparable to the photon frequency, \ie $\Omega_e / \gamma \equiv q_e B / (m_e \gamma) \sim \omega $, where $\gamma$ is the boost factor of the electrons and positrons in the plasma rest frame (which we take to be one). At this point, two additional effects may arise. First, new propagating modes will appear, implying the dispersion relation given in \Eq{eq:nr_dspR} may no longer be valid. We have verified numerically, however, that at the distances where this occurs the dispersion relation is to a very high degree given by the free space solution, \ie $\omega^2 = k^2$. Additionally, the fact that the dispersion relation given in \Eq{eq:nr_dspR} is void of poles and resonances precludes the possibility of mode mixing\footnote{As the propagating Langmuir-O mode approaches the branch cut at $\omega \sim \omega_p$, mode tunneling may be possible, but this is naturally a suppressed process~\cite{swanson2012plasma}.} (in which the propagating wave jumps from one dispersion relation to another)~\cite{fuchs1981theory,swanson1985radio,swanson2012plasma}. 
The second novel effect is resonant cyclotron absorption, in which the propagating wave excites an electron or ion to a higher Landau level. In general, this happens when $\omega - k_{||} v_{||} - \Omega_e / \gamma = 0$ (see \eg~\cite{blandford1976scattering,melrose1979propagation,luo2001cyclotron}). For a non-relativistic plasma the second term can be dropped, and the resonance occurs when $\omega \sim \Omega_e$.  We emphasize here two crucial points. First, while the cyclotron resonance never has a significant impact on the total luminosity, it can significantly suppress the flux coming from the torus and lobes (note that the luminosity is unaffected only because the dominant contribution comes from the throats, where the optical depth is always small). Second, the cyclotron resonance becomes important for magnetic field strengths characteristic of magnetars, which implies (contrary to prior belief) that stronger magnetic fields are not always better suited for axion searches. We outline the computation of the optical depth $\tau$ and illustrate both of these points in Appendix~\ref{sec:cyc_res}.

\medskip
Despite the simplistic form of \Eq{eq:nr_dspR}, solving the ray-tracing equations is complicated by the non-trivial functional dependencies of $\omega_p$ and $\tilde{\theta}$, which are inherently determined by the magnetosphere model and the direction of photon propagation. In order to solve Eqs.~(\ref{eq:raytrace1}\,--\,\ref{eq:raytrace3}), one needs to be able to efficiently and accurately determine the gradients of each of these quantities without adding crippling computational overhead. For this reason we  employ the auto-differentiation package \texttt{ForwardDiff}~\cite{RevelsLubinPapamarkou2016}, which enables fast derivative calculations with minimum loss in accuracy, and can be easily generalized to work for generic dispersion relations and magnetosphere models in future work\footnote{This package, and auto-differentiation in more general terms, works by augmenting the algebra into a space of multidimensional dual numbers -- this new algebra allows one to iteratively apply the chain rule to find derivatives at any order and to arbitrary working precision. Importantly, this is different from symbolic differentiation, which is extremely computationally inefficient, and numerical differentiation, which is subject to rounding and discretization errors.  }. 

Throughout {this} work we adopt the GJ  model of the magnetosphere~\cite{Goldreich:1969sb}, which is derived by finding the minimum charge density allowed by a stable self-consistent solution to Maxwell's equations in the presence of a strong rotating magnetic field. The charge density and magnetic field are given by
\begin{gather}
n_{GJ}(\vec{r}) = \frac{2 \vec{\omega}_{NS} \cdot \vec{B}}{e} \, \frac{1}{1 - \omega_{NS}^2 r^2 \sin^2\theta} \, , \\
B_r = B_s \left( \frac{r_{NS}}{r} \right)^3 \, (\cos\theta_m \cos\theta + \sin\theta_m \sin\theta \cos\psi) \nonumber \, , \\
B_\theta = \frac{B_s}{2} \left( \frac{r_{NS}}{r} \right)^3 \, (\cos\theta_m \sin\theta - \sin\theta_m \cos\theta \cos\psi) \nonumber \, , \\
B_\phi = \frac{B_s}{2} \sin\theta_m \sin\psi \, .
\end{gather}
Here, $\vec{\omega}_{NS} = \omega_{NS} \hat{z}$ is the rotational frequency of the neutron star, $B_s$ is the magnetic field strength at the surface of the neutron star, $\theta_m$ is the misalignment angle between $\vec{B}$ and $\omega_{NS}$, $\psi(t) = \phi - \omega_{NS} t$, and $r_{NS}$ is the radius of the neutron star.  We assume throughout that $n_{e^-} = n_{GJ}$; since electrons and positrons have the same effect, taking $n_{e^+} = 0$ does not influence our results (so long as the net charge density is given by $n_{GJ}$).  It is worth mentioning that realistic magnetospheres may have large charge multiplicities $M$ (defined by $M\equiv n_e / n_{GJ}$), relativistic streaming plasma's, and deviations from the dipole magnetic field configuration -- we leave a thorough investigation of these effects to future work. It is also worth mentioning that non-active neutron stars are expected to be comprised almost entirely of electrons and ions, with a strong spatial separation between the charged populations (see \eg~\cite{Safdi:2018oeu}). In this case, the results presented below are likely to only be meaningful for photons originating from the electron-dominated region, as the plasma mass for ions is suppressed relative to that of electrons. 

In \Fig{fig:SphericalMap}, we provide an illustration outlining the ray-tracing technique presented here. In this image, we show how the trajectories of photons sourced on the conversion surface of a neutron star can be accurately traced to asymptotic distances (note that in the true procedure, we track photons to distances $r \sim R_{LC}$, which extends far beyond the distances illustrated in \Fig{fig:SphericalMap}). Close inspection of the region near the conversion surface shows evidence of strong reflections and refraction for many of these randomly selected trajectories. After propagating photons to the light cylinder, one may map their sky location back to the conversion surface in order to better understand the sensitivity of the flux, at a given viewing angle, to modeling uncertainties in the magnetosphere. This mapping is illustrated in \Fig{fig:SphericalMap}, where we assign a color to the conversion surface based on the sky location of the sourced photons at the light cylinder; here, red identifies the point of genesis for photons that propagate to $\theta \sim 0$ (\ie the north pole) at $r \sim R_{LC}$, while blue indicates photons that finish at $\theta \sim \pi$ (\ie the south pole). One can see that photons sourced on the top of the torus (the azimuthal feature) or the northern lobe predominantly propagate upward, while photons sourced on the bottom of the torus or the southern lobe predominantly propagate downward. While this agrees with naive expectations, a more complicated structure can be seen on the sides of the lobes and in the throats (the areas between the lobes and the torus), where axion-photon conversion can be much more efficient. We present a more systematic illustration of the photon refraction and reflection using two-dimensional projections in \App{sec:photonDisp}.
We emphasize that throughout this work we adopt a flat Minkowski metric. Using the Schwarzschild metric would induce corrections proportional to $r_s / r$, with $r_s$ being the Schwarzschild radius (note that the rotational frequency of neutron stars is never sufficiently large so as to necessitate the use of the Kerr metric~\cite{ravenhall1994neutron}). At the surface of the neutron star, this amounts to a maximal correction of the metric of $\sim 10\%$. While future high precision measurements may require the inclusion of these effects, we neglect them here for computational efficiency.

\section{Photon Production Rate} \label{sec:rate} 
The previous section outlined the procedure for propagating photons produced from the point of conversion in the magnetosphere to the light cylinder. We now describe how to properly source these photons from an arbitrary phase space distribution. In this section we outline a novel approach based on MC integration which is easily generalizable to a broad range of problems in which the phase space density is non-trivial. 
The total rate at which photons are produced from axion-photon conversion is given by
\begin{equation}\label{eq:rateEq}
R_{a \rightarrow \gamma} = \int d^3 v \, \int dS \, n_{a}(\vec{r}) \, f(\vec{r}, \vec{v}) \, |\vec{v} \cdot \hat{n}| \, P_{a \rightarrow \gamma}(\vec{r}, \hat{v}) \, ,
\end{equation}
where $S$ is the conversion surface defined by the unit normal vector $\hat{n}$, $n_a$ is the axion number density, $f(\vec{r}, \vec{v})$ is the velocity phase space distribution, and $P_{a\rightarrow \gamma}$ is the axion-photon conversion probability given by~\cite{Hook:2018iia,Leroy:2019ghm} (see \App{sec:converP} for a complete derivation)
\begin{equation}\label{eq:gen_rate}
P_{a\rightarrow \gamma} = \frac{\pi}{2 v_c^2} \left( \frac{g_{a\gamma\gamma} \, B}{\sin\theta}\right)^2   \, |\partial_\ell  k_{\gamma}|^{-1} \frac{1}{\sin^2\theta}\, .
\end{equation}
Here, $g_{a\gamma\gamma}$ is the axion-photon coupling, $B$ the magnetic field strength, $v_c$ the axion velocity at the conversion surface, and $\partial_\ell k_{\gamma}$ the derivative of the momentum of the photon along the path length at the conversion surface. In previous work, the derivative of $k_{\gamma}$ has been truncated at leading order in velocity, yielding $\partial_\ell k_{\gamma} \sim  \partial_\ell \omega_p / (v_c \sin^2\tilde{\theta})$ (see Appendix of~\cite{Hook:2018iia}); if one further assumes radial trajectories, the derivative can be approximated by $|\partial_\ell k_\gamma| \sim 3 m_a / (2 r_c v_c)$, where $r_c$ is the radial distance of the conversion surface (see \eg\cite{Hook:2018iia,Battye:2019aco,Leroy:2019ghm}). Importantly, many trajectories are not radial and can have derivatives that vary significantly from this simplified assumption -- in some cases by many orders of magnitude. In addition, the higher order correction in velocity carries a term proportional to $\partial_\ell \tilde{\theta}$, with $\tilde{\theta}$ the angle between $\vec{k}$ and the magnetic field, which may in some circumstances compensate for the velocity suppression and become the dominant term. We leave the details outlining the derivation of this expression (keeping the next-to-leading order velocity expansion intact) to \App{sec:converP}. Furthermore, we emphasize that auto-differentiation can be used to directly compute the derivatives of interest without having to resort to simplified approximation schemes.

Equation~(\ref{eq:gen_rate}) has been derived under the assumption that axions and photons can be described by one-dimensional plane waves (see Appendix~\ref{sec:converP}). Obtaining the generalized equations describing the evolution of the mixing in three dimensions is of great interest, but we deem this to be beyond the scope of the current work. We do, however, note that the resonant conversion will be de-tuned if photons strongly deviate from their assumed straight line trajectories on sufficiently short timescales. Here, we attempt to account for this effect through a numerical re-scaling of the conversion length as follows.

Axion-photon conversion will remain efficient while the phase overlap, defined by $\phi(z) = \int_0^z d\ell \, (k_a - k_\gamma)$ (where $k_a$ and $k_\gamma$ are the axion and photon momenta), remains small. For photons that travel along straight line trajectories (on scales where the conversion is active), there exists a direct mapping between the conversion length $L_c \equiv \sqrt{\pi / |\partial_\ell k_\gamma|}$ and the de-phasing induced over that length scale -- in particular, for straight line trajectories this amounts to $\phi(L_c) \simeq \pi/2$.
If we now take an axion trajectory oriented along the $z$-axis, the one-dimensional plane wave approximation amounts to making the substitution: $e^{i\vec{k}\cdot \vec{r}} \rightarrow e^{ikz}$. Should refraction occur, one can see that the projection of the phase onto the $z$-axis is actually given by $e^{ikz\cos\alpha}$, with $\alpha$ defining the angle between $k_\gamma(\ell)$ and $\hat{z}$. This is the source of the refraction induced de-phasing. We approximate the impact of axion-photon de-phasing by replacing $k_\gamma \rightarrow k_\gamma \cos\alpha$ in the phase $\phi$, and computing the modified conversion length $L_c^\prime$ such that $\phi(L_c^\prime) = \pi/2$. Since $P_{a\rightarrow \gamma}$ is proportional to $L_c^2$, we then re-scale the conversion probability by a factor of $(L_c^\prime / L_c)^2$. We believe this procedure represents a reasonable approximation to the impact of pre-mature refraction induced de-phasing. A more detailed description of this procedure is outlined in Appendix~\ref{sec:converP}.


\begin{figure*}
	\includegraphics[width=.325\textwidth, trim={0cm 0cm 0cm 0cm}, clip]{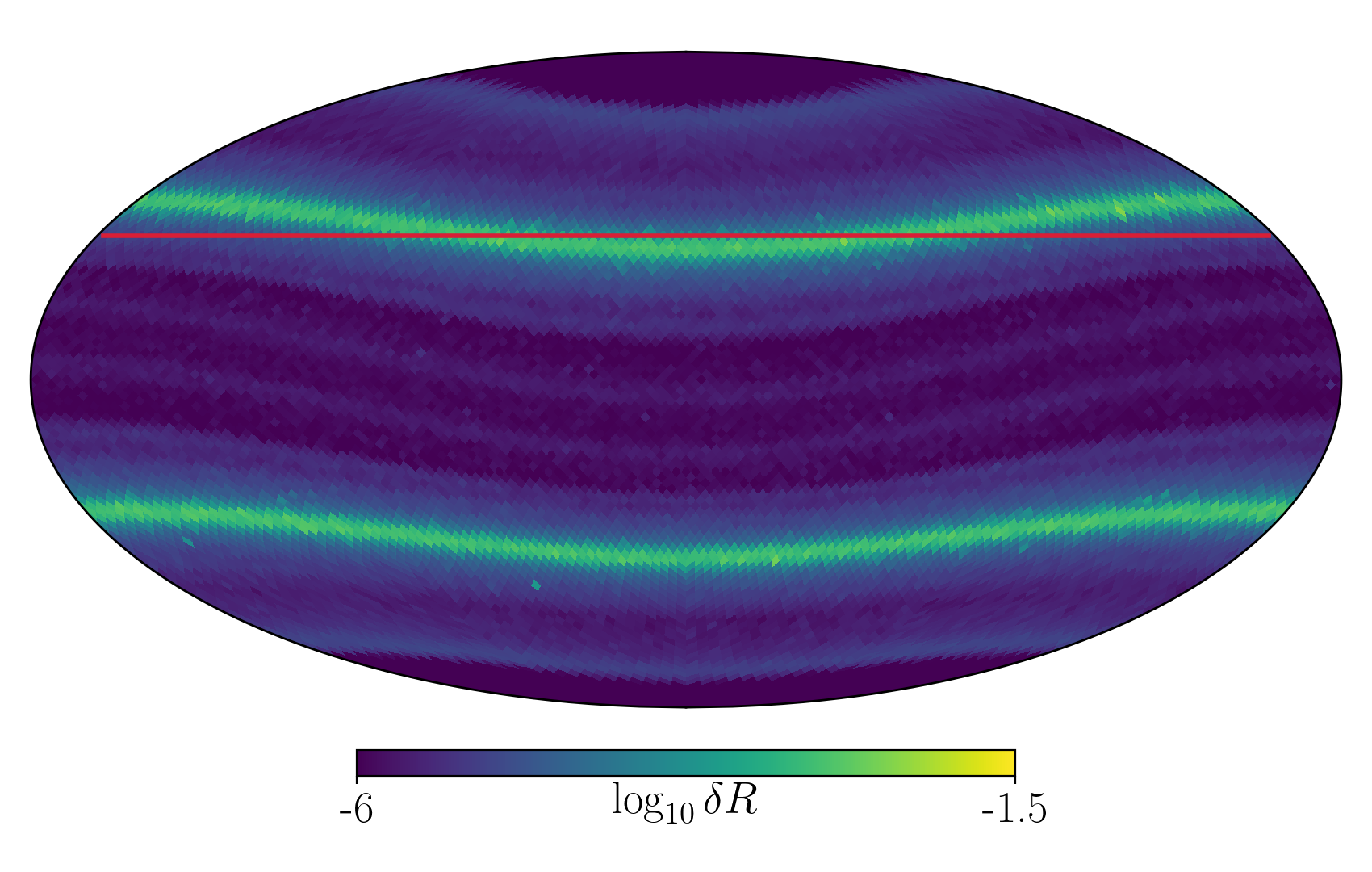}
	\includegraphics[width=.325\textwidth, trim={0cm 0cm 0cm 0cm}, clip]{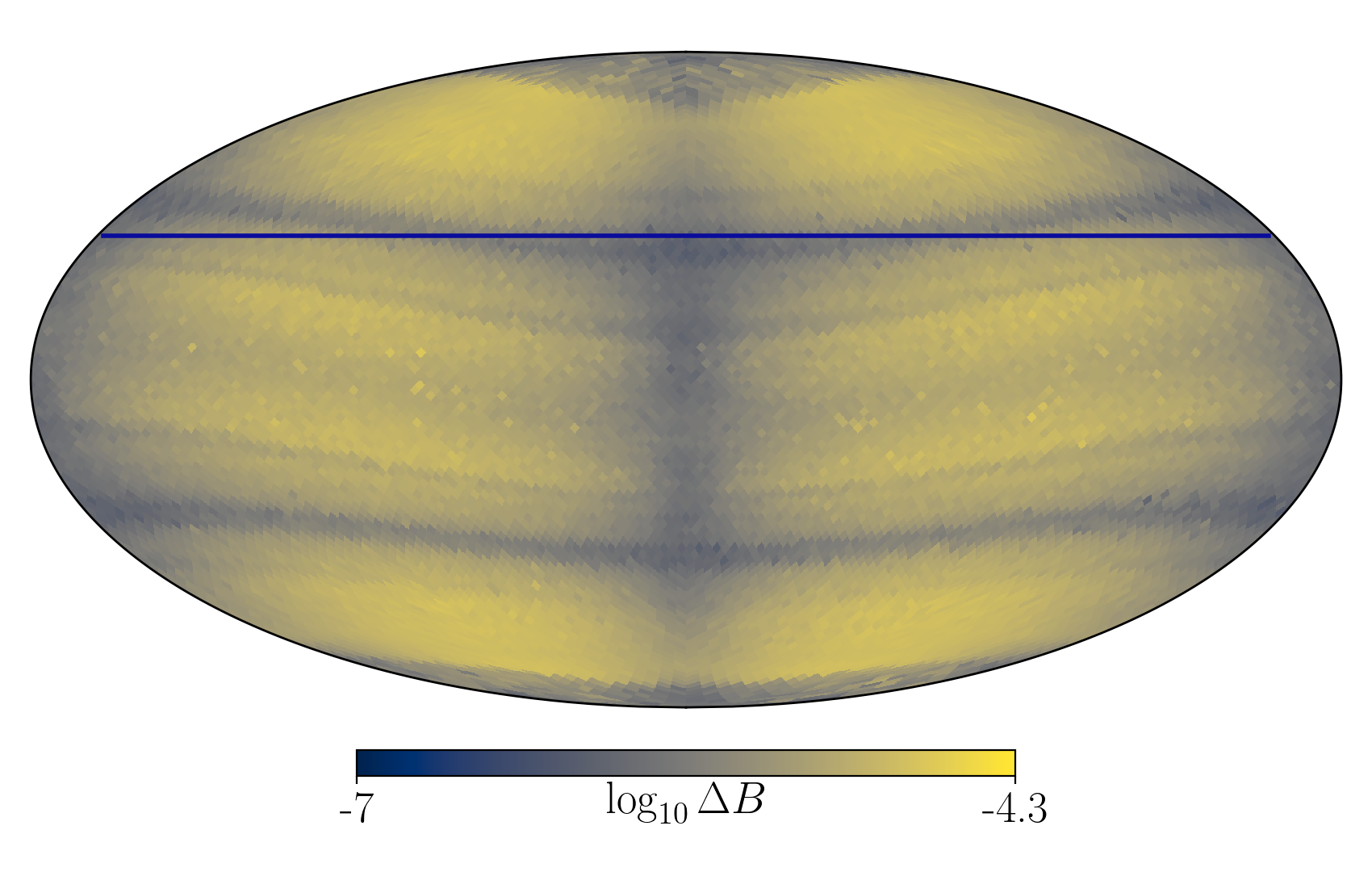}
	\includegraphics[width=0.325\textwidth, trim={0cm 0cm 0cm 0cm}, clip]{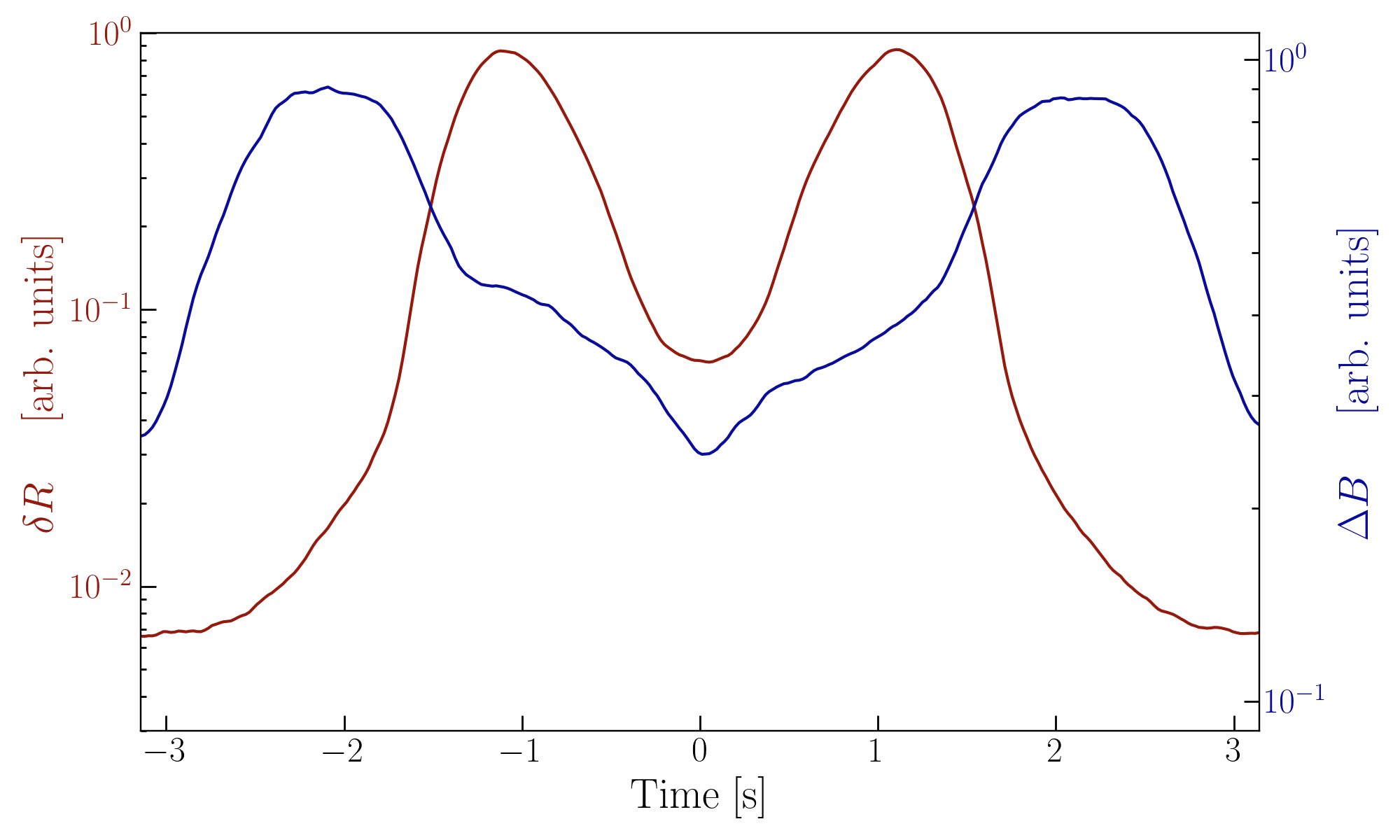}
	\caption{\label{fig:sky_map_1} Fractional rate (left, see \Eq{eq:fracR}), and characteristic line width (center, see \Eq{eq:DelB}) as a function of angular position in the sky after having traveled a distance $r \sim R_{LC}$, with $R_{LC}$ the radius of the light cylinder. Results adopt the fiducial GJ model and $m_a = 10^{-6}$ eV (see text). The time dependence of the flux (red) and line width (blue) at a fixed viewing angle of $\theta = 57^\circ$ can be seen by tracing horizontal lines across each of the maps; this process is illustrated in the right panel.}
\end{figure*}

Equation~(\ref{eq:rateEq}) can be computed under arbitrary circumstances via a MC integration. We outline here the procedure of this calculation for the specific scenario where dark matter in a virialized halo falls onto the neutron star isotropically:
\begin{enumerate}
\item Sample a direction $\hat{v}$ from an isotropic distribution in $(\theta_v, \phi_v)$.
\item Define a disk $\mathcal{A}$ perpendicular to $\hat{v}$ and centered at the origin with radius $R_\mathcal{A} \geq R_{s, \, {\rm max}}$, where $R_{s, \, {\rm max}}$ is the maximum radius of the conversion surface. 
\item Randomly sample a point uniformly from the disk $\vec{r}_{\mathcal{A}}$, and define the vector passing through $\vec{r}_{\mathcal{A}}$ in the direction $\hat{v}$ as $\vec{b}$. \footnote{We found that convergence can be increased by using importance sampling to preferentially map out the phase space at small radii where conversion probabilities are large.}
\item Extend the vector $\vec{b}$ forward and backward to radii $r \geq R_{s, \, {\rm max}}$, and compute the plasma frequency along the vector. The points where $\omega_p \simeq m_a$ define the potential points of conversion $\vec{p}_i$. Notice that the aforementioned defines a generic procedure for performing an arbitrary surface integral~\cite{Detwiler:2008bs}. Performing the integration over the direction of the axion velocity can be done by randomly sampling from a unit sphere, and re-weighting the samples by $\hat{v} \cdot \hat{n}$.\footnote{Technically this sampling step is unnecessary, as one could just as well take $\hat{v}$ to be in the direction $\hat{b}$ (in this case the angular weight is built into the sampling itself). However, due to the fact that the conversion probability diverges in the limit $\hat{v} \cdot \hat{n} \rightarrow 0$, re-sampling improves convergence. }
\item At this point only the integration over $v$ remains. Unfortunately, sampling uniformly from the local speed distribution is non-trivial. We therefore choose to use importance sampling, where we instead sample from the asymptotic speed distribution
\begin{equation}
f_\infty(v_\infty) = \frac{4\pi}{(\pi v_0^2)^{3/2}} \, v_\infty^2 \, e^{-v_\infty^2 / v_0^2} \, ,
\end{equation}
and re-weight the samples by the ratio $f(\vec{r}, \vec{v}) / f_\infty(v_\infty)$.
\end{enumerate}
The net result is that one can express \Eq{eq:rateEq} as 
\begin{equation} \label{eq:rateMC}
R_{a\rightarrow \gamma} \simeq \frac{1}{N} \sum_{j=1}^N \, \frac{\pi R_{\mathcal{A}}^2 \, \rho_\infty}{m_a} \sum_{\vec{p}_i \in N_j} \frac{(v_{\rm min}^2 + v_\infty^2)}{v_\infty} \, P_{a\rightarrow \gamma}(\hat{v}_{\vec{p}_i}) \, , 
\end{equation}
where $N$ is the number of samples drawn, $v_{\rm min} = \sqrt{2 G M_{NS} / r}$, $v_\infty$ is the sampled asymptotic speed (which we obtain using inversion sampling), and $\vec{p}_i$ are the points of conversion in the sample $N_j$. Notice that the local and asymptotic speed are related via $v^2 = v_\infty^2 + v_{\rm min}^2$. In this derivation we have adopted the local phase space distribution obtained by applying Liouville's theorem to the asymptotic velocity distribution, which can be approximately expressed as~\cite{Hook:2018iia}
\begin{equation}
 f(\vec{r}, \vec{v}) \simeq \frac{1}{(\pi v_0^2)^{3/2} } \exp\left(\frac{2 GM_{NS}}{r v_0^2} \right) \, \exp\left(\frac{-v^2}{v_0^2} \right) \, ,
\end{equation}
for $v \geq v_\text{min}$. We emphasize that this procedure can be straightforwardly generalized to more complex anisotropic phase space distributions which may arise in, for example, in-falling axion miniclusters~\cite{Edwards:2020afl}, binary neutron star -- black hole mergers~\cite{Edwards:2019tzf}, and dark matter streams. 

\section{Application to GJ Model} \label{sec:GJ}
In the previous sections we defined how to generate photon seeds on the conversion surface with position and momentum $(\vec{x}_i, \vec{k}_i)$, properly weighted by the phase space density and conversion probability, and how to properly trace the position, momentum, and energy transfer of each photon as it propagates away from the neutron star. In this section we present the results of repeating this procedure over many millions of trajectories in order to generate detailed maps of the photon distributions \footnote{We defer a discussion of the number of photons required for convergence, and the typical associated computational time, to \App{sec:converge}.}. With these maps, one can understand in detail the isotropy, time-dependence, viewing angle dependence, and line width of the radio signal generated from the conversion of axion dark matter around neutron stars.  

We present the results of our fiducial model in this section (obtained by taking $M_{NS} = 1 \, M_\odot$, $r_{NS} = 10$ km, $\omega = 1 \, {\rm s}^{-1}$, $B_s = 10^{14}$ G, and $\theta_m = 0.2$), and discuss the implications of changing these fiducial values in Appendix~\ref{sec:paramD}. Photons are all sourced at time $t=0$, and propagated to distances near the light cylinder; we have verified explicitly that increasing this threshold by orders of magnitude has no impact on the results. We then translate the position of these photons into a healpix sky map, and define the fractional rate in a pixel as
\begin{equation}\label{eq:fracR}
\delta R_i \equiv	\frac{R_{\rm pixel}}{R_{\rm total}} = \frac{\sum_{\gamma_i \in {\rm pixel}} R_{\gamma_i} \, e^{-\tau_{\gamma_i}}}{\sum_i R_{\gamma_i } \, e^{-\tau_{\gamma_i}}} \, ,
\end{equation}
where $R_{\gamma_i}$ is given by \Eq{eq:rateMC}, and $\tau_{\gamma_i}$ is the optical depth of the photon. A map of the fractional rate is illustrated in the left panel of \Fig{fig:sky_map_1}. In addition to the positions of each photon, we also track the energy exchanged with the plasma during photon propagation as this may induce line broadening. 
Defining the final state energy of a single photon as $E_{\gamma_i} = \delta \omega_{\gamma_i} + E_{a_i, \infty}$, where $E_{a_i, \, \infty}$ is the asymptotic energy of the axion sourcing $\gamma_i$ prior to in-fall and $\delta \omega_{\gamma, \, i}$ being the time integration of \Eq{eq:raytrace3}, one can write the characteristic width of the spectral line in a pixel as
\begin{equation}\label{eq:DelB}
\Delta B = \sqrt{ \frac{\sum_{\gamma_i \in {\rm pixel}} R_{\gamma_i}\, e^{-\tau_{\gamma_i}} \,  \times (E_{\gamma_i} - \left<E_{\gamma}\right>)^2 }{ \left<E_{\gamma}\right>^2 \sum_{\gamma_i \in {\rm pixel}} R_{\gamma_i} \, e^{-\tau_{\gamma_i}} } }\, .
\end{equation}
Here $\left< E_{\gamma}\right>$ is the typical final state photon energy, which we set to $m_a$ (accurate to a very high degree when averaged over the full sky).  We plot this quantity in the central panel of \Fig{fig:sky_map_1}. This image illustrates a number of important features. First, there is a strong anisotropy expected in the signal -- depending on the position in the sky, the time-averaged flux may vary by many orders of magnitude. Second, the time variation of the flux, which can be read off the figure by tracing horizontal lines over the period of oscillation (we illustrate this process using a red band in the left panel of \Fig{fig:sky_map_1}), depends crucially on the observing angle, but can be significant in many cases. We present a more detailed analysis and discussion of the time dependence in \App{sec:timeD}. Finally, the line dispersion greatly exceeds the dispersion introduced by the asymptotic axion velocity dispersion (by $\sim1.5$ orders of magnitude in the fiducial model presented); that is to say, the line width is strongly dominated by energy exchanges with the plasma. 

In general, $\Delta B$ does not uniquely characterize the line width, but rather the energy variance in the flux about the axion mass. Large values of $\Delta B$ can thus manifest either as energy shifts in the location of $\left< E_{\gamma}\right>$ or as an actual broadening of the line about $\left< E_{\gamma}\right> \sim m_a$. We have verified that for most pixels the shift in the central value of the line (away from $m_a$) is typically on the order of $33\% \times \Delta B$, implying the dominant influence is broadening, with the shift of the line playing a small but not necessarily negligible role.    

\begin{figure}
	\includegraphics[width=0.49\textwidth, trim={0cm 0.5cm 1cm 1.5cm}, clip]{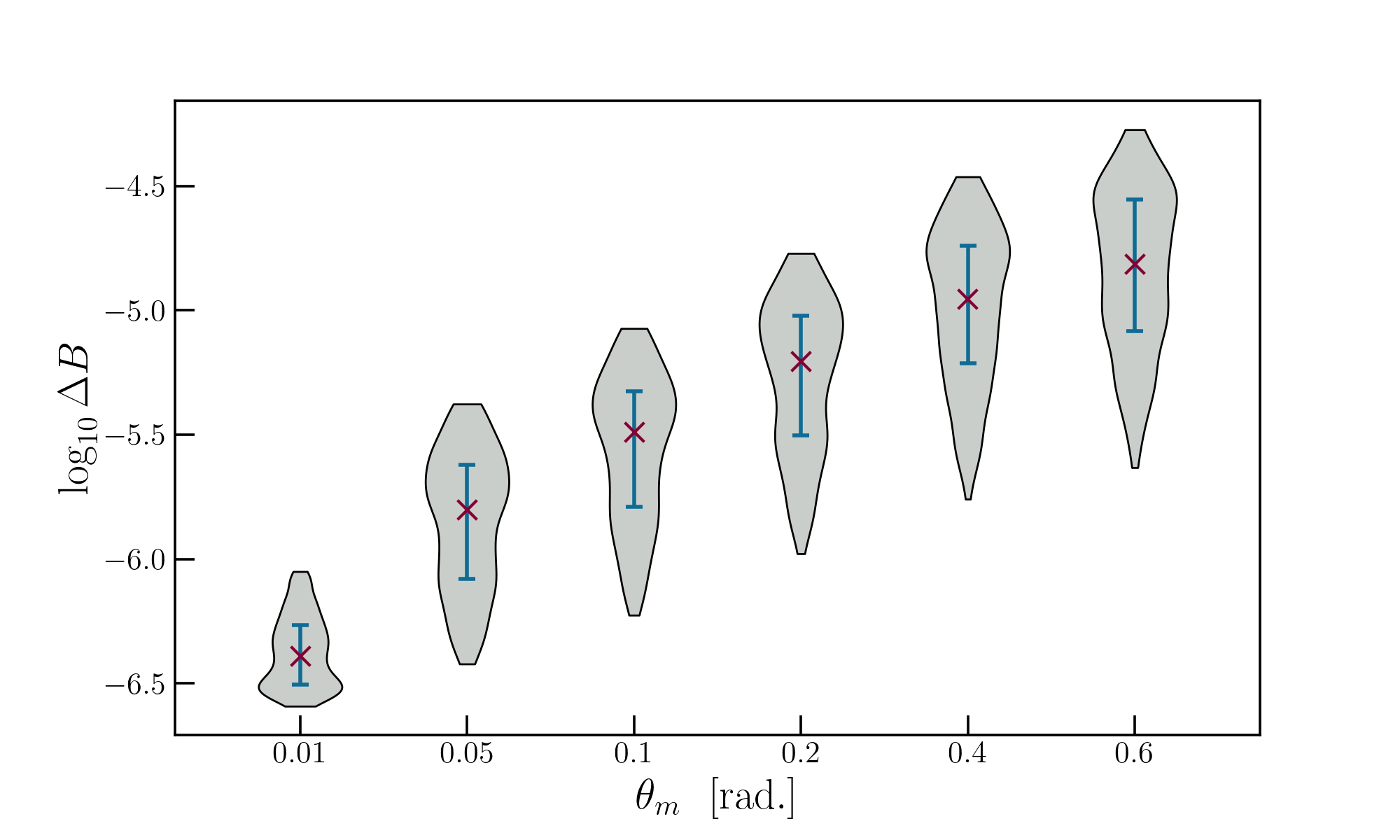}
	\includegraphics[width=0.49\textwidth, trim={0cm 0.5cm 1cm 1.5cm}, clip]{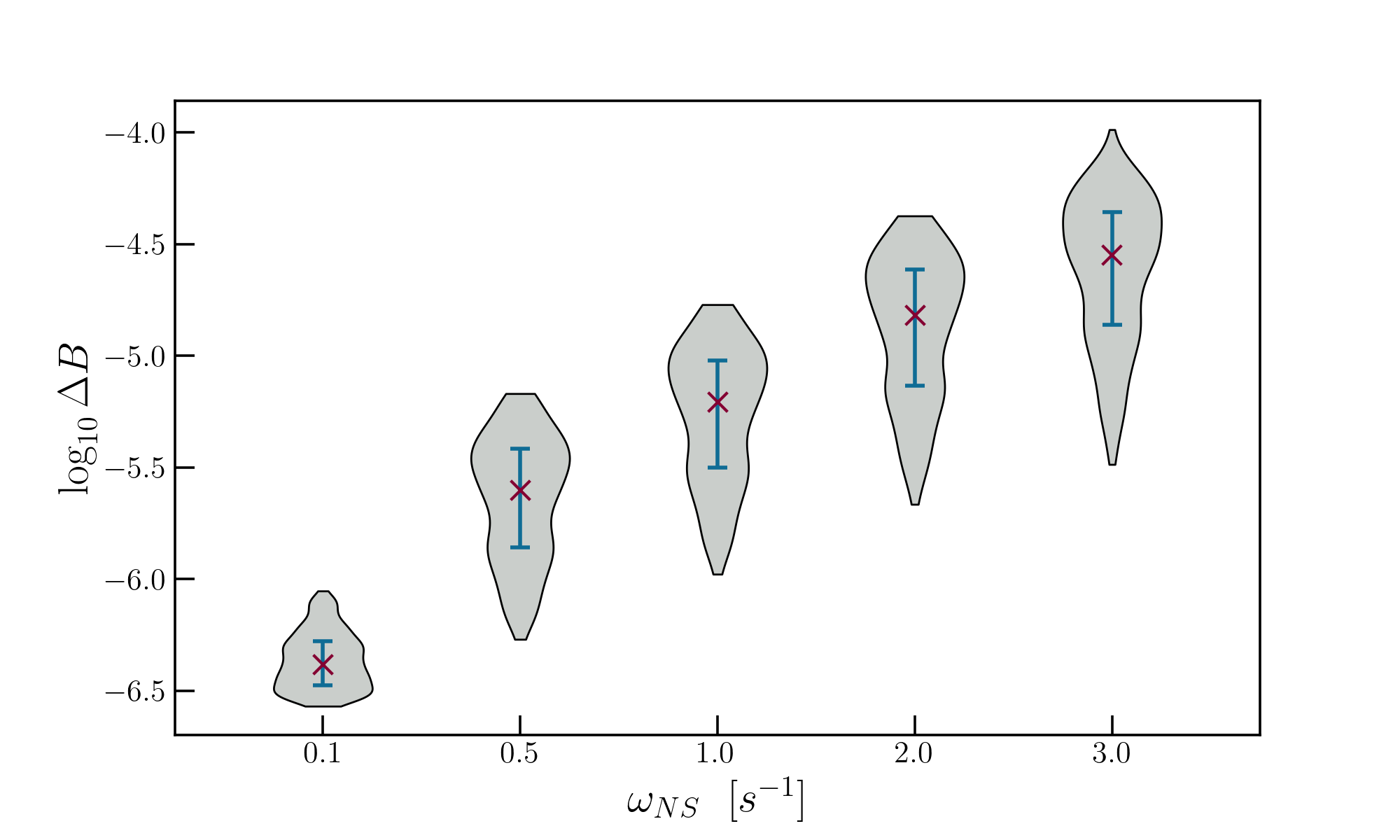}
	\caption{\label{fig:VioDisp}Distribution of spectral line widths $\Delta B$ (see \Eq{eq:DelB}) in each pixel for an $N_{\rm side} = 8$ healpix map, varying $\theta_m$ (top) and $\omega_{NS}$ (bottom). Median value and 25/75\% quartiles are shown in red and blue, respectively. }
\end{figure}

If one neglects systematic uncertainties, the flux sensitivity of a radio telescope is given by the ideal radiometer equation~\cite{1984bens.work..234D} which depends inversely on the square root of the bandwidth~\cite{1984bens.work..234D} -- a large dispersion will therefore reduce the sensitivity of a line search and complicate the identification of the signal. The time dependence of both the bandwidth (blue) and flux (red) for a fixed viewing angle, taken here to be $\theta = 57^\circ$, is shown in the right panel of \Fig{fig:sky_map_1} (using arbitrary units).   While \Fig{fig:sky_map_1} is produced under fixed magnetosphere parameters, the results for the most part do not change qualitatively under variations of these parameters (although the magnitude of the time variation is sensitive to the characteristic scale of the conversion radius, which depends on the magnetosphere parameters) -- perhaps the only exception being $\theta_m$, which can modify the isotropy of the flux. Quantitatively, however, the total luminosity, relative luminosity (\ie the variation in the luminosity between the highest and lowest flux), and the line width may all be affected by the choice of $B_s$, $\omega_{NS}$, $\theta_m$,  and $m_a$.  We defer a discussion of how variations in these parameters affect the luminosity to \App{sec:paramD}, and focus here only on the impact of the line width.\footnote{We highlight here the importance of the line width as this has been among the most debated aspects of the expected radio signal.  } 

In \Fig{fig:VioDisp} we plot the pixel-to-pixel probability distribution functions of the line width $\Delta B$ as a function of $\theta_m$ (top) and $\omega_{NS}$ (bottom),  keeping the remaining parameters fixed to their fiducial values. The median and 25/75$\%$ quartiles are shown with red crosses and blue vertical lines, respectively.  Importantly, the width also depends on the characteristic scale of the conversion surface, which may be altered by changing $m_a$ and $B_s$ (in addition to $\omega_{NS}$).  Comparing with the minimum expected line width (which can be obtained by assuming the width is given by the asymptotic axion velocity distribution, and is roughly of the order $\Delta B \sim 10^{-6}$), one can see that the true width across a majority of the pixels greatly exceeds previous expectations. We note here that the broadening of the line width has been previously discussed separately for reflection and refraction in Refs.~\cite{Battye:2019aco} and \cite{Foster:2020pgt} respectively. Our analysis inherently accounts for these effects, and roughly appears to agree with the linear scaling of $\Delta B$ with $\omega_{NS}$ as described in~\cite{Foster:2020pgt}.

The ultimate goal of this work is to properly assess the consequences of including non-trivial plasma phenomena for axion-based radio surveys. In order to address this point, we estimate the sensitivity of the Square Kilometer Array (SKA) to axion-photon conversion in the magnetosphere of the magnetar PSR J1745-2900, situated only $\sim 0.1$pc from the Galactic Center~\cite{Kennea:2013dfa,Mori:2013yda,Shannon:2013hla,Eatough:2013nva}. 
The flux density observed by a radio telescope in a  bandwidth $\Delta f$ is given by
\begin{equation}
S(t, \theta, \phi) = \frac{1}{D^2 \, \Delta f} \,  \frac{dP}{d\Omega}(\theta, \phi, t)\ \, ,
\end{equation}
where $dP/d\Omega$ is the differential power per unit solid angle and $D$ the distance to the neutron star (taken here to be 8.5 kpc). In the idealized case that the noise is purely thermal, the signal-to-noise ratio (SNR) of the time-averaged flux density $\overline{S}$ is given by the radiometer equation
\begin{equation}
{\rm SNR} = \overline{S}(\theta, \phi) \frac{\sqrt{2 \tau_{\rm obs} \, \Delta f}}{{\rm SEFD}} \, ,
\end{equation}
where $\tau_{\rm obs}$ is the observational time and SEFD is the system equivalent flux density (taken here to be SEFD = 0.098 Jy~\cite{skatele}). Should there exist a strong time-dependence (and assuming one is able to both look for, and characterize, the strongly time-dependent signal), the SNR can be further enhanced by a factor of $\sqrt{\sigma_\tau^2}$, where $\sigma_\tau^2$ is the time-variance of the flux~\cite{Leroy:2019ghm}. Notice that since $dP/d\Omega$ is proportional to the axion-photon coupling squared, the derived limit on $g_{a\gamma\gamma}$ clearly scales as $\propto 1 / \sqrt{\mathrm{SNR}}$. 

Measurements of PSR J1745-2900 have inferred a surface magnetic field strength $B_s \sim 1.6 \times 10^{14}$ G and a rotational period $P_{NS} \sim 3.76 \, {\rm s}$~\cite{Kennea:2013dfa,Mori:2013yda}.\footnote{In addition, we take the mass and radius of the neutron star to be $1 \, M_\odot$ and 10 km, respectively.} We further assume that the dark matter density in the Galactic Center is well-described by a Navarro-Frenk-White (NFW)  profile~\cite{navarro1996structure}; importantly, adopting a cored Burkert dark matter profile~\cite{Nesti:2013uwa} (or assuming the existence of a dark matter spike~\cite{lacroix2018dynamical}) may suppress (enhance) the estimated flux density by a factor of $\sim 10^4$ ($\sim 10^4$).

\begin{figure}
	\includegraphics[width=0.45\textwidth, trim={0cm 0cm 0cm 0cm},clip]{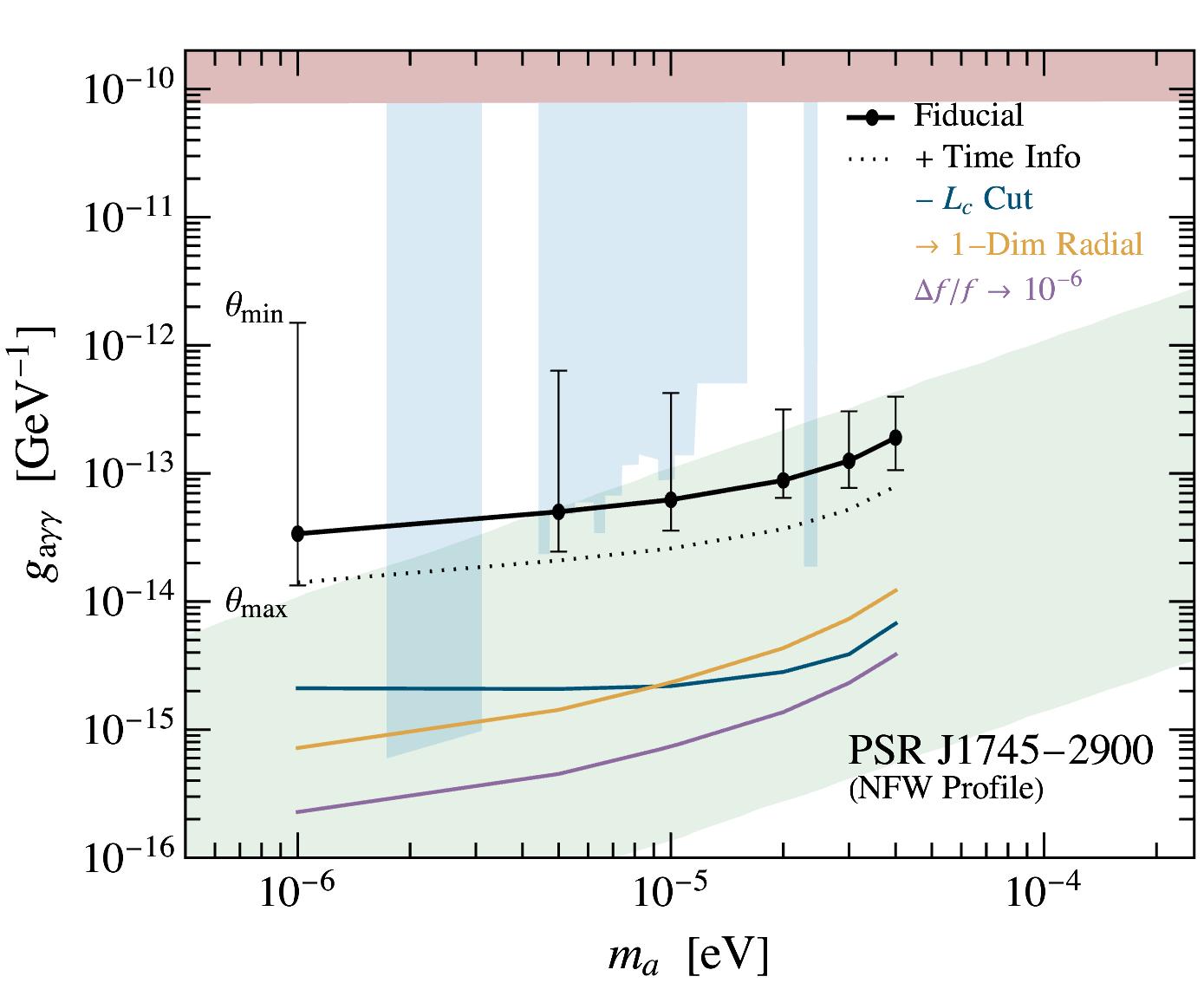}
	\caption{\label{fig:BndCompare}Projected sensitivity to axion dark matter from SKA observations of PSR J1745-2900 (obtained requiring $\mathrm{SNR}=5$), assuming an NFW dark matter profile, 100 observing hours, and $\mathrm{SEFD}=0.098\,\mathrm{Jy}$. Our fiducial sensitivity, obtained by marginalizing over the viewing angle and time-averaging, is shown with a thick black line.  This can be compared with the sensitivity that one would have assuming conservatively (optimistically) the viewing angle $\theta_{\rm min}$ ($\theta_{\rm max}$) with the minimum (maximum) radiated power (shown using vertical error bars). We also estimate the gain in sensitivity that would occur should one assume a strongly time-varying flux (specifically we take the time variation shown in \Fig{fig:sky_map_1}) -- this is illustrated by the black dotted line. Our fiducial sensitivity can be compared to what would have been obtained had we: $(i)$ not performed the conversion length cut (blue), $(ii)$ assumed axions travel on radial trajectories and treating the conversion using the 1-dimensional formalism (in addition to the $L_c$ cut; labeled `1-Dim Radial', shown in yellow), and $(iii)$ taken  a bandwidth $\Delta f/ f = 10^{-6}$ (in addition to both of the aforementioned; purple). Existing bounds from axion haloscopes~\cite{Hagmann:1998cb,Asztalos:2001tf,Du:2018uak} (blue) and helioscopes~\cite{Anastassopoulos:2017ftl} (red) are compared with the QCD axion band (green), taken from~\cite{DiLuzio:2016sbl}. Notice that in this scenario cyclotron absorption affects the bound by no more than $10\%$, and has been included only in the black curves. }
\end{figure}

In \Fig{fig:BndCompare}, we show the projected sensitivity (defined here using a threshold of SNR = 5) of SKA to axions in the mass range $m_a \in [10^{-6},  4\times 10^{-5}] \, {\rm eV}$,\footnote{We adopt here a lower mass threshold consistent with the minimum frequency currently used in radio searches for axions (see \eg~\cite{Darling:2020plz,Darling:2020uyo,Foster:2020pgt}), and an upper threshold roughly consistent with the maximum observable frequency by SKA-Mid. For axion masses near the upper threshold, the conversion surface is very near to the surface of the neutron star itself -- thus telescopes operating at higher frequencies are unlikely to offer further improvement in this direction. } assuming 100 hours of observing time and a bandwidth $\Delta f / f = 10^{-4}  $ (which represents a conservative estimate based on the results of~\Fig{fig:VioDisp}, although we note that this choice is slightly too conservative at large masses).  In our fiducial analysis we assume $\theta_m \sim 0.2$, a time-averaged flux density,  and we marginalize over the viewing angle $\theta$. The fiducial sensitivity is shown in \Fig{fig:BndCompare} using black dots (an interpolation is shown by the thick black line). We note that the marginalization over $\theta$ significantly biases the expected sensitivity toward lower axion-photon couplings; we illustrate this point by plotting vertical bars which represent the sensitivity from the maximally optimistic $\theta_{\rm max}$ and pessimistic $\theta_{\rm min}$ viewing angles. We also show the extent to which the sensitivity could be improved should the signal be strongly time-varying (black dotted; specifically, we assume a time-varying flux consistent with the result shown in the right panel of \Fig{fig:sky_map_1}).

In order to illustrate how each of the effects we have studied influences the expected sensitivity, we systematically remove each effect one by one, and re-assess the sensitivity estimate at each step. We begin by discarding the re-scaling of the conversion length, which was needed to account for the premature axion-photon de-phasing induced by refraction on scales below the conversion length. The result of removing this cut is shown in \Fig{fig:BndCompare} (blue line, labeled `$L_c$ Cut'). At small masses this amounts to roughly an order of magnitude correction in the sensitivity at all masses. Next, rather than computing the conversion probability using $\partial_\ell k_\gamma$ for each individual photon, we use the simplified radial trajectory approximation adopted in previous works $\partial_\ell k_\gamma \sim 3 m_a / (2 r_c v_c)$, and adopt the 1-dimensional formalism in which the axion velocity is assumed to be perpendicular to the conversion surface (\ie two of the assumptions that entered into~\cite{Hook:2018iia}). This cut, labeled `1-Dim Radial', results in a factor of two reduction in the bound at large masses 
and strengthens the bound by a factor of $\sim 3$ at small masses. The strong mass dependence can at least partially be understood as a geometric effect, arising from the fact that the fraction of the conversion surface with a radially aligned normal vector increases as the conversion surface shrinks towards the surface of the neutron star (\ie for larger axion masses). Consequently, the radial approximation becomes worse, and thus the discrepancy grows, at smaller axion masses. Finally, if the width of the radio line were set by the asymptotic axion velocity (rather than interactions with the plasma),  the SNR could be enhanced by choosing a narrower bandwidth closer to $\Delta f / f \sim 10^{-6} $; making this assumption further increases the sensitivity uniformly by a factor of 3.2 (shown by the purple line).  Cumulatively, our fiducial sensitivity is suppressed by a factor $\sim 10^2$ relative to previous estimates,  and a more conservative treatment of the viewing angle may lead to a further suppression by up to one or two orders of magnitude.  Finally, we note that the magnetic field of PSR J1745-2900 is not sufficiently large for cyclotron absorption to become important -- the effect is never larger than a factor of two suppression in the differential power, and on average is much smaller. We include this effect in the computation of the black lines, but do not include it in the colored lines (note that it is hardly visible after marginalizing over $\theta$).

While the cumulative suppression in the sensitivity presented above may at first glance appear devastating for indirect axion searches, it is important to emphasize that there is still hope. In particular, we point out that the largest suppression has arisen from the uncertainty in the magnetar viewing angle. Indirect searches using neutron star populations can directly circumvent this by observing neutron stars across a wide array of possible angles (some of which are likely aligned such that the brightest flux points toward Earth). Slowly rotating and highly aligned neutron stars also experience a rather minimal amount of line broadening, implying one may still be able to use narrow bandwidths if the properties of the magnetosphere are reasonably well understood. Removing the aforementioned effects, and including the enhancement from strong time variability, one may still have sensitivity to a broad range of the axion parameter space.  Importantly, we reiterate that the importance of each effect studied here is only valid in the GJ model. Further studies are required in order to understand how the conclusions found here change under reasonable assumptions of the electron and positron charge distributions.

{\bf{Brief comment on Battye et al. (2021):} }\\
The preprint of Ref.~\cite{Battye:2021xvt} appeared online shortly after the publication of this work, with the goal of addressing similar questions to those discussed above. Ref.~\cite{Battye:2021xvt} adopts a backward ray-tracing technique, as presented in Ref.~\cite{Leroy:2019ghm}, to follow photons from asymptotic distances to the conversion surface -- the authors improve, however, upon Ref.~\cite{Leroy:2019ghm} by (1) using the cold plasma dispersion relation $\omega^2 = \omega_p^2 + k^2$ (rather than the free space dispersion relation), (2) propagating photons in the Schwarzschild metric, and (3) computing line broadening effects (in a manner akin to what has been performed here). Due to the differences in methodology and treatment of the system, it is not trivial to make a direct comparison between these results. Future comparisons between the algorithm developed here and that in Ref.~\cite{Battye:2021xvt} will hopefully shed light on the importance of each effect and on the most promising approach moving forward.

\section{Conclusions} \label{sec:con}
In this work we have presented a novel tool and formalism that allows for a complete treatment of axion-photon conversion in neutron stars, tracing  individual trajectories of millions of photons from their genesis to distances comparable to the light cylinder. This procedure allows for an accurate determination of the anisotropy and time-dependence of the flux as well as the width of the radio line resulting from axion-photon conversion in magnetospheres. Furthermore, our formalism is capable of accounting for axion-photon de-phasing induced from photon refraction near the point of production. Our formalism is fast, highly accurate, and easily generalizable to arbitrary magnetosphere models. 

The results presented here illustrate four crucial points. First, the radio flux is strongly anisotropic, implying a strong inherent sensitivity to the viewing angle. In addition, photon refraction is likely to induce a premature de-phasing of the axion resonance, which will strongly suppress the flux. Next, our results indicate that the width of the line for typical neutron star parameters will be determined by the energy transfer between photons and the plasma, and is typically far greater than previously appreciated. Third, the extent to which one expects a strongly time-dependent signal crucially depends on the misalignment angle and the viewing angle from the neutron star. Finally, for sufficiently large magnetic field strengths, resonant cyclotron absorption may no longer be negligible, implying stronger magnetic fields do not always make ideal targets for indirect axion searches. It is now of ample importance to re-evaluate existing search strategies and constraints whilst taking these effects into full account, which we leave to upcoming work.

We have assumed throughout this paper a simplified magnetosphere model that fails to account for plasma inhomogeneities, charge separated plasma, and relativistic streaming plasma substructure. While it is unlikely that the detailed nature of the magnetosphere will be available in the near future with sufficient resolution to capture the necessary details on the scales of interest, the methods presented here will allow for many of these uncertainties to be treated systematically in a meaningful and quantitative manner. Crucially, our results represent a complete end-to-end model of the conversion process, which significantly reduces the theoretical uncertainty of the resulting flux and allows for reliable inferences of the expected radio signal. 

\medskip
\textbf{\textit{Acknowledgments --}}  The authors would like to thank Ben Safdi, Joshua Foster, Andrea Caputo, Edoardo Vitagliano, and Jamie McDonald for their useful comments and discussions on the manuscript. This work has also greatly benefited from discussions with Sebastian Baum, Matthew Lawson, M. C. David Marsh, and Alexander Millar. 
This work has received funding from the European Research Council (ERC) under the European Union's Horizon 2020 research and innovation programme (Grant agreement No. 864035 - Un-Dark). T.E. acknowledges support by the Vetenskapsr{\aa}det (Swedish Research Council) through contract No.  638-2013-8993 and the Oskar Klein Centre for Cosmoparticle Physics. T.E was also supported in part by the research environment grant `Detecting Axion Dark Matter In The Sky And In The Lab (AxionDM)' funded by the Swedish Research Council (VR) under Dnr 2019-02337.  The computations/data handling were/was partially enabled by resources provided by the Swedish National Infrastructure for Computing (SNIC) at HPC2N partially funded by the Swedish Research Council through grant agreement no. 2018-05973. We acknowledge the use of the Python scientific computing packages NumPy~\cite{numpy,Harris:2020xlr} and SciPy~\cite{scipy}, as well as the graphics environment Matplotlib~\cite{Hunter:2007}. Finally, we acknowledge the use of the Julia auto-differentiation package \texttt{ForwardDiff}~\cite{RevelsLubinPapamarkou2016}.
\bibliography{Notes}
\clearpage
\newpage
\onecolumngrid
\begin{center}
\textbf{\large Axion-Photon Conversion in Neutron Star Magnetospheres:} \\ 
\vspace{0.1in}
{ \it \large Supplemental Material}\\ 
\vspace{0.05in}
{}
\end{center}
\onecolumngrid
\appendix
\section{Photon Dispersion Equation}\label{sec:photonDisp}
In this section we outline the derivation of the generalized dispersion relation relevant for the propagation of radio waves near the surface of neutron stars. First, we derive the dispersion relation for a cold and non-relativistic plasma (for a detailed overview, see \eg~\cite{swanson2012plasma} and references therein), and then generalize the derivation to the case where electrons and positrons have non-negligible boosts in the plasma rest frame. 

We begin by writing down the linear equations of motion for non-relativistic charged particles in the presence of an external electromagnetic field, given by
\begin{eqnarray}\label{eq:particleM}
m_s \frac{d \vec{v}_s}{dt} = q_s (\vec{E} + \vec{v}_s \times \vec{B}) \, ,
\end{eqnarray}
where we have introduced the subscript $s$ to denote the species (\eg electron, positron, proton, etc). The relevant Maxwell equations can be expressed as\footnote{In this section we will work with Gaussian units and take $c=1$ (the latter being adopted throughout the text).}
\begin{gather} \label{eq:maxw1}
\nabla \times \vec{E} = -\frac{\partial \vec{B}}{\partial t} \, , \\ \label{eq:maxw2}
\nabla \times \vec{B} = \left( 4\pi \vec{J} + \frac{\partial \vec{E}}{\partial t} \right) \, .
\end{gather}
Here the current is given by $\vec{J} \equiv \sum_s q_s \, n_s \, \vec{v}_s$ and $n_s$ is the number density. We can now assume an isotropic and homogeneous plasma, and take a plane wave ansatz for the fields as $\vec{E} = \vec{E}_1 \, e^{i(\vec{k} \cdot \vec{r} - \omega t)}$, $\vec{B} = \vec{B}_0 + \vec{B}_1 \, e^{i(\vec{k} \cdot \vec{r} - \omega t)}$, and $\vec{v} = \vec{v}_1 \, e^{i(\vec{k} \cdot \vec{r} - \omega t)}$, where $\vec{B}_0$ is the static background magnetic field. We furthermore take $\vec{B}_0$ to be in the $\hat{z}$ direction and assume that the magnitude $B_0 \gg B_1$. Using Eqs.~(\ref{eq:particleM}-\ref{eq:maxw2}), one can solve for the various components of the current density
\begin{gather}\label{eq:current1}
 J_{\pm} = i \epsilon_0 \sum_s \frac{\omega_{p, s}^2}{m_s \, (\omega \mp \, \Omega_s)} \, E_\pm \, , \\
 J_z = i \epsilon_0 \sum_s \frac{\omega_{p, s}^2}{\omega} \, E_z \, .
\end{gather}
Here, we have introduced the $E_\pm$ notation to define $E_\pm \equiv E_x \pm i E_y$ (and similarly for the velocity and current vectors). We have also defined the cyclotron frequency of species $s$ as $\Omega_s \equiv q_s B_0 / m_s$; for the plasma of interest here, the dominant species are electrons and positrons, and thus $\Omega_s = \mp \Omega_e$. 

The dielectric tensor $\boldsymbol\epsilon$ is generically defined as
\begin{equation}
\boldsymbol\epsilon = \mathbb{I} - 4\pi\frac{\boldsymbol\sigma}{i \omega} \, ,
\end{equation}
where the conductivity $\boldsymbol\sigma$ can be inferred from the current equation $\vec{J} = \boldsymbol\sigma \cdot \vec{E}$. With the current as defined in \Eq{eq:current1}, one can express the dielectric tensor as
\begin{equation}
\begin{gathered}
\epsilon_{xx} = \epsilon_{yy} = 1 - \sum_s \frac{\omega_{p,s}^2}{\omega^2 - \Omega_s^2} \, , \\
\epsilon_{xy} = -\epsilon_{yx} = \sum_s \frac{\Omega_s \, \omega_{p,s}^2}{\omega (\omega^2 - \Omega_s^2)} \, , \\
\epsilon_{zz} = 1 - \sum_s \frac{\omega_{p,s}^2}{\omega^2} \, , \\
\epsilon_{xz} = \epsilon_{zx} = \epsilon_{yz} = \epsilon_{zy} = 0 \, .
\end{gathered}
\end{equation}
Maxwell's equations given in \Eq{eq:maxw1} can be expressed in terms of the dielectric tensor as
\begin{gather}
i \vec{k} \times \vec{E} = i \omega \vec{B} \, , \\
i \vec{k} \times \vec{B} = i \omega {\boldsymbol{\epsilon}} \cdot \vec{E} \, .
\end{gather}
Combining these equations yields the so-called wave equation, given by
\begin{eqnarray} \label{eq:waveE}
\vec{n} \times (\vec{n} \times \vec{E}) + {\boldsymbol{\epsilon}} \cdot \vec{E} = 0 \, ,
\end{eqnarray}
where $\vec{n} = \vec{k} / \omega$ is the index of refraction. Without loss of generality, we can take $\vec{k}$ to lie in the x-z plane; in this case, the dispersion relations of propagating electromagnetic modes can be directly determined by solving
\begin{equation}\label{eq:det}
\begin{vmatrix}
n^2\cos^2\theta - \epsilon_{xx} & -\epsilon_{xy} & -n^2\cos\theta\sin\theta - \epsilon_{xz}  \\  \\
-\epsilon_{yx} & n^2 - \epsilon_{yy} & -\epsilon_{yz} \\  \\
-n^2\cos\theta\sin\theta - \epsilon_{zx} & -\epsilon_{zy} & n^2\sin^2\theta - \epsilon_{zz} 
\end{vmatrix} = 0 \, ,
\end{equation}
where we have defined $\theta$ as the angle between $\vec{k}$ and $\vec{B}_0$. In the low frequency and large magnetic field limit, which equates to $\lvert \Omega_e \rvert \gg \omega, \omega_p$, one finds the following three dispersion relations 
\begin{gather}
\omega^2 = k^2 \, , \\
\omega^2 = \frac{1}{2}\left(k^2 + \omega_p^2 - \sqrt{k^4 + \omega_p^4 - 2 k^2 \omega_p^2 (1 - 2 \cos^2\theta)} \right) \, , \\
\omega^2 = \frac{1}{2}\left(k^2 + \omega_p^2 + \sqrt{k^4 + \omega_p^4 - 2 k^2 \omega_p^2 (1 - 2 \cos^2\theta)} \right) \, .
\end{gather}
These correspond to the magnetosonic-t, Alfv\'{e}n, and Langmuir-O mode, respectively (see \eg~\cite{gedalin1998long}). 

\medskip

The dispersion relations above are derived assuming a non-relativistic plasma. In order to determine whether our calculations are robust against this assumption, we now derive generalized dispersion relations for charged species with arbitrary phase space distributions. This discussion largely follows the work of \cite{gedalin1994nonlinear,gedalin1998long,lyutikov1998waves,gedalin2001low} and the references therein. As before, we will work in the low frequency and strong magnetic field limit. In addition to taking $\lvert \Omega_e \rvert \gg \omega, \omega_p$, the assumption of a strong magnetic field allows one to focus on plasma distribution functions that are one-dimensional (oriented along the magnetic field lines). 

We begin by generalizing the equations of motion for charged particles in an external electromagnetic field. In full generality, the evolution of the distribution function of a species $f_s$ is given by the Vlasov equation~\cite{gedalin1998long,lyutikov1998waves,gedalin2001low}
\begin{eqnarray}
\frac{\partial f_s}{\partial t} + \vec{v}\frac{\partial f_s}{\partial \vec{r}} + \frac{q_s}{m_s}(\vec{E} + \vec{v} \times \vec{B}) \frac{\partial f_s}{d \vec{u}} = 0 \, ,
\end{eqnarray}
where $\vec{u} = \vec{v}\gamma$ is the proper velocity and $\gamma = 1/\sqrt{1-v^2}$. Note that we still take $\vec{B} = \vec{B}_0 + \vec{B}_1$, with the background field $\vec{B}_0$ being oriented along the $z$-axis and much larger than $\vec{B}_1$. The charge density and current density can be defined in terms of the distribution functions using
\begin{gather}
\rho = \sum_s q_s  \int d^3 u \, f_s(\vec{u}) \, , \\
\vec{J} = \sum_s q_s  \int d^3 u \, \vec{v} \, f_s(\vec{u}) \, .
\end{gather}
In order to solve Maxwell's equations, it is useful to expand the phase space distribution in cylindrical velocity coordinates as
\begin{equation}
f_s = \sum_{n = -\infty}^{\infty} \, f_{s, n}(u_\perp, u_z) e^{-i n u_\phi} \, .
\end{equation}
By rewriting the Vlasov equation, this expansion allows one to write a single equation governing the evolution of each mode $f_{s,n}$
\begin{equation}\label{eq:vlasov_modes}
(L_n + i n \tilde{\Omega}) f_n + \sum_{\sigma = \pm 1} G^{n-\sigma}_\sigma f_{n-\sigma} = 0 \, ,
\end{equation}
where the subscript $s$ has been omitted for clarity. The operators $L_n$ and $G^n_{\sigma}$ are given by
\begin{gather}
L_n = \frac{\partial}{\partial t} + v_z \frac{\partial}{\partial z} + \alpha E_z \frac{\partial}{\partial u_z}  + i n \alpha \gamma^{-1} B_z \, , \\
G_\sigma^n = \frac{v_\perp}{2} \nabla_\sigma + \frac{\alpha}{2} (E_\sigma d_{\sigma n} + i \sigma B_\sigma r_{\sigma n}) \, .
\end{gather}
In the above we have introduced the definitions $\alpha = q/m$, $\tilde{\Omega} = \Omega_s / \gamma$, $E_\sigma = E_x + i \sigma E_y$ (the same relation holds for B and $\nabla$), and we have defined
\begin{gather*}
d_{\sigma n} = \frac{\partial}{\partial u_\perp } - \frac{\sigma n}{u_\perp} \, , \\
r_{\sigma n} = v_z d_{\sigma n} - v_{\perp} \frac{\partial }{\partial u_z} \, .
\end{gather*}  
One can also write the charge and current densities in terms of the moments $f_n$, which yield
\begin{gather}
\rho = \sum_s q_s  \int du_\perp du_z \, u_\perp f_{s, 0} \, , \\
J_z = \sum_s q_s  \int du_\perp du_z \, u_\perp v_z f_{s, 0} \, , \\
J_\sigma = J_x + i \sigma J_y = \sum_s q_s  \int du_\perp du_z \, u_\perp v_\perp f_{s, \sigma} \, .
\end{gather}
In this notation, Maxwell's equations can be expressed as
\begin{gather}
\frac{i}{2} \sum \sigma \nabla_\sigma E_{-\sigma} = -\frac{\partial}{\partial t}B_z \, , \\
\frac{\partial}{\partial z}E_\sigma - \nabla_\sigma E_z = \frac{\partial}{\partial t}(i \sigma B_\sigma) \, , \\
\frac{-1}{2} \sum \nabla_\sigma (-i\sigma B_{-\sigma}) = \frac{\partial}{\partial t} E_z + 4\pi J_z \, , \\
\frac{\partial}{\partial z} (i\sigma B_\sigma) - i\sigma \nabla_\sigma B_z = \frac{\partial}{\partial t} E_\sigma + 4\pi J_\sigma \, , \\
\frac{1}{2} \sum \nabla_\sigma E_{-\sigma} +  \frac{\partial}{\partial z }E_z = 4\pi \rho \, , \\
\frac{i}{2} \sum \sigma \nabla_\sigma  (-i\sigma B_{-\sigma})  + \frac{\partial}{\partial z }B_z = 0 \, . 
\end{gather}

Up until now the analysis is still completely general and while the above equations are no easier to solve than the original versions, our notation has clarified that the current density is fully defined by knowing only $f_0$ and $f_\sigma$. In order to identify these terms, we can exploit the low frequency approximation in which a small parameter $\xi \sim \omega\gamma / |\Omega_s| \ll 1$ is introduced through the substitution $\Omega_s \rightarrow \Omega_s / \xi$. With this substitution, \Eq{eq:vlasov_modes} can be written as
\begin{equation}\label{eq:vlasov_long}
\begin{gathered}
L_0 f_0 = - \sum_\sigma G^\sigma_{-\sigma} f_\sigma \, , \\
f_n = \frac{\xi}{i n \tilde{\Omega}} \Bigg[ -L_n f_n - \sum_\sigma G_\sigma^{n - \sigma} f_{n - \sigma} \Bigg] \, \textrm{for} \, \vert n \rvert \geqslant 1 \, .
\end{gathered}
\end{equation}
Due to the equilibrium distribution being gyrotropic~\cite{gedalin1998long}, all $f_n$ with $\lvert n \rvert \geqslant 1$ should vanish if $\vec{E} \rightarrow 0$, $\vec{B} \rightarrow 0$ and $\nabla \rightarrow 0$. Looking at the second of the above equations, this fact tells us that these $f_n$ cannot contain terms dependent on negative powers of $\xi$. Furthermore, the overall form of \Eq{eq:vlasov_long} shows that all terms in a given $f_n$ have to be at least one order of $\xi$ higher than the lowest order term in $f_{n - 1}$. As a result, it is possible to represent the phase space distribution in the following way
\begin{equation}\label{eq:psd_xi}
f_n = \sum_{m = \vert n \rvert}^\infty \xi^m \, f_n^{(m)} \, .
\end{equation}
One thus needs to start by determining $f_0$, and all higher order $f_n$ can subsequently be found through Eq. \eqref{eq:vlasov_long}. We can further expand in the weak turbulence limit, $\eta \sim E / B_0 \ll 1$~\cite{gedalin1998long}, and keep only terms up to orders $\xi^2$ and $\eta$, which is enough for our purposes. This allows us to write $f_0$ and $f_\sigma$ in the general forms
\begin{gather}
f_0 = F_0(u_\perp, u_z) + \eta \sum_{n=0}^2 \xi^n \, f_0^{(n)} \, , \\
f_\sigma = \eta \sum_{n=1}^2 \xi^n \, f_\sigma^{(n)} \, .
\end{gather}
Additionally, working up to order $\xi^2$, and plugging \Eq{eq:psd_xi} into \Eq{eq:vlasov_long} results in
\begin{equation}
\begin{gathered}
L_0 f_0 = \sum_\sigma G^\sigma_{-\sigma} \frac{\xi}{i \sigma \tilde{\Omega}} \Bigg(1 - \frac{\xi L_\sigma}{i \sigma \tilde{\Omega}}\Bigg) G^0_\sigma f_0 \, , \\
f_\sigma = \frac{-\xi}{i \sigma \tilde{\Omega}} G^0_\sigma f_0 + \Bigg(\frac{\xi}{i \sigma \tilde{\Omega}}\Bigg)^2 L_\sigma G^0_\sigma f_0 \, .
\end{gathered}
\end{equation}
If we now switch to Fourier space and assume space-time dependencies given by $\textrm{exp}(i\vec{k}\cdot\vec{r} - i \omega t)$, the above equations can be combined to derive explicit expressions for both $f_0$ and $f_\sigma$. They are given by
\begin{gather}
f_0 = F_0(u_\perp, u_z) - \frac{\alpha k_\perp v_\perp}{2 \tilde{\Omega} \zeta} E_y \mu_0 F_0 + \frac{i\alpha k_\perp v_\perp}{2 \tilde{\Omega}^2 \omega} E_x \mu_0 F_0 + \frac{i \alpha k_\perp^2 v_\perp v_z}{2 \tilde{\Omega}^2 \zeta} E_z \mu_0 F_0 + \frac{i \alpha k_\perp v_\perp}{2 \tilde{\Omega}^2} E_x \mu_0 F_0 \, , \\
f_\sigma = \frac{i \alpha}{2 \sigma \tilde{\Omega}} [E_\sigma + E_z (k_\perp v_z / \zeta)]\mu_0 F_0 + \frac{\alpha k_\perp^2 v_\perp^2}{4 \sigma \tilde{\Omega}^2 \zeta} E_y \mu_0 F_0 + \frac{i\alpha \zeta}{2 \tilde{\Omega}^2}[E_\sigma + E_z (k_\perp v_z/\zeta)]\mu_0 F_0 \, .
\end{gather}
Here we used that $\vec{B} = \vec{k} \times \vec{E} / \omega$ and introduced the parameters $\zeta = \omega - k_z v_z$ and $\mu_0 = [\zeta (\partial / \partial u_\perp) + k_z v_\perp (\partial / \partial u_z)] / \omega$. Notice that the artificial parameter $\xi$ previously introduced to justify the expansion of \Eq{eq:vlasov_long} has also been set to $1$, justifying the validity of the above procedure.

At this point, one can directly compute the current density, and subsequently the dielectric tensor. Pulling out the density normalization from $F_{s,0}$, such that $\int d^3 u F_{s,0} = 1$, the generalized dielectric tensor is given by
\begin{equation}\label{eq:diel_tensor}
\begin{gathered}
\epsilon_{xx} = 1 - \sum_s \frac{\omega_{p, s}^2}{2 \omega \Omega_s^2} \int du_{\perp} du_z \, u_{\perp}^2 \gamma \zeta \mu_0 F_{s, 0} \, , \\
\epsilon_{yy} = \epsilon_{xx} + \sum_s \frac{\omega_{p, s}^2 k_{\perp}^2}{4 \omega \Omega_s^2} \int du_{\perp} du_z \, u_{\perp}^4 \gamma^{-1} \zeta^{-1} \mu_0 F_{s, 0} \, , \\
\epsilon_{zz} = 1 + \sum_s \frac{\omega_{p, s}^2}{\omega} \int du_{\perp} du_z \, u_{\perp} v_z \zeta^{-1} \frac{\partial F_{s, 0}}{\partial u_z} - \sum_s \frac{\omega_{p, s}^2 k_{\perp}^2}{2 \omega \Omega_s^2} \int du_{\perp} du_z \, u_{\perp}^2 u_z^2 \gamma^{-1} \zeta^{-1} \mu_0 F_{s, 0} \, , \\
\epsilon_{xy} = -\epsilon_{yx} = i\sum_s \frac{\omega_{p, s}^2}{2 \omega \Omega_s} \int du_{\perp} du_z \, u_{\perp}^2 \mu_0 F_{s, 0} \, , \\
\epsilon_{xz} = \epsilon_{zx} = -\sum_s \frac{\omega_{p, s}^2 k_{\perp}}{2 \omega \Omega_s^2} \int du_{\perp} du_z \, u_{\perp}^2 u_z \mu_0 F_{s, 0} \, , \\
\epsilon_{yz} = -\epsilon_{zy} = i\sum_s \frac{\omega_{p, s}^2 k_{\perp}}{2 \omega \Omega_s} \int du_{\perp} du_z \, u_{\perp}^2 u_z \zeta^{-1} \mu_0 F_{s, 0} \, , \\
\end{gathered}
\end{equation}
Focusing on leading order terms in the low frequency limit, Eq. \eqref{eq:diel_tensor} reduces to
\begin{equation}\label{eq:diel_tensor2}
\begin{gathered}
\epsilon_{xx} = \epsilon_{yy} = 1 + \mathcal{O}\left(\frac{\omega_{p, s}^2}{\Omega_s^2}\right) \, , \\
\epsilon_{zz} =  1 + \sum_s \frac{\omega_{p, s}^2}{\omega} \int du_{\perp} du_z \, u_{\perp} v_z \zeta^{-1} \frac{\partial F_{s, 0}}{\partial u_z} + \mathcal{O}\left(\frac{\omega_{p, s}^2}{\Omega_s^2}\right) \, , \\
\epsilon_{xy} = -\epsilon_{yx} = \mathcal{O}\left(\frac{\omega_{p, s}^2}{\omega\Omega_s}\right) \, , \\
\epsilon_{xz} = \epsilon_{zx} =  \mathcal{O}\left(\frac{\omega_{p, s}^2}{\Omega_s^2}\right) \, , \\
\epsilon_{yz} = -\epsilon_{zy} = \mathcal{O}\left(\frac{\omega_{p, s}^2}{\omega\Omega_s}\right) \, . \\
\end{gathered}
\end{equation}
Some of these corrections contain thermal terms that can partially remove the suppression. Near the neutron star surface where the thermal corrections may be large, however, one expects the cyclotron frequency to be $\Omega_{e} \sim 4 \times 10^{6}$ eV, which is about $10$ orders of magnitude larger then the maximum frequency we are interested in. It therefore seems safe to neglect the higher order terms in Eq. \eqref{eq:diel_tensor2}, even in the presence of large thermal corrections. This means that in the end we will only need to evaluate one term, namely
\begin{equation}
I \equiv \sum_s \frac{\omega_{p, s}^2}{\omega} \int du_{\perp} du_z \, u_{\perp} v_z \zeta^{-1} \frac{\partial F_{s, 0}}{\partial u_z} \, .
\end{equation}

\begin{figure}
	\includegraphics[width=0.5\textwidth, trim={0cm 0cm 0cm 0cm},clip]{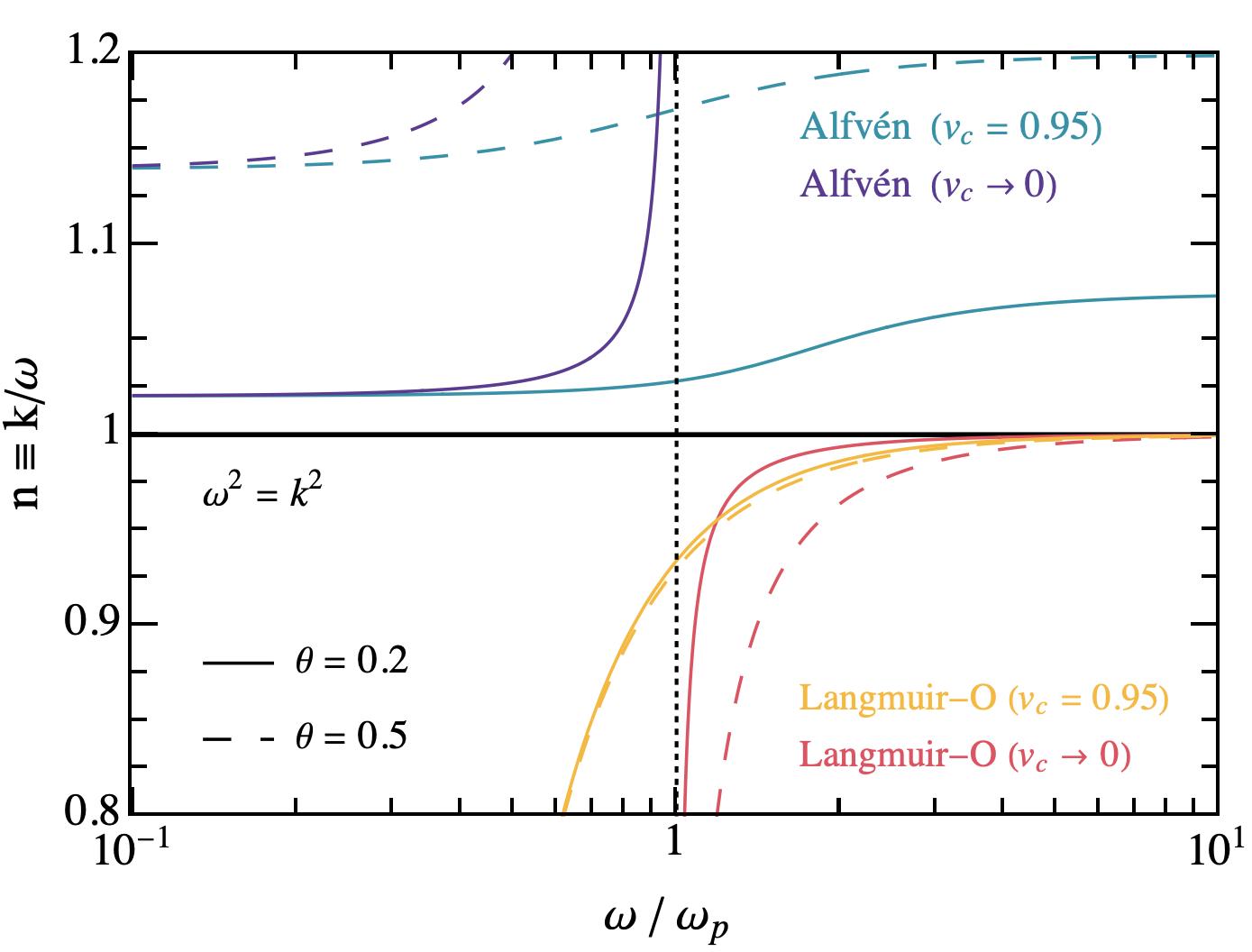}
	\caption{\label{fig:modes} Comparison of the three electromagnetic modes present in the infinite magnetic field limit (note that the magnetosonic-t mode in this limit is synonymous with the free space mode, $\omega^2 = k^2$). The Alfv\'{e}n and Langmuir-O modes are shown for waves propagating at an angle $\theta = 0.2$ (solid) and $0.5$ (dashed) radians with respect to the magnetic field. Both modes are shown taking the limit of a non-relativistic plasma (red, purple) and for a Waterbag distribution with cut-off velocity $v_c = 0.95$ (see \Eq{eq:waterbag}).  The vertical dotted line identifies the branch cut in the non-relativistic Langmuir-O mode arising at $\omega = \omega_p$.  }
\end{figure}

To this end, we start by taking the thermal plasma distribution to be one-dimensional, \ie $F_{s, 0}(u_{\perp}, u_z) = \tilde{F}_{s, 0}(u_z) \delta(u_\perp)/u_\perp$ with normalization $\int du_z \tilde{F}_{s, 0} = 1$. Computing this remaining function yields
\begin{equation}\label{eq:I}
I(\omega, \theta, n) = \sum_s \frac{\omega_{p, s}^2}{\omega^2} \int du_z \, \frac{1}{n_\parallel} \frac{1}{1 - n_\parallel v_z} \frac{d \tilde{F}_{s, 0}}{d u_z} \, ,
\end{equation}
where we have defined $n_\parallel = k_z / \omega = n \cos\theta$. To further simplify this expression one must define the shape of the one-dimensional distribution function. Since we are interested in studying the general behavior or the dispersion relation, we adopt an intuitively straightforward distribution for both electrons and positrons called the Waterbag distribution~\cite{arons1986wave}. This assumes a linear flat distribution in velocities up to a cut-off $u_{c}$, \ie
\begin{equation}\label{eq:waterbag}
\tilde{F}_{s, 0}  = \frac{1}{2 u_{c, s}} \, \Theta(u_{c, s}^2 - u_z^2) \, .
\end{equation}
With this assumption, the integral in \Eq{eq:I} becomes
\begin{equation}
I(\omega, \theta, n) = -\sum_s \frac{\omega_{p, s}^2}{\omega^2}\frac{1}{\gamma_{c, s}(1 - n_{||}^2 \, v_{c, s}^2)} \, ,
\end{equation}
with $\gamma_c$ and $v_c$ being the cut-off gamma factor and velocity respectively. We can now directly solve for the dispersion relations using \Eq{eq:det}. In this case, the magnetosonic-t modes remain unaltered by boosts, but the roots associated with the Alfv\'{e}n and Langmuir-O are now obtained by solving the following
\begin{equation}
\omega^2 = \frac{(n_{||}^2 - 1) \, \cos^2\theta}{n_{||}^2 - \cos^2\theta} \sum_s \frac{\omega_{p, s}^2}{\gamma_{c, s}(1 - n_{||}^2 \, v_{c, s}^2)} \, .
\end{equation}
One can nicely see that in the limit $v_c \rightarrow 0$ and $\gamma_c \rightarrow 1$, the non-relativistic dispersion relations for the Alfv\'{e}n and Langmuir-O modes are recovered. An alternative limit of interest is that of a single relativistic species. In this case, the dispersion relation of the Langmuir-O mode is given by
\begin{equation}\label{eq:disp_rel}
\omega^2 = \frac{1}{2\gamma_c^2} \left(k^2 (\gamma_c^2 + \cos^2\theta(\gamma_c^2 - 1)) + \gamma_c \omega_p^2 + \sqrt{k^4 (\gamma_c^2 - \cos^2\theta(\gamma_c^2 - 1))^2 - 2 k^2 \omega_p^2 \gamma_c (\cos^2\theta  + \gamma_c^2 (\cos^2\theta - 1)) + \gamma_c^2 \omega_p^4}\right) \, .
\end{equation}
Notice that in the limit where $\cos \theta \ll 1$, photons follow the non-relativistic dispersion relation with the replacement $\omega_p \rightarrow \omega_{p} / \sqrt{\gamma_c}$. In addition, in the limit $\gamma \rightarrow \infty$, one recovers the free space distribution $\omega^2 = k^2$. This leads to the important observation that the inclusion of a boosted plasma is likely to reduce the importance of the various effects studied in this work. 

\begin{figure}
	\includegraphics[width=0.32\textwidth]{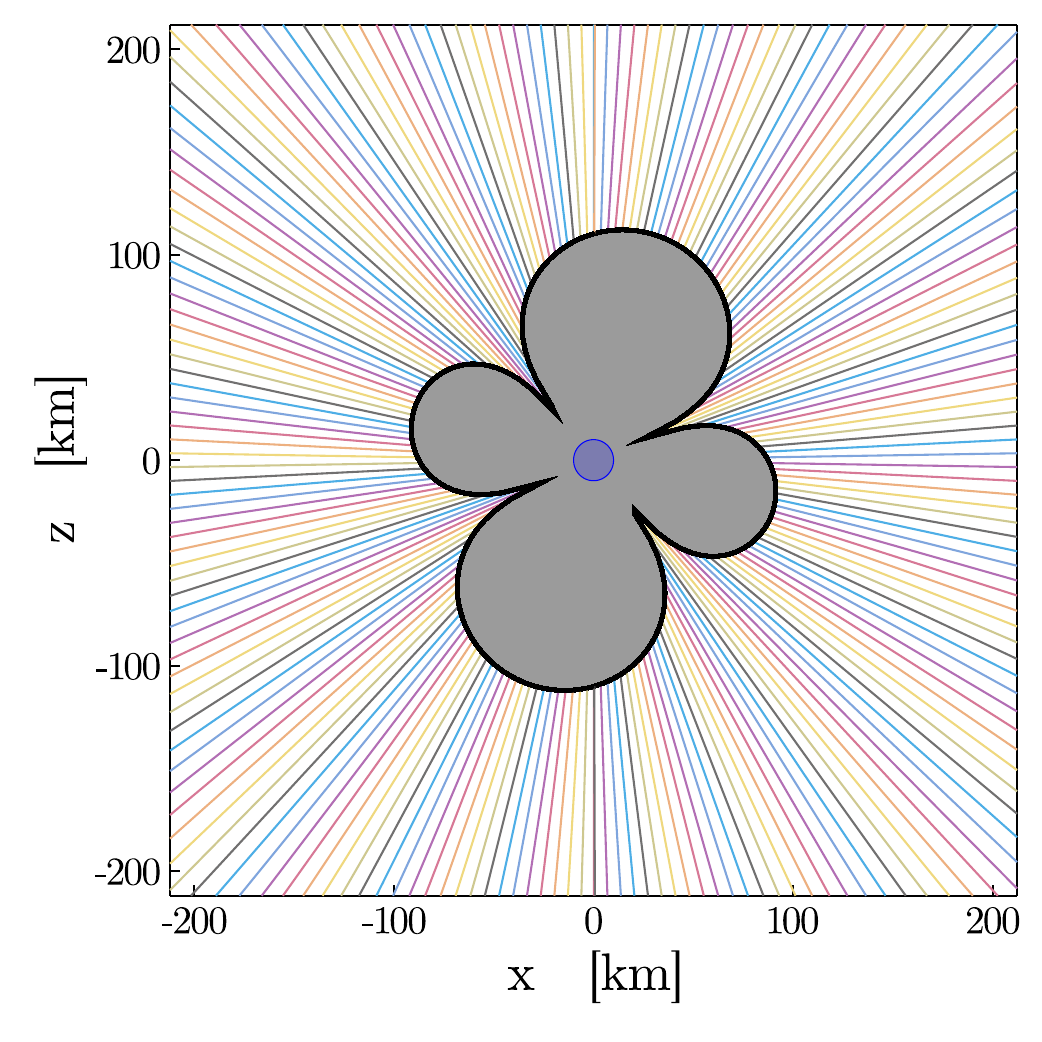}
	\includegraphics[width=0.32\textwidth]{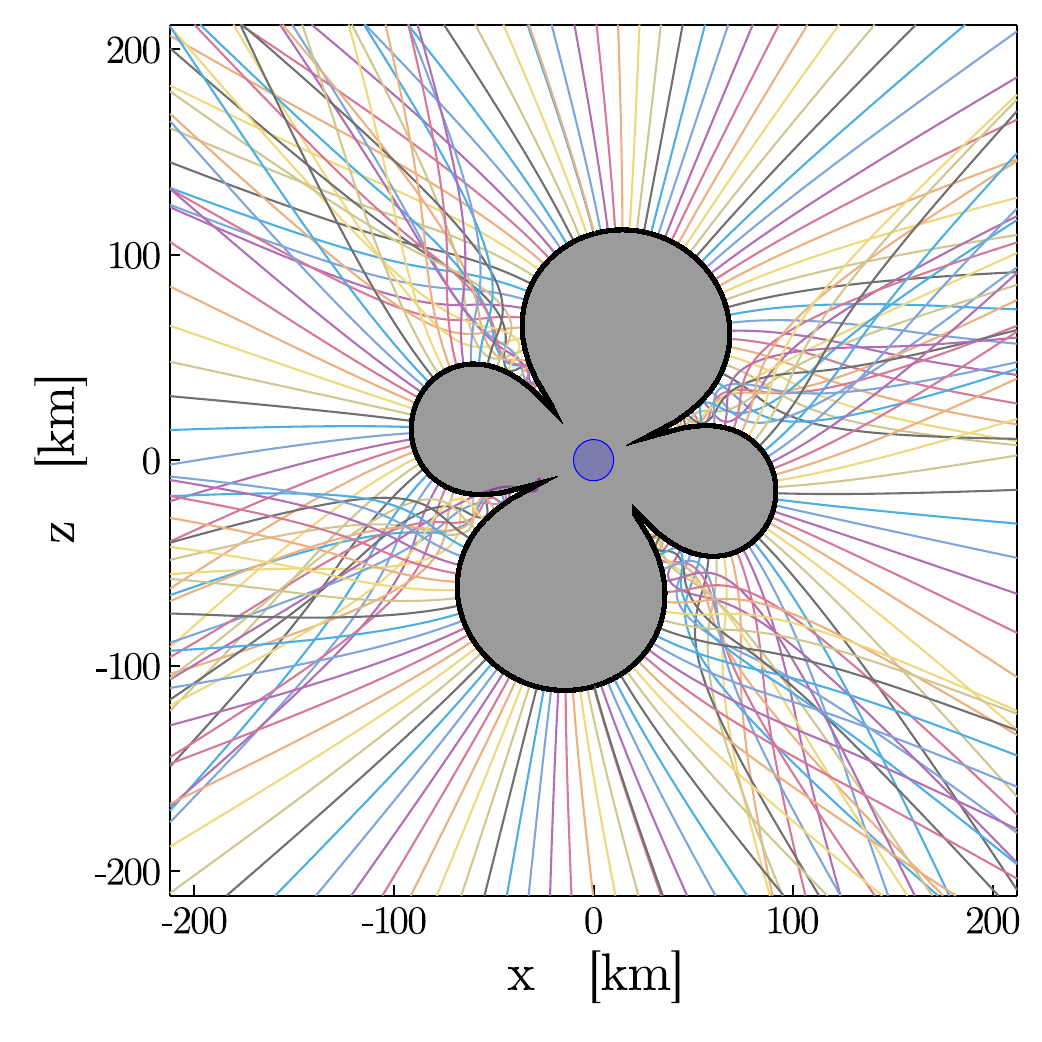}
	\includegraphics[width=0.32\textwidth]{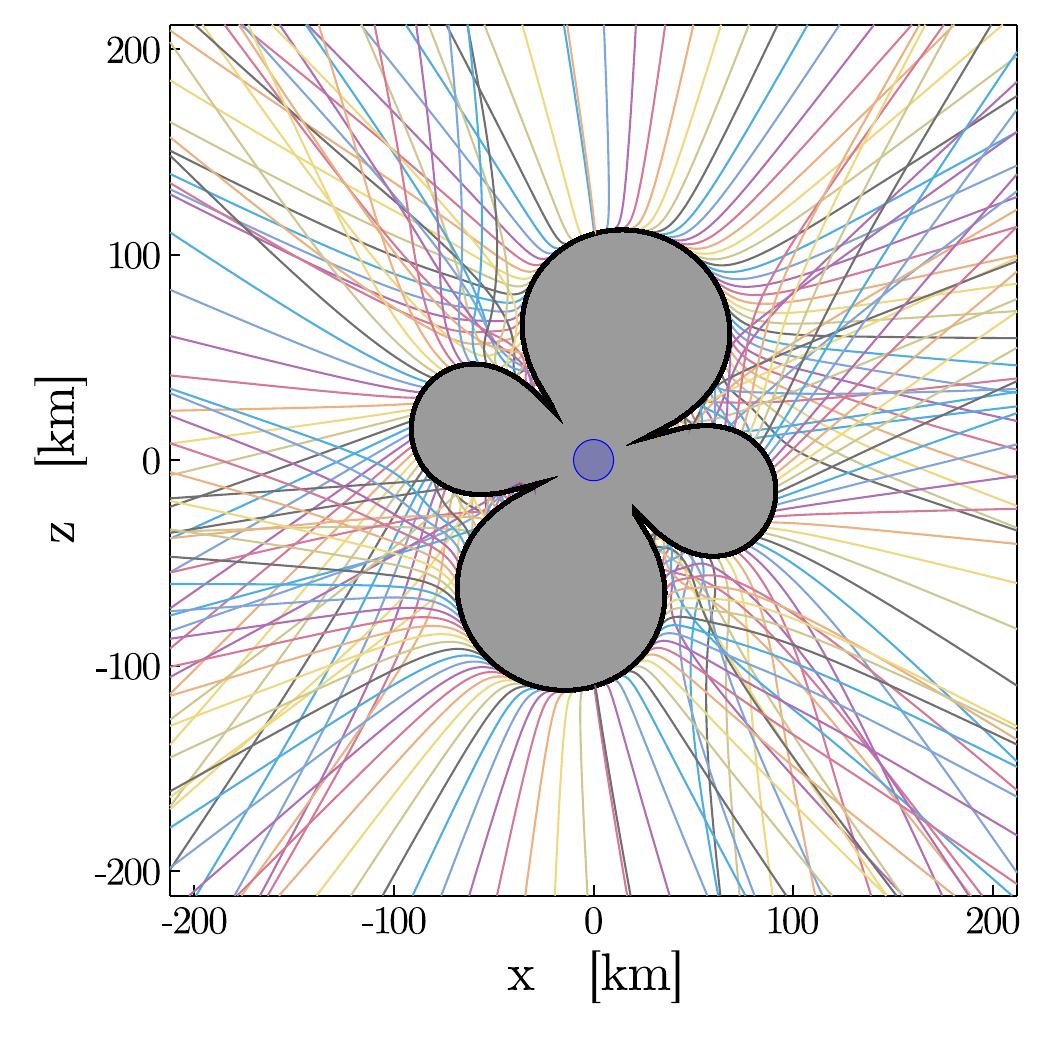}
	\caption{\label{fig:Squig_Ex} Result of ray-tracing photon trajectories from source location at the conversion surface (black contour) of a neutron star, assuming a GJ model of the magnetosphere. Photons are sourced in the $x-z$ plane (with $\vec{k} = 10^{-3} m_a \hat{r}$), allowed to propagate in three dimensions, and projected back into the $x-z$ plane. The three panels correspond to dispersion relations $\omega^2 = k^2$, $\omega^2 = k^2 + \omega_p^2$, and \Eq{eq:nr_dspR}. 	}
\end{figure}

We demonstrate the behavior of the various modes that appear in the strong magnetized limit in \Fig{fig:modes}. Here, we plot the spectral index as a function of the frequency (in units of $\omega_p$) for each of the modes at angles of $\theta = 0.2$ and $0.5$ radians with respect to the magnetic field. For both the Alfv\'{e}n and Langmuir-O modes, we show the impact of taking a Waterbag distribution with cut-off velocity $v_c = 0.95$; the primary effect is to lower (raise) the branch cut of the Langmuir-O (Alfv\'{e}n) mode, such that a larger range of frequencies can propagate. The vertical dotted line highlights the branch cut that appears in the Langmuir-O mode at $\omega = \omega_p$. Photons sourced from axion conversion on the Langmuir-O mode trajectories will tend to travel to the right in the phase diagram as they escape the magnetosphere, and will quickly tend toward to the free space dispersion relation. Importantly, \Fig{fig:modes} illustrates that modes do not cross, and thus remain well-defined and isolated in the strong field limit.

In order to illustrate the importance of adopting the correct dispersion relation -- and in tracking the individual photon trajectories -- we show in \Fig{fig:Squig_Ex} the evolution of photons in the $x$-$z$ plane propagating away from a neutron star conversion surface using either the free space dispersion relation $\omega^2 = k^2$ (left), cold plasma dispersion relation $\omega^2 = k^2 + \omega_p^2$ (center), or the non-relativistic highly-magnetized plasma dispersion relation given in \Eq{eq:nr_dspR} (right). All photons are sourced with $\vec{k} = 10^{-3} m_a \, \hat{r}$, where $\hat{r}$ is the unit vector directed from the origin toward the point of genesis. One can see that the angular inhomogeneities in the plasma frequency of the GJ model generate enormous anisotropic features in the photon trajectories. The impact of accounting for the magnetized nature of the plasma is more subtle, however; this effect is most apparent near the magnetic poles, where photon trajectories can experience strong refraction.

\section{Cyclotron Resonance}\label{sec:cyc_res}

\begin{figure}
	\includegraphics[width=0.6\textwidth]{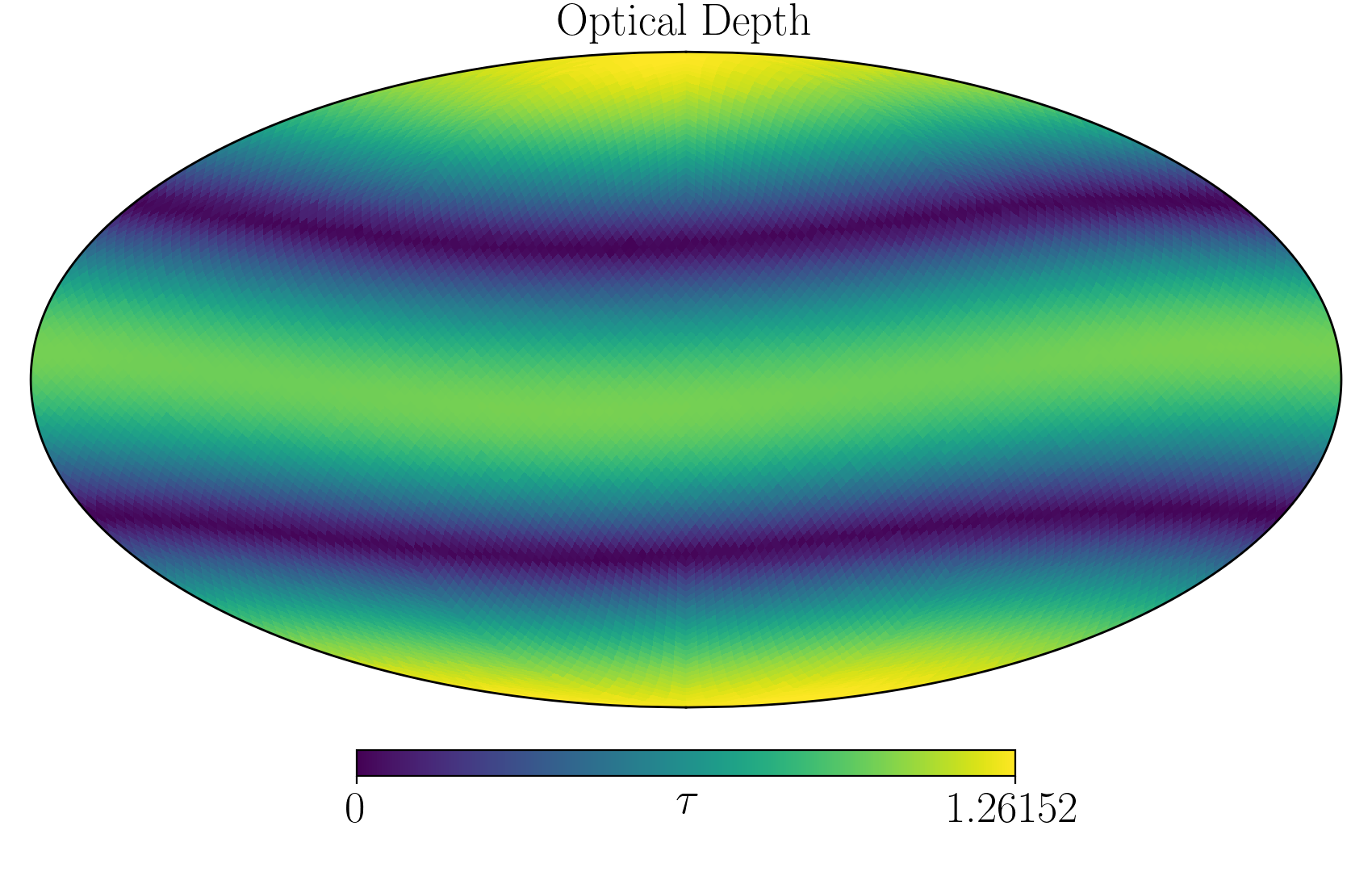}
	\caption{\label{fig:tau}Optical depth $\tau$ for neutron star with $B_s = 5 \times 10^{14}$ G, $\omega_{NS} = 1 \, {\rm s}^{-1}$, $\theta_m = 0.2$, and for an axion mass $m_a = 10^{-6}$ eV.  Since the rate is suppressed by $e^{-\tau}$, one can see that for these parameters the optical depth is beginning to suppress the flux coming from the lobes and torus, but leaves the flux coming from the throats effectively untouched. }
\end{figure}

The dominant absorption process of low energy radio photons escaping the magnetosphere occurs via the cyclotron resonance, in which electrons in the ambient plasma are excited to higher Landau orbitals. For strong magnetic fields and relativistic plasmas, this occurs when~\cite{blandford1976scattering,melrose1979propagation,luo2001cyclotron}
\begin{equation}\label{eq:cycl}
\omega - k_{||} v_{||} - \Omega_e / \gamma = 0 \, ,
\end{equation}
where $k_{||} = k \cos\tilde{\theta}$,  $v_{||}$ is the velocity of the plasma in the direction of the magnetic field, and $\Omega_e$ is the electron cyclotron frequency. Since the magnetic field falls off as $B \propto r^{-3}$, one expects all photons to cross the cyclotron resonance  -- for non-relativistic plasmas this occurs at a fixed distance, while for relativistic distributions this may take place over an extended region.
We show here that, in the simplified context of the GJ model, the cyclotron resonance is not necessarily negligible. We begin by noting that for non-relativistic plasmas the second term in \Eq{eq:cycl} can be neglected, and $\gamma$ can be set to one. The optical depth of photons scattering of ambient electrons and positrons is given by
\begin{equation}
\tau = \int \, d\ell \, \sigma \, n_e \, ,
\end{equation}
where the cross section (assuming $\omega \ll m_e$) for non-relativistic cyclotron absorption is given by~\cite{blandford1976scattering}
\begin{equation}
\sigma = (2\pi)^2 \, r_e \, \delta \left(\omega - \Omega_e \right) \, .
\end{equation}
Here, $r_e \equiv \alpha / m_e$ is the classical electron radius. Assuming the photon frequency is approximately constant, we find
\begin{equation}
\tau = \pi  \left(\frac{\omega_p^2}{\omega} \right) \left. \frac{\omega }{|\partial_\ell \Omega_e|} \right|_{\ell = \ell_c} \, .
\end{equation}
If one assumes outward radial trajectories this can be approximated as
\begin{equation}
\tau \sim \frac{\pi}{3} \left(\frac{\omega_p^2}{\omega} \right) \, \ell_c \, .
\end{equation}
Here, $\ell_c$ is the location of the resonance along the path. Importantly, in the above we have assumed that the resonance condition is met within the light cone with radius $R_{LC} = 1 / \omega_{NS}$, otherwise we set the optical depth to zero. 
We illustrate the optical depth $\tau$ as a function of sky position in \Fig{fig:tau} for a strong magnetic field model with $B_s = 5 \times 10^{14}\,$G. In this case, photons originating from either the bulge or the torus have an optical depth $\tau \sim 1$, and thus experience a suppression of the flux on the level of $\sim 60\%$. This effect can be far more extreme for larger magnetic fields and larger rotational speeds.

\section{Revisiting the Conversion Probability} \label{sec:converP}
An important result of this paper is the correction of the conversion probability adopted in previous works (see main text). There are two problems which have gone overlooked: $(i)$ the angular contribution to the conversion length, and $(ii)$ the de-phasing of the photon and axion wave functions which results from non-linear photon propagation. In order to be clear on the origin of each effect, we review the derivation of the conversion probability before highlighting where the novel effects enter.
The equations detailing the propagation of electromagnetic waves in the presence of axions, at leading order in the axion-photon coupling constant $g_{a\gamma\gamma}$, are given by
\begin{gather}\label{eq:couples}
-\nabla^2 \vec{E} + \nabla(\nabla \cdot \vec{E}) = - \partial_t^2 (\boldsymbol\epsilon \cdot \vec{E}) - g_{a\gamma\gamma} \vec{B}_0 \partial_t^2 a \, , \\
(\partial_t^2 - \nabla^2 + m_a^2)a = g_{a\gamma\gamma} \vec{E} \cdot \vec{B}_0 \, \nonumber ,
\end{gather}
where $m_a$ is the axion mass, $\boldsymbol\epsilon$ is the dielectric tensor, and $\vec{B}_0$ is the local magnetic field. These expressions are derived from Maxwell's equations coupled to axions, assuming a constant background magnetic field $\vec{B}_0$. If we furthermore work in the high magnetization and low frequency limit, \ie $\lvert \Omega_e \rvert \gg \omega, \omega_p$, the dielectric tensor takes the form (see \App{sec:photonDisp} for derivation)
\begin{equation}\label{eq:displ}
\boldsymbol\epsilon = R_\theta^{xz} \cdot \begin{pmatrix}
1 & 0 & 0 \\
0 & 1 & 0 \\
0 & 0 & \epsilon_{zz}
\end{pmatrix} \cdot R_{-\theta}^{xz} \, ,
\end{equation}
where $\epsilon_{zz} = 1 - \omega_p^2/\omega^2$ and the rotation matrix is given by
\begin{equation}
R_\theta^{xz} = \begin{pmatrix}
\cos\theta & 0 & \sin\theta \\
0 & 1 & 0 \\
-\sin\theta & 0 & \cos\theta 
\end{pmatrix} \, .
\end{equation}
Without loss of generality we have taken the external magnetic field to lie in the first quadrant of the $x$-$z$ plane, at an angle $\theta$ from the $z$-axis. Moreover, we choose the photon and axion to travel along the $z$-axis. Notice that this is a different orientation relative to Appendix~\ref{sec:photonDisp}, where previously we had chosen the magnetic field to be oriented along the $z$-axis and taken the photon momentum to be in the $x$-$z$ plane. Nevertheless, the choice is irrelevant as our results only depend on $\theta$. After computing the dielectric tensor in Eq.~\eqref{eq:displ}, it becomes clear that the photon polarization perpendicular to the external field ($E_y$) fully decouples. Furthermore, we find that the solution for $E_z$ can be re-expressed in terms of $E_x$. Assuming an oscillatory time dependence with frequency $\omega$, one then obtains the following equation for the mixing between the axion and the $E_x$ component of the electric field
\begin{equation}\label{eq:ap_mixing}
-\partial_z^2 \begin{pmatrix}
E_x \\
a
\end{pmatrix} = \begin{pmatrix}
\frac{\omega^2 - \omega_p^2}{1 - \frac{\omega_p^2}{\omega^2} \cos^2 \theta} & \frac{\omega^2 g_{a\gamma\gamma} B_0 \sin \theta}{1 - \frac{\omega_p^2}{\omega^2} \cos^2 \theta} \\
\frac{g_{a\gamma\gamma} B_0 \sin \theta}{1 - \frac{\omega_p^2}{\omega^2} \cos^2 \theta} & \omega^2 - m_a^2
\end{pmatrix} \cdot \begin{pmatrix}
E_x \\
a
\end{pmatrix} \, .
\end{equation}
Importantly, the general solution thus contains both a transverse and a longitudinal component, as expected for a Langmuir-O mode propagating at an oblique angle with respect to $\vec{B}_0$. The longitudinal component will naturally be damped (evolving into a fully transverse O mode) as the wave propagates away from the neutron star.

\begin{figure}
	\includegraphics[width=0.8\textwidth, trim={10cm 18cm 0cm 0cm}, clip]{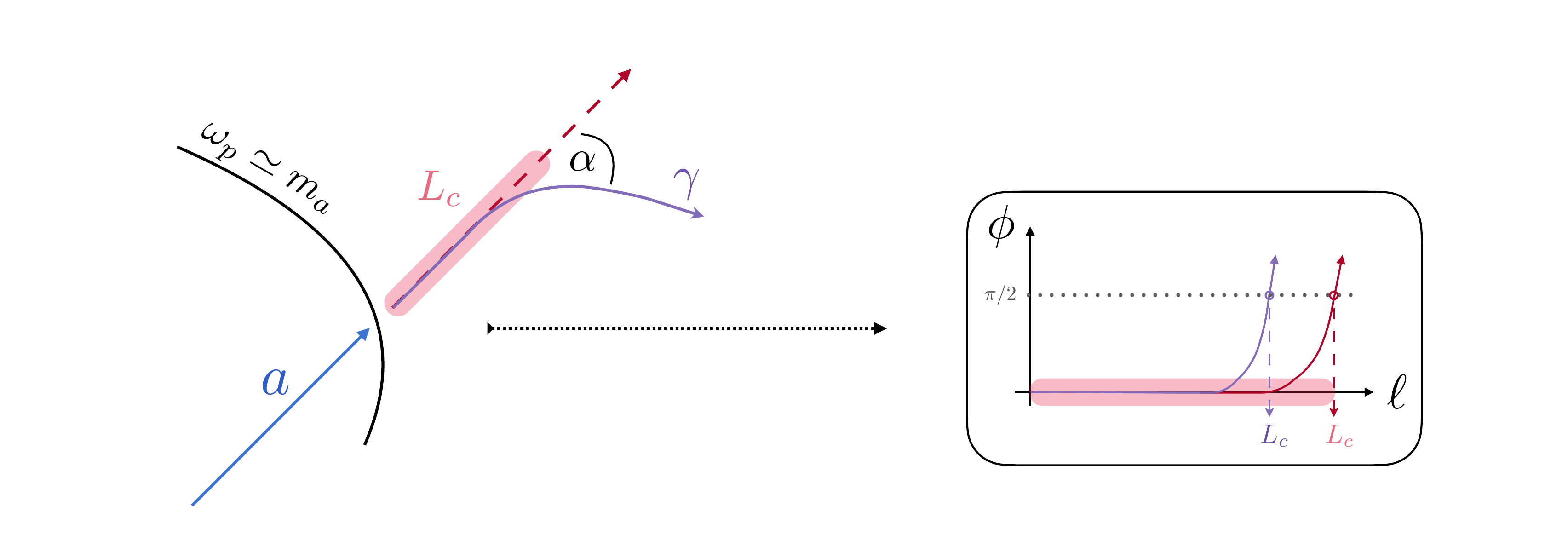}
	\caption{\label{fig:dephase} Illustration of de-phasing introduced from non-linear axion trajectories. In the context of the one-dimensional formalism, the conversion length is computed from the de-phasing induced along straight-line trajectories. Should the axion trajectory deviate from this trajectory by an angle $\alpha$, additional de-phasing will be introduced and the conversion length will be shortened. }
\end{figure}

After changing variables to the vector potential $A_x = - E_x / i \omega$, it is possible to adopt a plane wave ansatz and express both fields as
\begin{equation}\label{eq:ansatz}
a(z,t) = a_0 \ e^{i\xi(z) - i \omega t}, \hspace{20pt} A_x(z,t) = \frac{A_{x,0}}{\sqrt{k_{\gamma}(z)}} \ e^{i\chi(z) - i \omega t} \, .
\end{equation}
Notice here that the photon momentum can be inferred from the dispersion relation derived in Appendix~\ref{sec:photonDisp}, and in the non-relativistic limit it is given by
\begin{equation}\label{eq:kgam}
k_\gamma(z) = \sqrt{\frac{\omega^2 - \omega_p^2(z)}{1 - \frac{\omega_p^2(z)}{\omega^2} \cos^2 \theta(z)}} \, .
\end{equation}
The seemingly arbitrary factor $k_{\gamma}^{-1/2}$ present in the photon amplitude emerges from the WKB approximation (see \eg~\cite{swanson2012plasma}). In the spirit of~\cite{Adler_2008}, we now expand the functions $\xi(z)$ and $\chi(z)$ in powers of the magnetic field as
\begin{equation}
\begin{gathered}
\xi(z) = \sqrt{\omega^2 - m_a^2}z + \xi^{(1)}(z) + \ldots \, , \\
\chi(z) = \int^z_0 dz' \  k_{\gamma}(z') + \chi^{(1)}(z) + \ldots \, ,
\end{gathered}
\end{equation}
where we have accounted for the possibility of a strong positional dependence in the photon momentum. Employing both this expansion and the WKB approximation allows us to solve \Eq{eq:ap_mixing} for $\chi^{(1)}$, yielding
\begin{equation}
\chi^{(1)}(z) = \frac{i \omega a_0}{2 A_{x,0}} \int_0^z dz' \frac{\beta(z')}{\sqrt{k_{\gamma}(z')}} \ e^{i \int_0^{z'} dz'' (\sqrt{\omega^2 - m_a^2} - k_{\gamma}(z''))} \, ,
\end{equation}
where 
\begin{equation}
\beta(z) \equiv \frac{g_{a\gamma\gamma} B_0(z) \sin \theta(z)}{1 - \frac{\omega_p^2(z)}{\omega^2} \cos^2 \theta(z)} \, .
\end{equation}
With this expression, one can write the photon field at a distance $z$ (given a negligible initial field value) as 
\begin{equation}\label{eq:Afield}
A_{x}(z,t) = -\frac{\omega a_0}{2\sqrt{k_{\gamma}(z)}} \ e^{i\int^z_0 dz' k_{\gamma}(z') - i \omega t} \int_0^z dz' \frac{\beta(z')}{\sqrt{k_{\gamma}(z')}} \ e^{i \int_0^{z'} dz'' (\sqrt{\omega^2 - m_a^2} - k_{\gamma}(z''))} \, .
\end{equation}
Since the energy flux of a plane wave scalar field scales like $f \propto k_a^2|a|^2$, with $k_a$ the momentum of the axion field, the conversion probability can be directly obtained from~\cite{Leroy:2019ghm,Millar:2021gzs}\footnote{While this manuscript was in review, Ref.~\cite{Millar:2021gzs} appeared, revisiting the derivation of the conversion probability in three dimensions. This work has identified two significant corrections: (i) the factor of $1 / \sin^2\theta$ appearing in \Eq{eq:pafull}, arising from the longitudinal component of the vector potential, and (ii) the role of off-diagonal derivatives in \Eq{eq:couples}. We include the former factor, but defer the inclusion of the second modification to future work as this would also require revisiting the de-phasing calculation.  }
\begin{equation}\label{eq:pafull}
P_{a\rightarrow \gamma} = \frac{k_\gamma(z)^2}{k_a^2}\frac{|A_x(z,t)|^2 + |A_z(z,t)|^2 }{|a(0,t)|^2} = \frac{1}{\sin^2\theta}\left| \frac{A_x(z,t)}{a(0,t)} \right|^2 \, . 
\end{equation}
Notice that because conversion only takes place when $k_\gamma(z) \simeq k_a$, the pre-factors approximately cancel. Equation~\eqref{eq:Afield} can be integrated using the stationary phase approximation, where the phase in the integral
\begin{equation} \label{eq:phase_simple}
\phi \equiv \int_0^{z} dz' \, (k_a -  k_\gamma(z')) \, 
\end{equation}
is expanded about  $\partial \phi / \partial z = 0$. Keeping the leading order term in this Taylor expansion and performing both integrals in \Eq{eq:Afield}, one arrives at
\begin{equation}
P_{a\rightarrow \gamma} = \frac{\pi}{2 v_c^2} \left(\frac{g_{a\gamma\gamma} \, B}{\sin\theta}\right)^2   \, |\partial_z  k_{\gamma}|^{-1} \frac{1}{\sin^2\theta}\, ,
\end{equation}
where $v_c$ is the axion velocity at the conversion point. Importantly, we periodically find that conversion lengths of individual axions can be quite large -- notice that this is actually quite problematic since our formalism is not valid in this regime. In order to avoid spurious features, we thus cut out photons whose conversion lengths exceed 1 km. This threshold is somewhat arbitrary, but any reasonable change in this threshold introduces effectively no change in the physical observables. 

Now, as mentioned in the main text, the second derivative of the phase (and thus also the conversion probability) derived in \cite{Hook:2018iia,Battye:2019aco,Leroy:2019ghm} has been truncated at leading order in velocity. The derivative of $\theta$, however, only appears at next-to-leading order. If the axion speed at the conversion surface were always small this would not be a problem, but for conversions near the radius of the neutron star the speed can be as large as $v_c \sim 0.5$. Consequently, the next-to-leading order correction can be rather significant, particularly for particles with non-radial orbits. Working up to second order in the velocity expansion, one finds
\begin{equation}
\frac{\partial k_\gamma}{\partial z} \simeq \frac{1}{v_c}\bigg[\bigg(\frac{1}{ \sin^2 \theta} - \frac{v_c^2}{\tan^2 \theta}\bigg) \frac{\partial \omega_p}{\partial z} + \frac{m_a v_c^2}{\tan \theta} \frac{\partial \theta}{\partial z}\bigg] \, .
\end{equation}
This is related to the conversion length $L_c$ via
\begin{equation}
L_c = \sqrt{\pi} \left|\frac{\partial k_{\gamma}}{\partial z} \right|^{-1/2} \, .
\end{equation}
Notice that if one keeps only the leading order in $v_c$, considers perpendicular propagation (\ie $\sin\theta = 1$), and adopts the radial trajectory approximation of $\partial \omega_p / \partial z$, this result reproduces the conversion probabilities derived in Refs.~\cite{Safdi:2018oeu,Leroy:2019ghm,Battye:2019aco}. Here, we improve upon these approximations by directly calculating both directional derivatives using auto-differentiation for each trajectory of interest, and find that deviations from the radial trajectory approximation can be significant.

\begin{figure*}
    \includegraphics[width=0.32\textwidth]{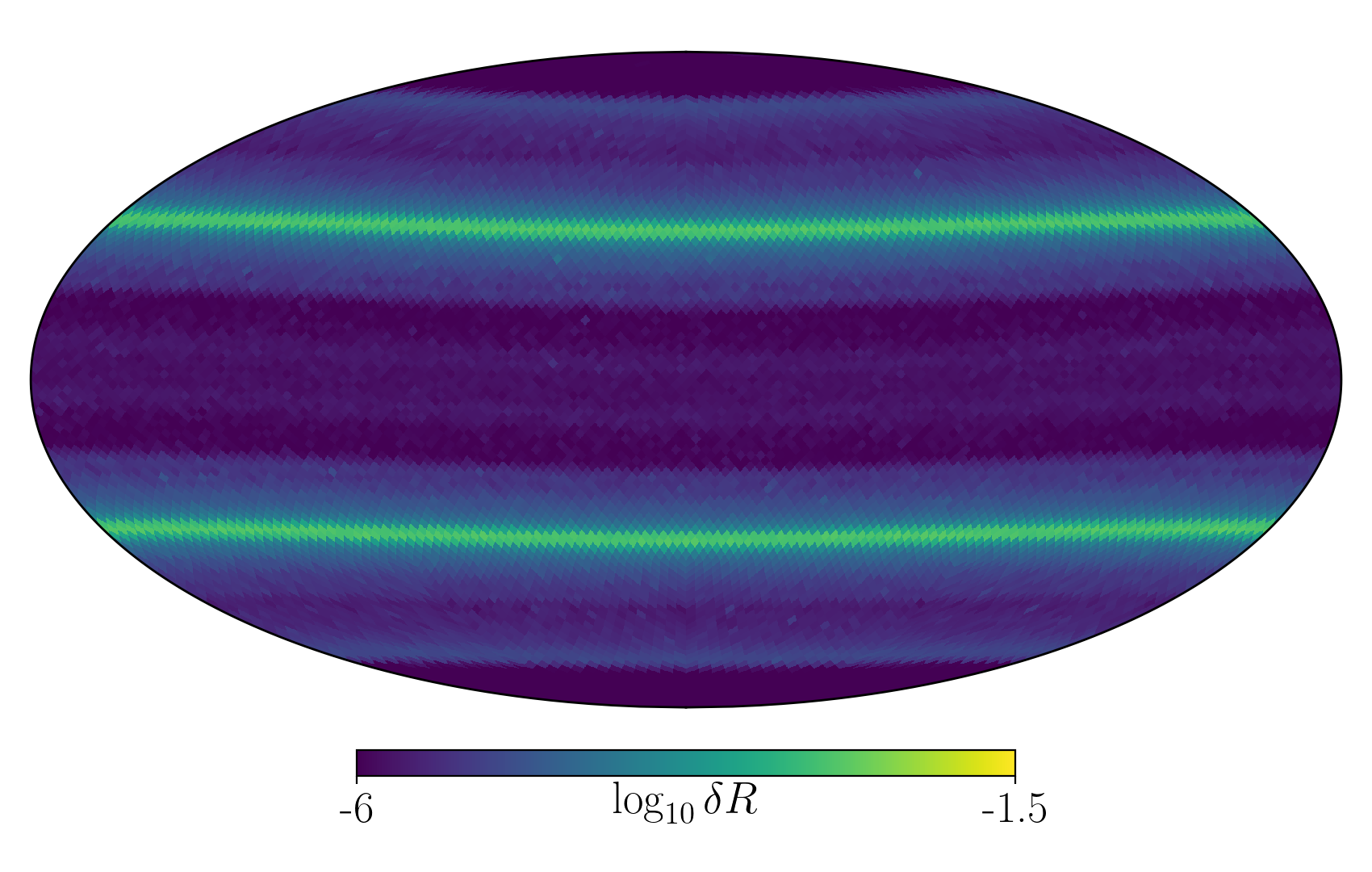}
	\includegraphics[width=0.32\textwidth]{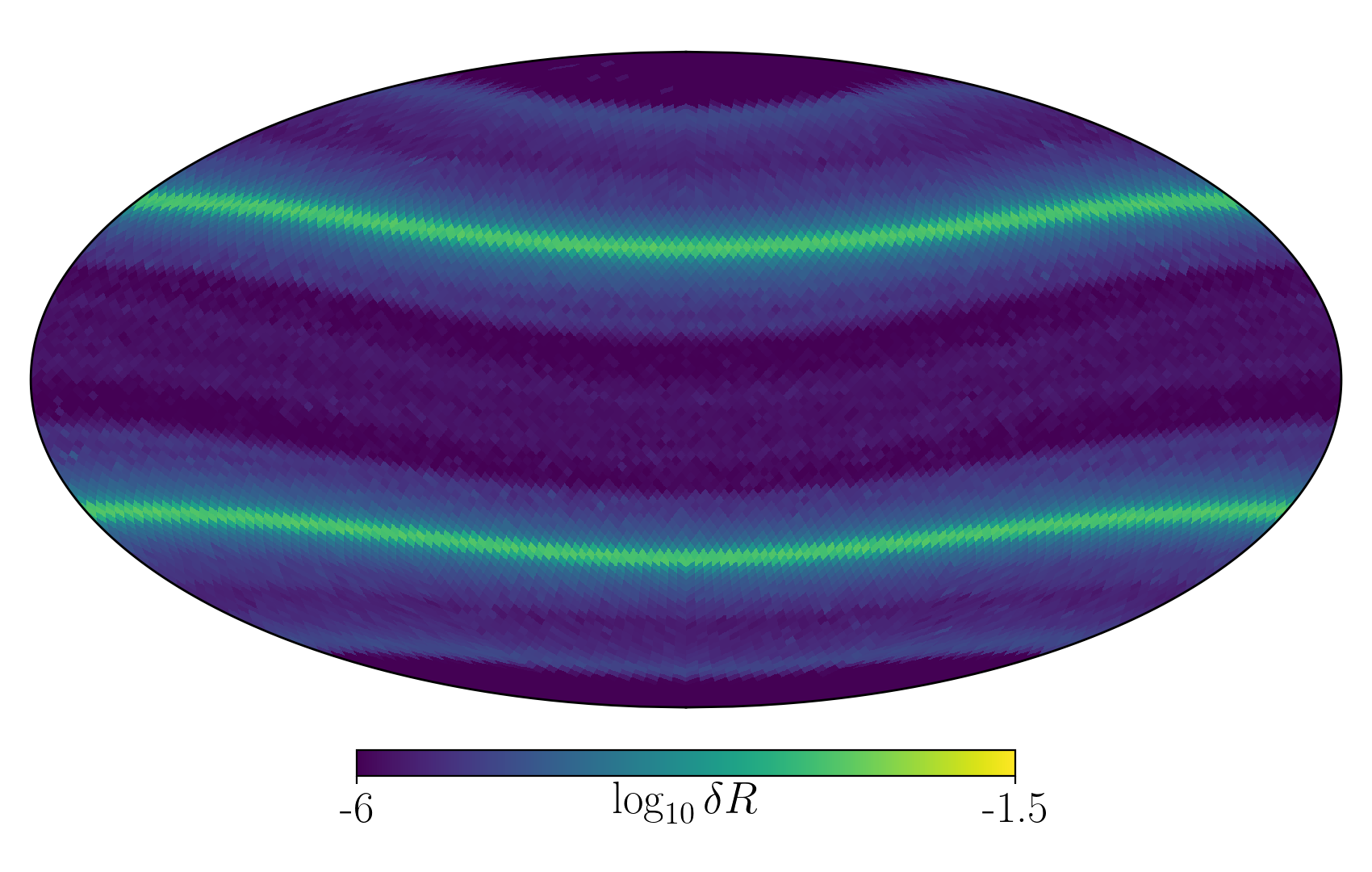}
	\includegraphics[width=0.32\textwidth]{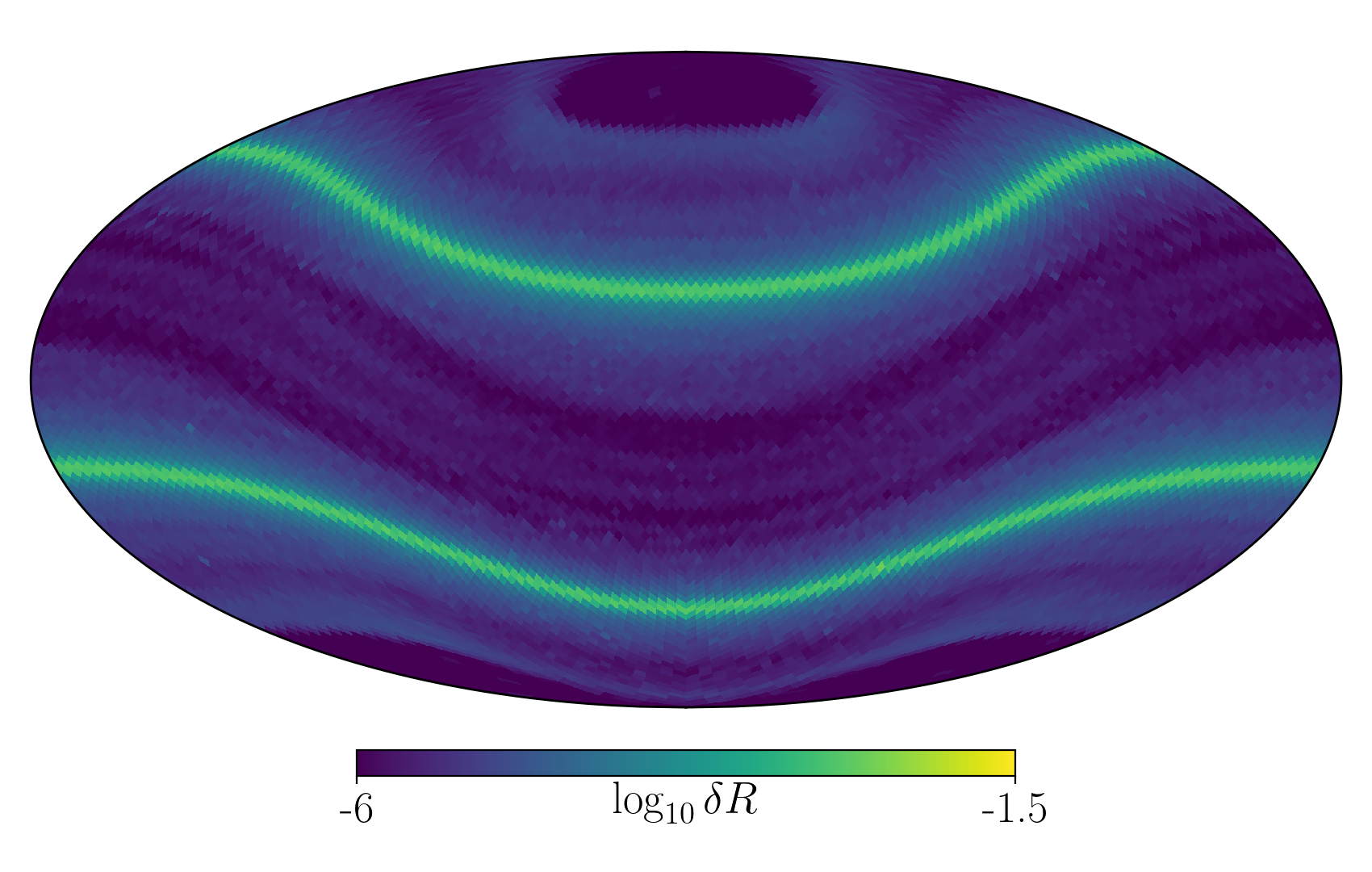}
	\includegraphics[width=0.32\textwidth]{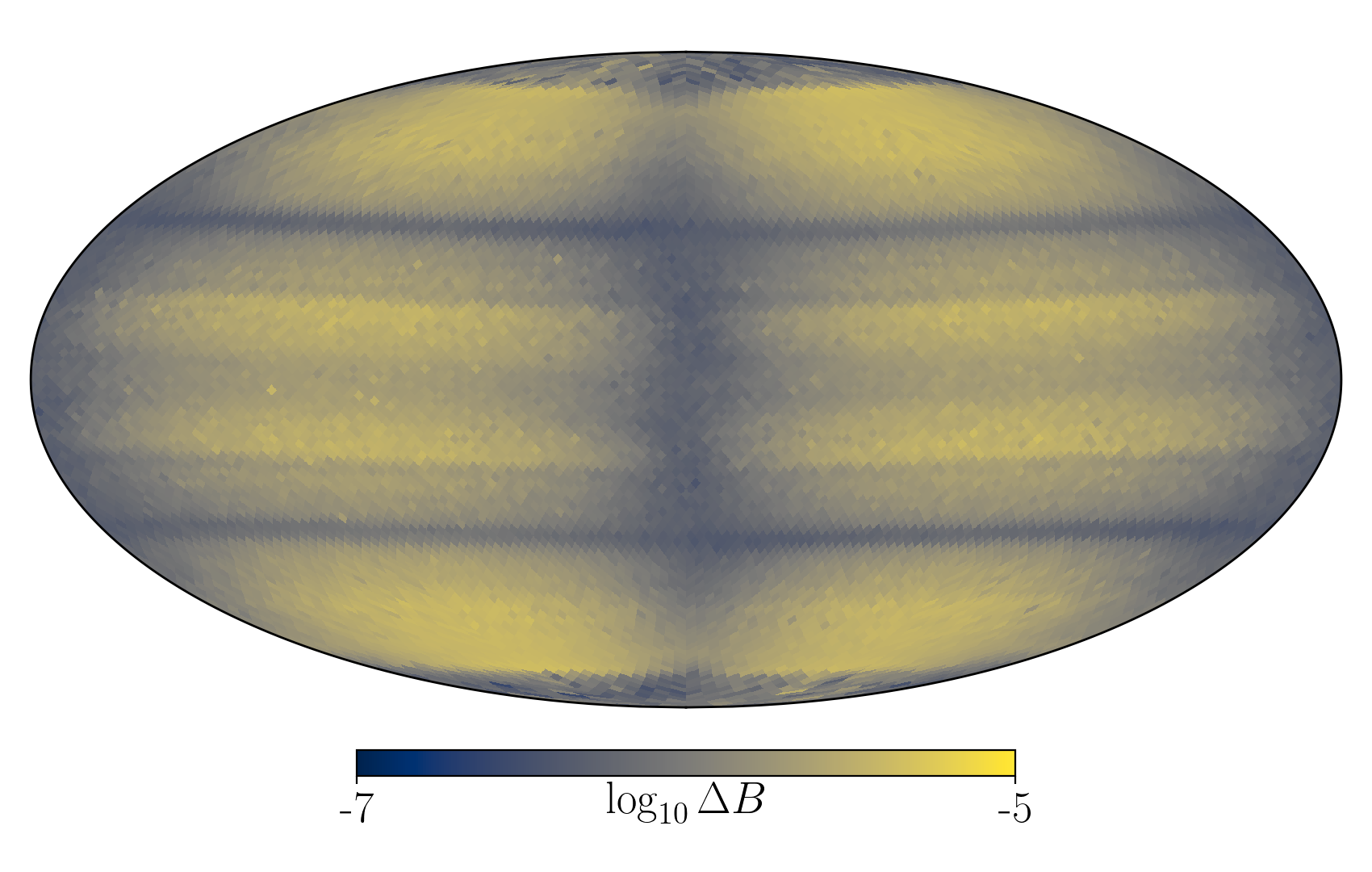}
	\includegraphics[width=0.32\textwidth]{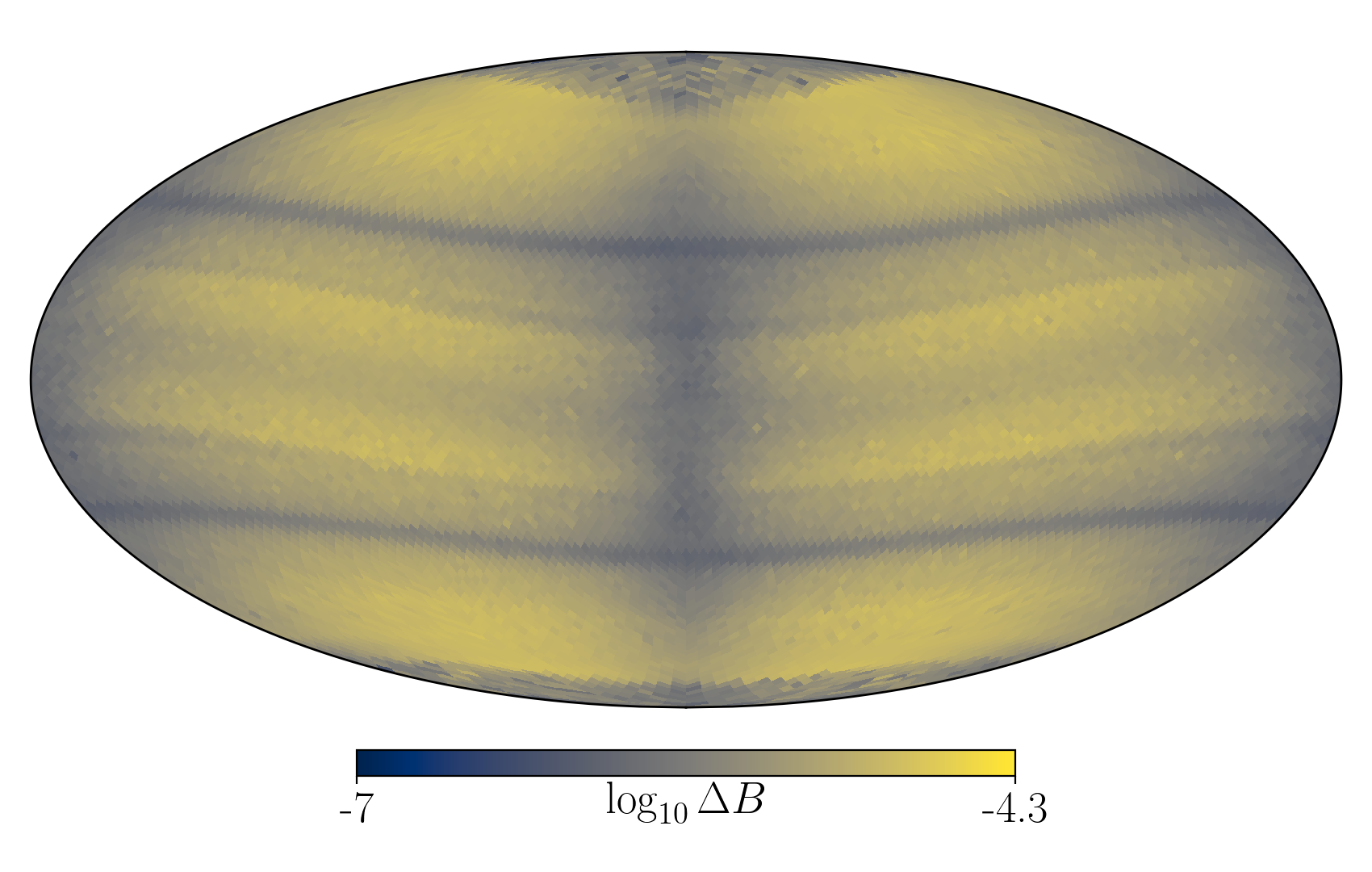}
	\includegraphics[width=0.32\textwidth]{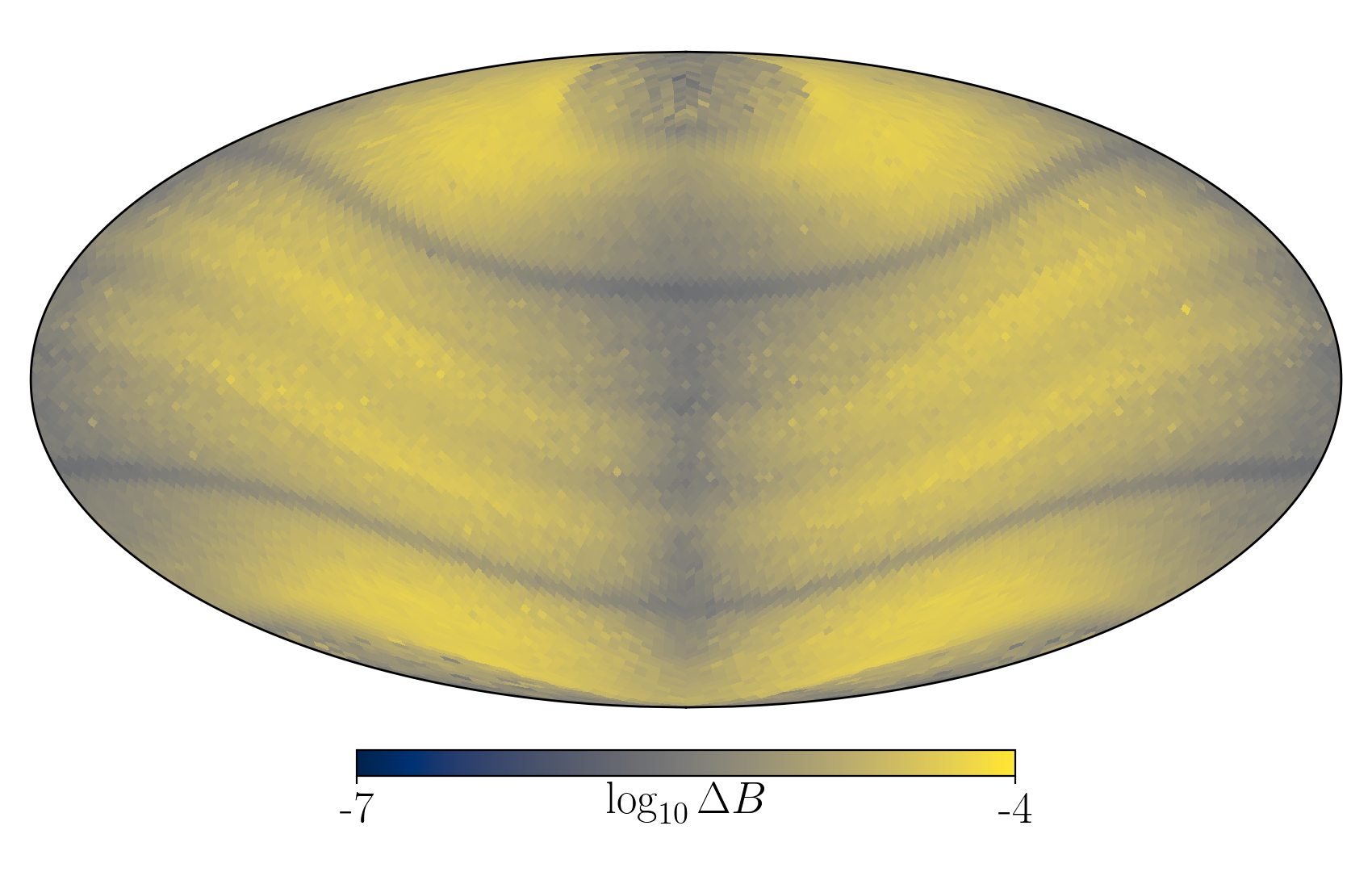}
	\caption{\label{fig:sky_map_theamM} Fractional rate (top) and line width (bottom) in each pixel at distances $r \sim R_{LC}$. Figures from left to right illustrate $\theta_m = 0.05$, $0.2$, and $0.6$ radians (note that scales are not equivalent).   }
\end{figure*}

\begin{figure}
	\includegraphics[width=0.4\textwidth, trim={1cm .2cm 2cm .5cm},clip]{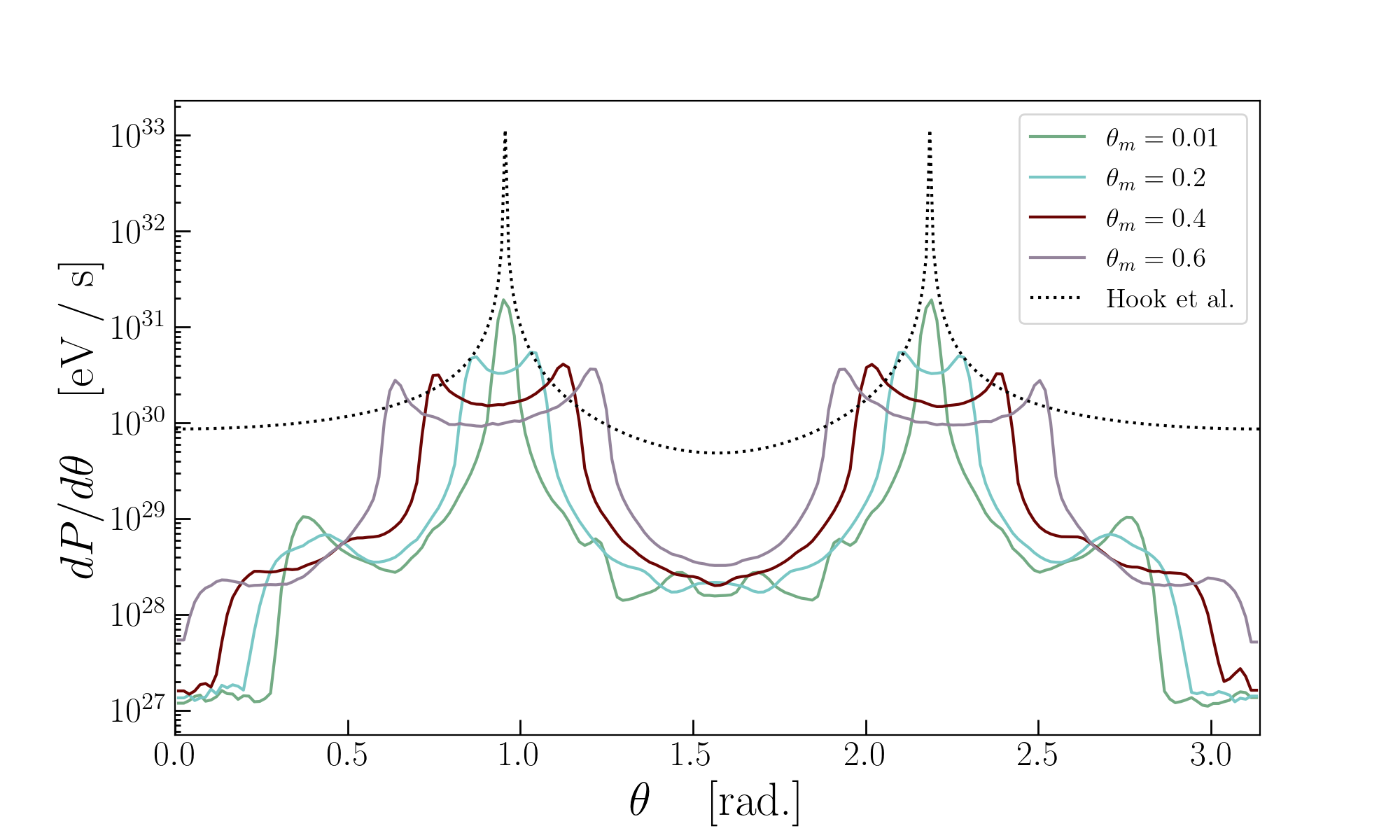}
	\includegraphics[width=0.4\textwidth, trim={1cm .2cm 2cm .5cm},clip]{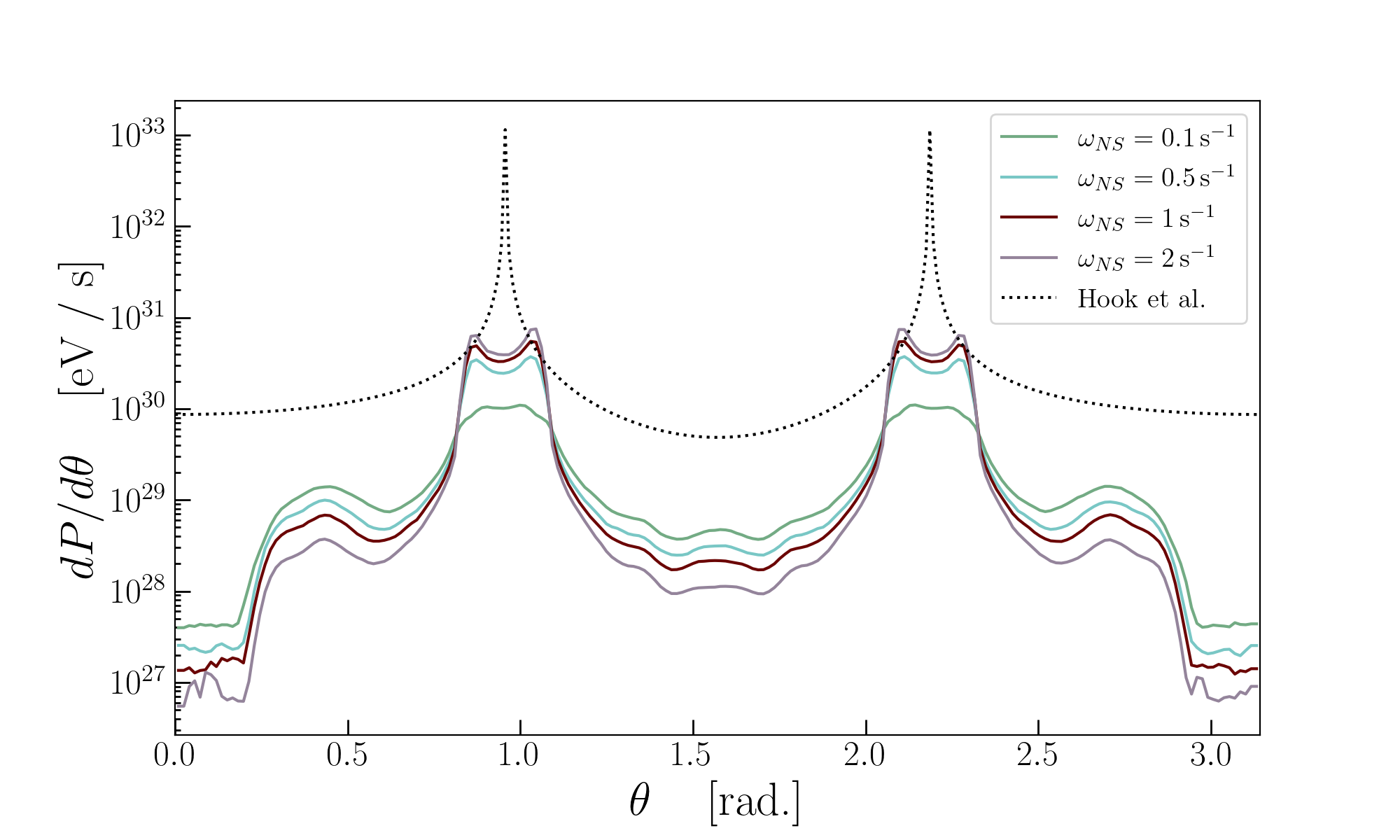}
	\includegraphics[width=0.4\textwidth, trim={1cm .2cm 2cm .5cm},clip]{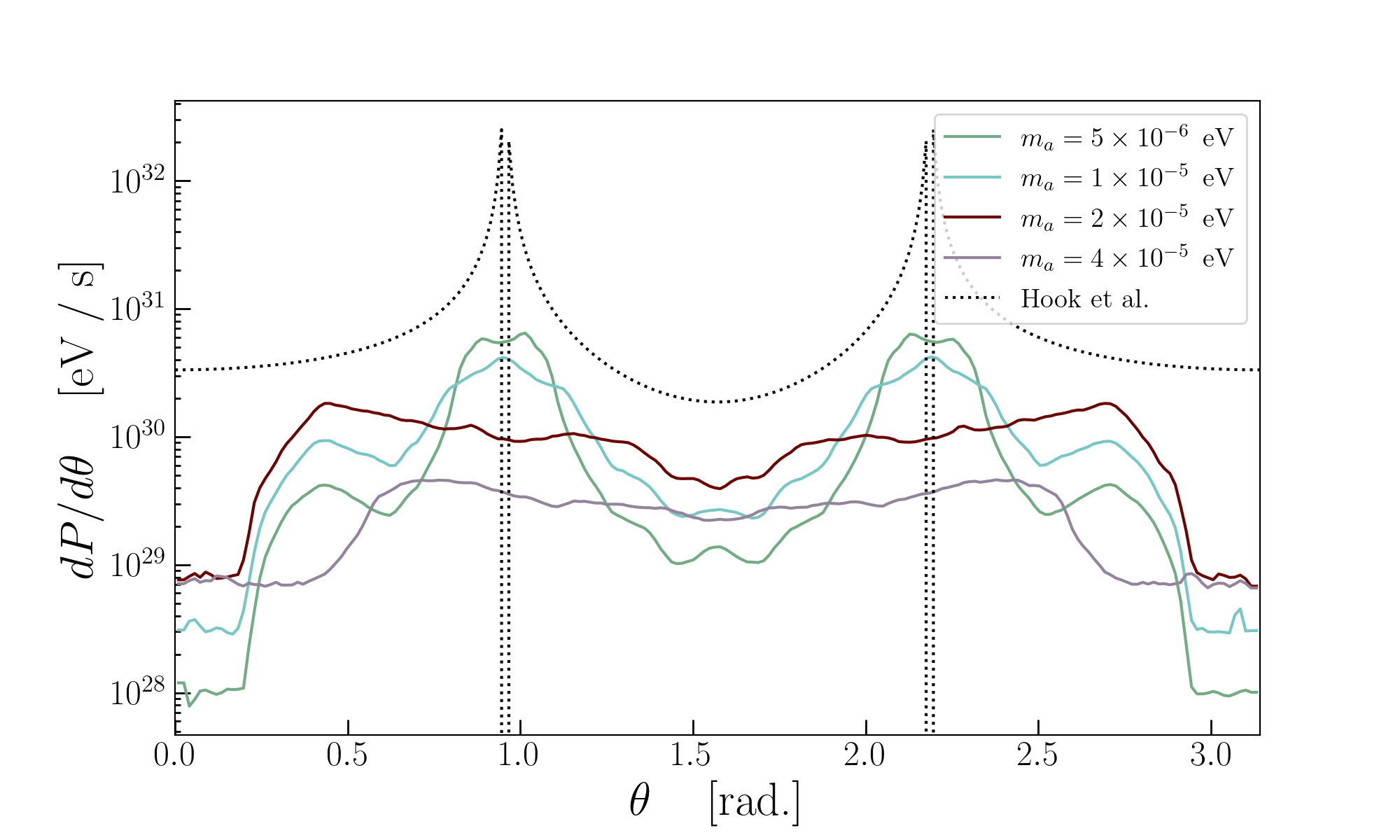}
	\caption{\label{fig:Luminosity} Comparison of time-averaged power radiated per unit viewing angle, $dP/d\theta$, varying $\theta_m$ (top-left), $\omega_{NS}$ (top-right), and $m_a$ (bottom). Results are shown in comparison with predictions from~\cite{Hook:2018iia}\footnote{Importantly, as pointed out in~\cite{Leroy:2019ghm}, Ref.~\cite{Hook:2018iia} is missing a factor of $v_c$ in the radiated power. We re-introduce this factor when illustrating `Hook et al.' results. } (in the case of varying $m_a$, we fix $m_a$ to $5\times10^{-6}$ eV), but taking $\theta_m = 0$ (black dashed). The axion-photon coupling is taken to be $g_{a\gamma\gamma} = 10^{-12} \, {\rm GeV}^{-1}$, and the magnetospheres are either fixed to the fiducial values used in the main text (the case for varying $\theta_m$ and $\omega_{NS}$) or the magnetar model (varying $m_a$).}
\end{figure}

The second concern that we must address is related to the fact that the three-dimensional mixing equations have not been solved (see \eg\cite{Millar:2021gzs} for recent progress in this direction). The derivation of the  conversion probability had assumed a plane wave solution proportional to $e^{ikz}$, which amounts to a one-dimensional simplification of the more general plane wave form $e^{i\vec{k} \cdot \vec{r}}$. If photons deviate strongly from this one-dimensional projection over distances $r \lesssim L_c$, the axion and photon will no longer oscillate in phase, and the conversion probability will be markedly reduced. 
In order to address the importance of this effect we begin by identifying the relationship between the conversion length and the phase difference of the axion and photon in the limit where photons propagate along straight trajectories perpendicular to the external field. In this case, the phase overlap of the axion and photon is given by \Eq{eq:phase_simple}, but now with $k_\gamma = \sqrt{\omega^2 - \omega_p^2}$. If one directly computes this phase from photon trajectories generated with the non-magnetized plasma dispersion relation (\ie $\omega^2 = k^2 + \omega_p^2$), one finds that the axion-photon phase difference $\phi$ for all photons (which have not undergone refraction) at the conversion surface is $\phi(L_c) \sim \pi/2$. As can be seen from the derivation above, the generalization to non-perpendicular directions of the phase involves the replacement of $k_\gamma$ by that given in \Eq{eq:kgam}. Nevertheless, this equation still assumes that photons travel on straight line trajectories (notice that this is apparent in the ansatz in \Eq{eq:ansatz}) -- additional de-phasing may enter as photons refract. In order to account for this effect we approximate the de-phasing by defining a new phase overlap
\begin{equation}
\phi^* \equiv  \int_0^{\ell} \, d\ell' \, (k_a - \cos\alpha \, k_\gamma(l')) \, ,
\end{equation}
where $\alpha$ is the angle between the initial axion momentum and the photon momentum after some time (and thus depends on time implicitly), and we have replaced the one-dimensional distance $z$ by the path length $\ell$. In our analysis we attempt to account for this additional de-phasing by identifying the distance each photon has traveled when  $\phi^* = \pi/2$; we associate this path length with the `corrected' conversion length $L_c^\prime$. Since the conversion probability is $\propto L_c^2$, we re-weight the MC weight by a factor of $(L_c^\prime / L_c)^2$. This procedure is illustrated schematically in \Fig{fig:dephase}. While a rigorous treatment of the three-dimensional mixing is required in order to robustly assess the impact of this effect, we believe that our prescription offers a reasonable estimate of its magnitude.

\begin{figure}
	\includegraphics[width=0.49\textwidth]{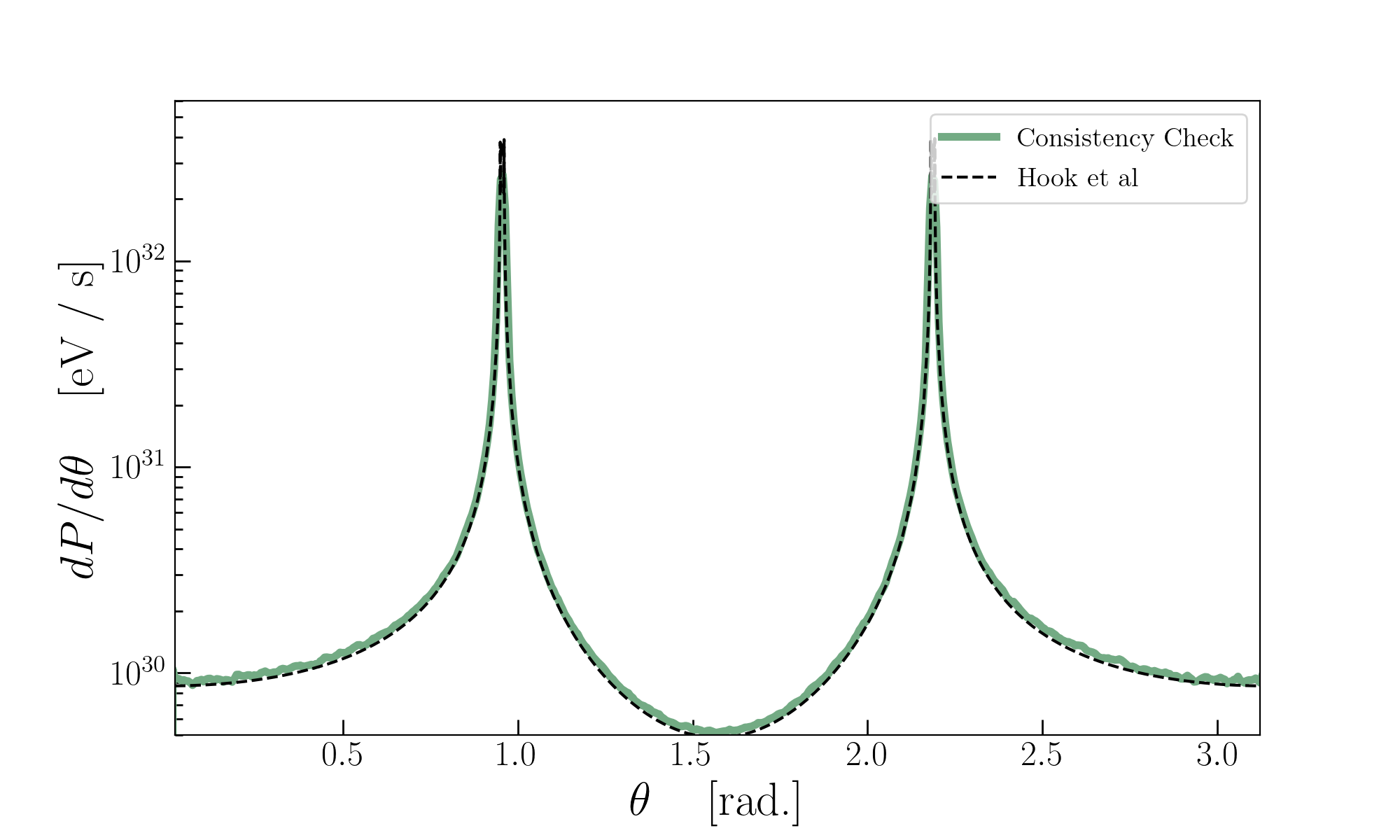}
	\includegraphics[width=0.49\textwidth]{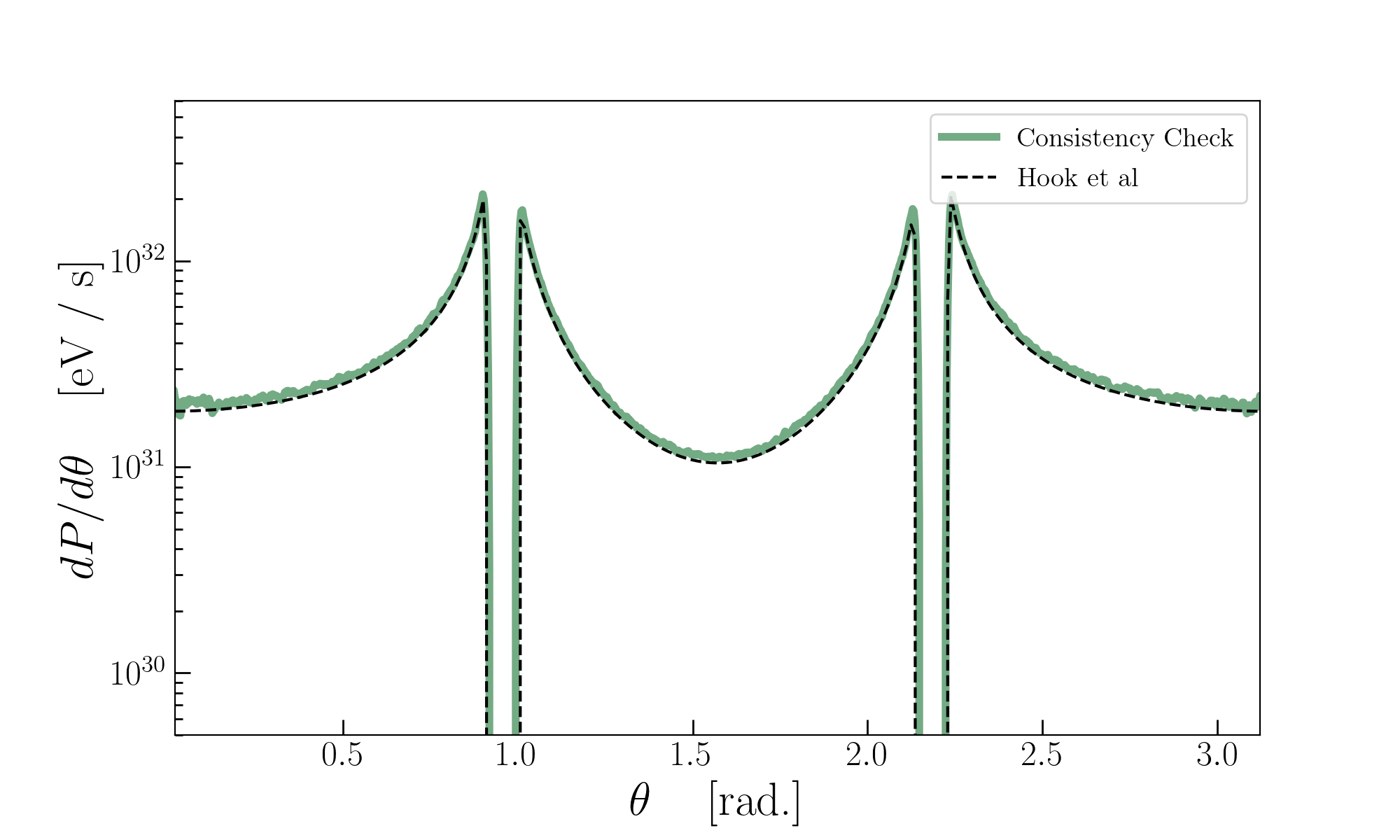}
\caption{\label{fig:HookC}  Comparison of the time-integrated differential power as approximated in ~\cite{Hook:2018iia} (black), with a run in which $(i)$ absorption and premature de-phasing is neglected, $(ii)$ the radial derivative approximation $\partial_\ell k_\gamma \rightarrow 3 m_a / (2 r_c v_c)$ is applied universally, $(iii)$ photons are assumed to travel radially outward, $(iv)$ the size of the surface element at each point is re-scaled to be consistent with the radial approximation, and $(v)$ the factor of $|\hat{v} \cdot \hat{n}|$ has been removed. The above assumptions should be roughly consistent with the approximation of \cite{Hook:2018iia}. Results are shown for the fiducial model used throughout this text (left) and the magnetar PSR J1745-2900 (right), but taking $\theta_m = 0.01$ and $m_a = 10^{-6}$ eV (left) and $10^{-5}$ eV (right), and $g_{a\gamma\gamma} = 10^{-12} \, {\rm GeV}^{-1}$.     }
\end{figure}

\section{Parameter Dependence of Sky Maps} \label{sec:paramD}
The results presented in the main text are shown for a fiducial set of parameters of the GJ model which are expected to be representative of neutron stars in the Milky Way. The purpose of this section is to address the sensitivity of the results presented to the choices of these parameters. In particular, we focus on the parameters likely to induce the largest effects; these include: the misalignment angle $\theta_m$, the rotational frequency of the neutron star $\omega_{NS}$, and the magnetic field strength at the surface $B_s$. We also discuss implications of changing the axion mass.

We begin in \Fig{fig:sky_map_theamM} by considering the impact of the misalignment angle on the isotropy (top) and line dispersion (bottom). Specifically we consider three choices of $\theta_m$ corresponding to $0.05$ (left), 0.2 (center), and 0.6 (right) radians. The trend in the both cases is immediately apparent -- larger values of $\theta_m$ induce stronger anisotropies, in particular with respect to photons produced in the throat, and larger line widths. Furthermore, large values of $\theta_m$ imply less sensitivity to the viewing angle, since the throats more broadly sweep through large areas of the sky. 

Since we have discussed the impact of varying the above parameters on line width in the main text, we focus next instead on their impact on the radiated power. We illustrate in  \Fig{fig:Luminosity} the time-averaged differential power per viewing angle $dP/d\theta$, as a function of viewing angle for various parameter choices. We plot for comparison the predicted power using the formalism of Ref.~\cite{Hook:2018iia} assuming $\theta_m = 0$ (all other parameters are either set to the fiducial values used in the main text or to those of the magnetar). We highlight two important results of these figures. Firstly, the flux is strongly reduced relative to previous predictions. Secondly, the sensitivity to the viewing angle is strongly increased (spanning, in some cases, nearly 5 orders of magnitude). 

In order to ensure the code developed for this work is reliable, we also present a consistency check in which we attempt to reproduce the results of Ref.~\cite{Hook:2018iia} by systematically removing the effects included here. We do this by $(i)$ removing all absorption, $(ii)$ removing the de-phasing cut on $L_c$, $(iii)$ taking the radial derivative approximation $\partial_\ell k_\gamma \rightarrow 3 m_a / (2 r_c v_c)$, $(iv)$ assuming photons sourced at some location $\vec{r}_i$ travel radially outward thereafter, $(v)$ rescaling the size of the surface element at each point by a factor of $|\hat{n} \cdot \hat{r}|$ (this comes from the fact that \cite{Hook:2018iia} has a self contradictory assumption of radial trajectories that are perpendicular to the normal), and $(vi)$ removing the factor of $|\hat{v} \cdot \hat{n}|$ . As shown in  \Fig{fig:HookC}, this procedure perfectly reproduces the results of Ref.~\cite{Hook:2018iia}. The panels display results for the fiducial model with $\theta_m = 0.01$ and $m_a = 10^{-6}$ eV (left) and $m_a = 10^{-5}$ eV (right) (and taking $g_{a\gamma\gamma} = 10^{-12} \, {\rm GeV}^{-1}$).  

\begin{figure}
	\includegraphics[width=0.49\textwidth]{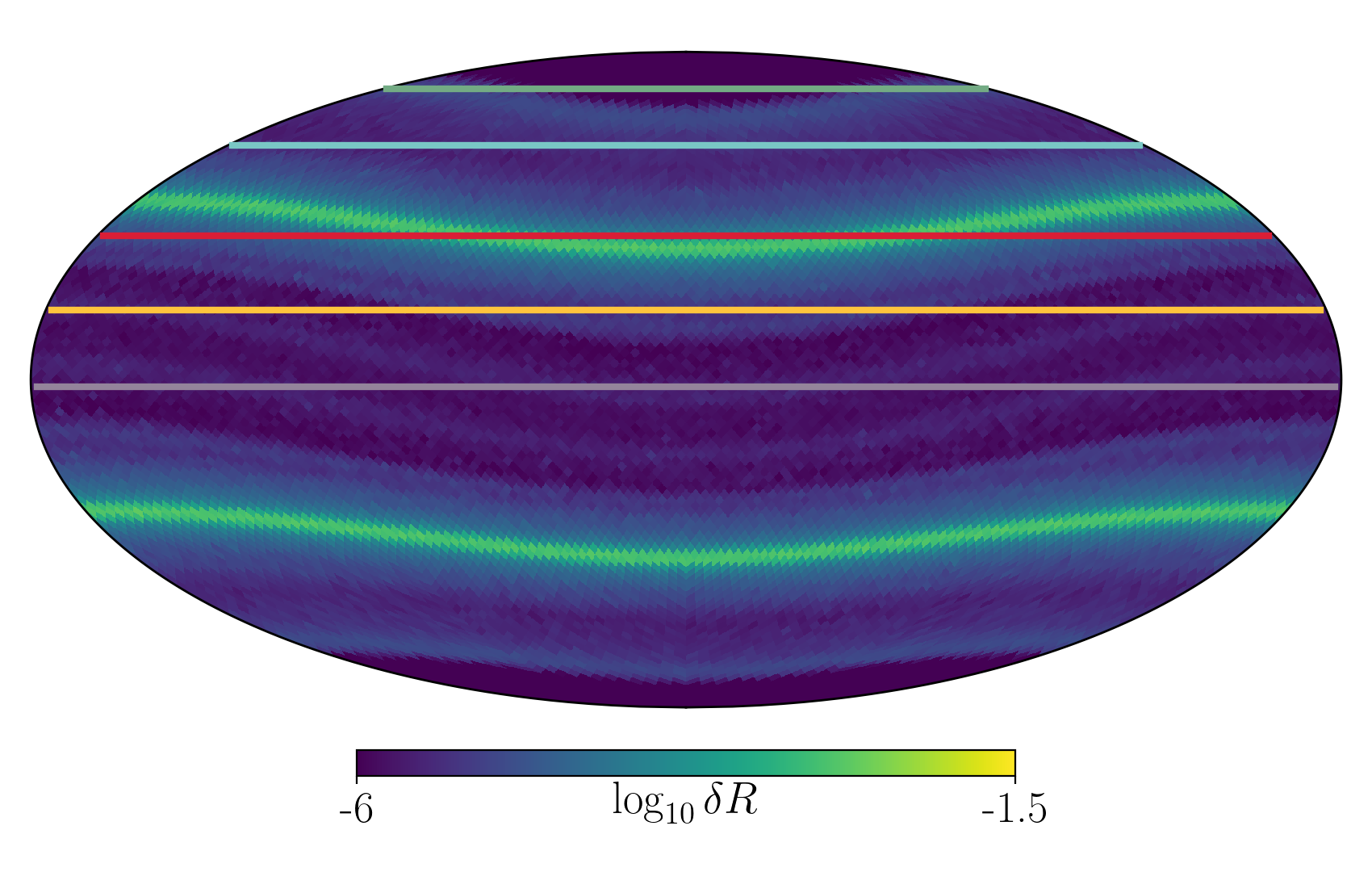}
	\includegraphics[width=0.49\textwidth]{{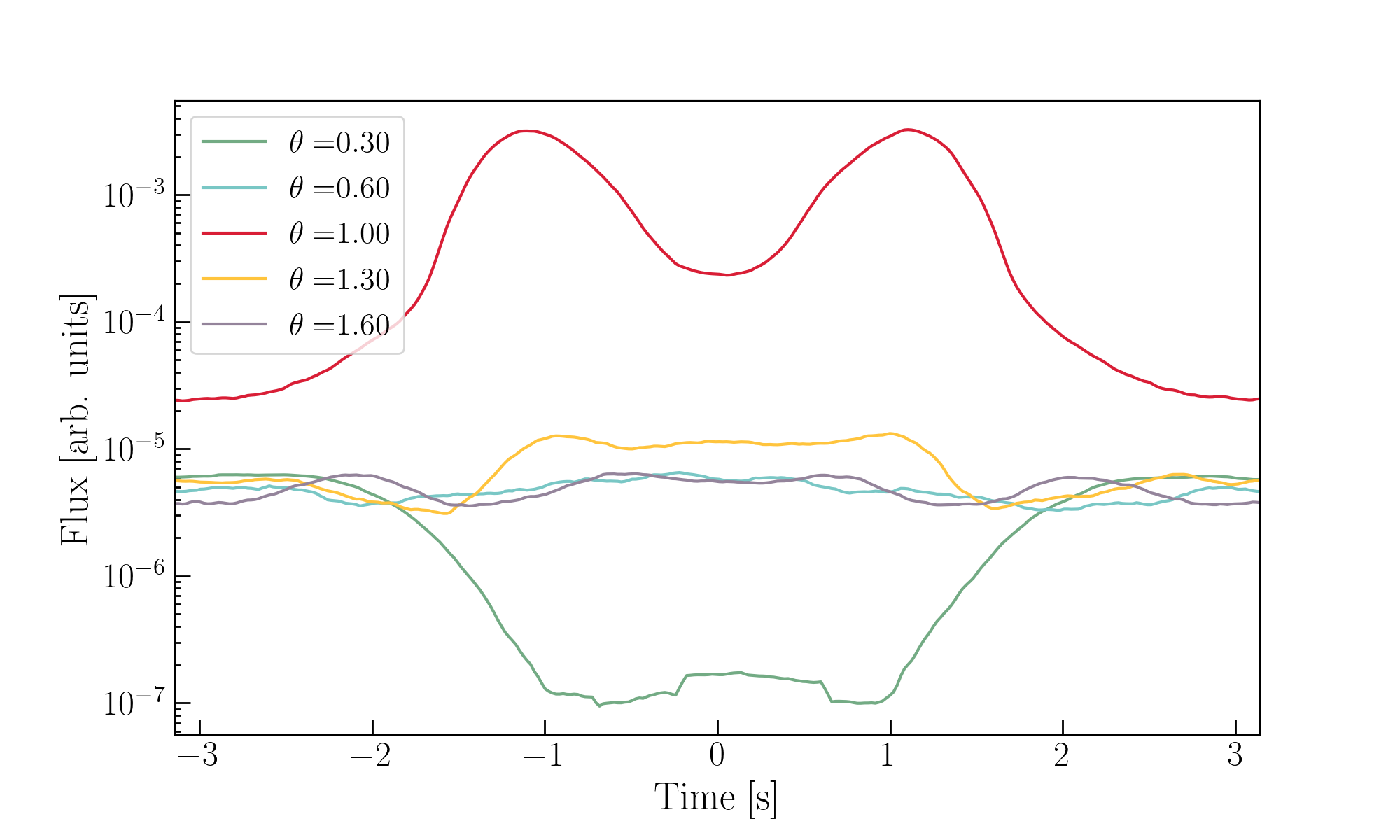}}
	\includegraphics[width=0.49\textwidth]{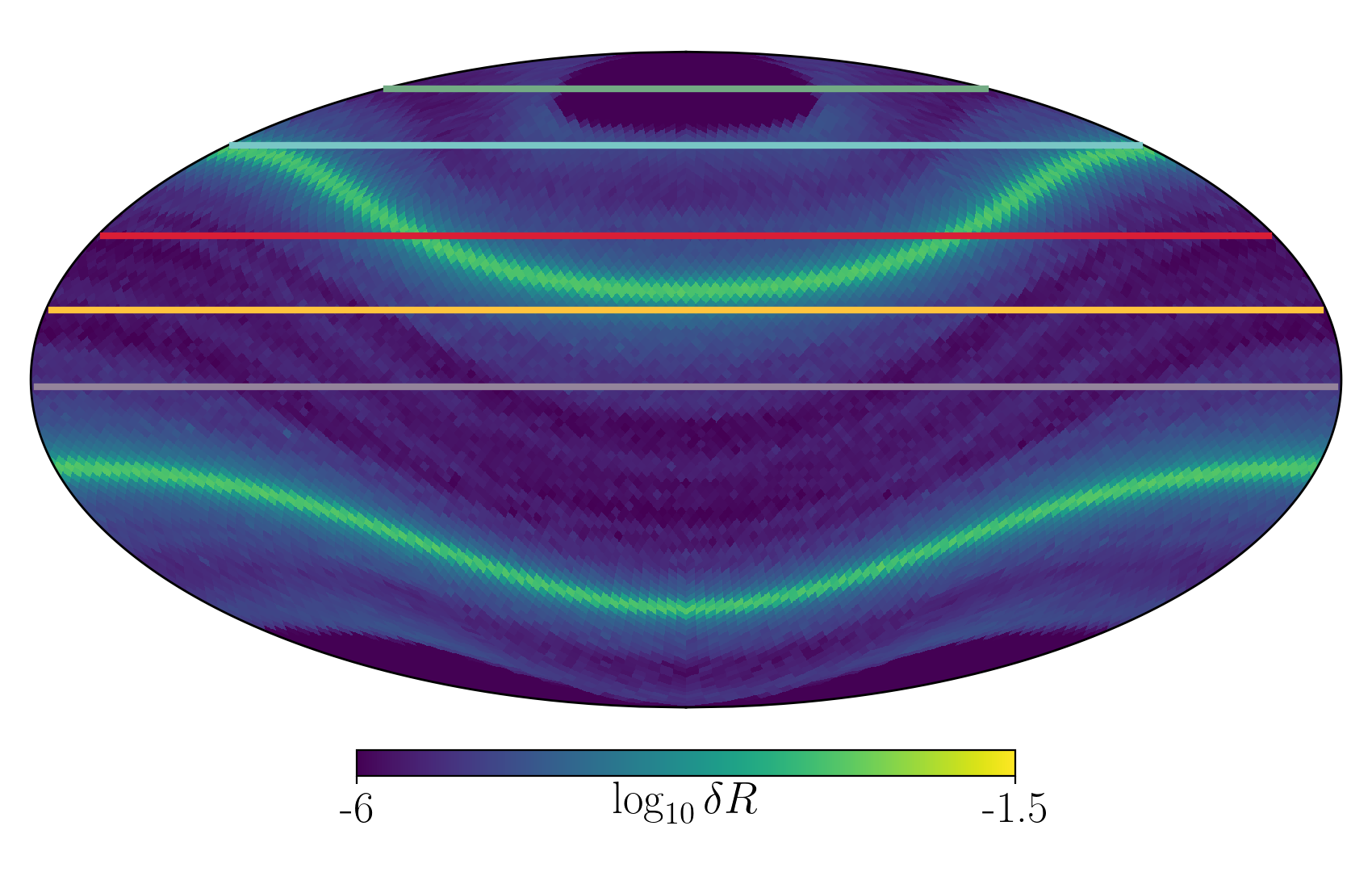}
	\includegraphics[width=0.49\textwidth]{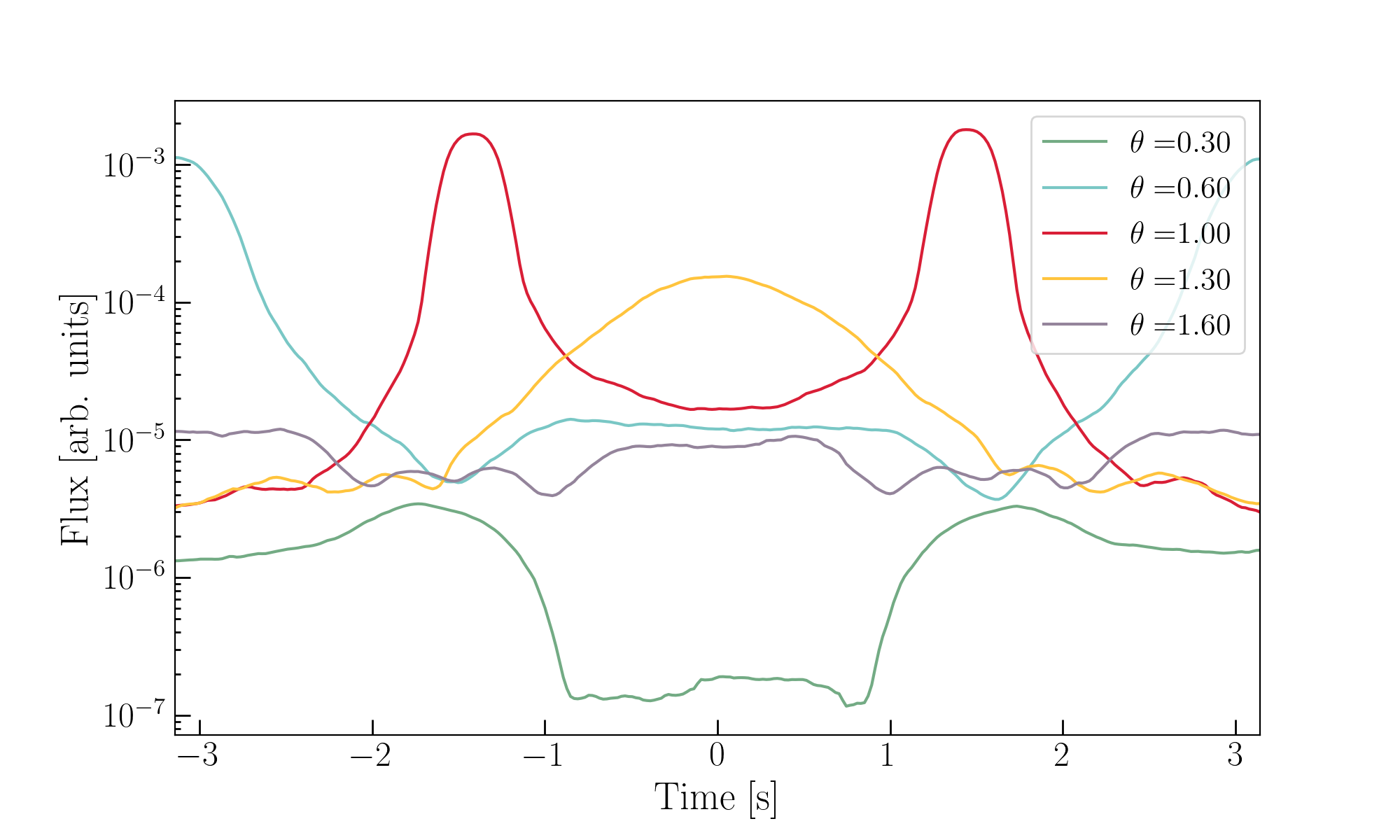}
	\caption{\label{fig:TimePlt} Time dependence of the flux for our fiducial model (top) and model with $\theta_m = 0.6$ (bottom). The five horizontal lines shown in each sky map (left) correspond to $\theta = 0.3, 0.6, 1, 1.3, 1.6$ radians, and are translated into time-dependent fluxes (right). Results are obtained by binning over photons with $\theta = \theta_i \pm 0.02$ radians. }
\end{figure}

\begin{figure}
	\includegraphics[width=0.6\textwidth]{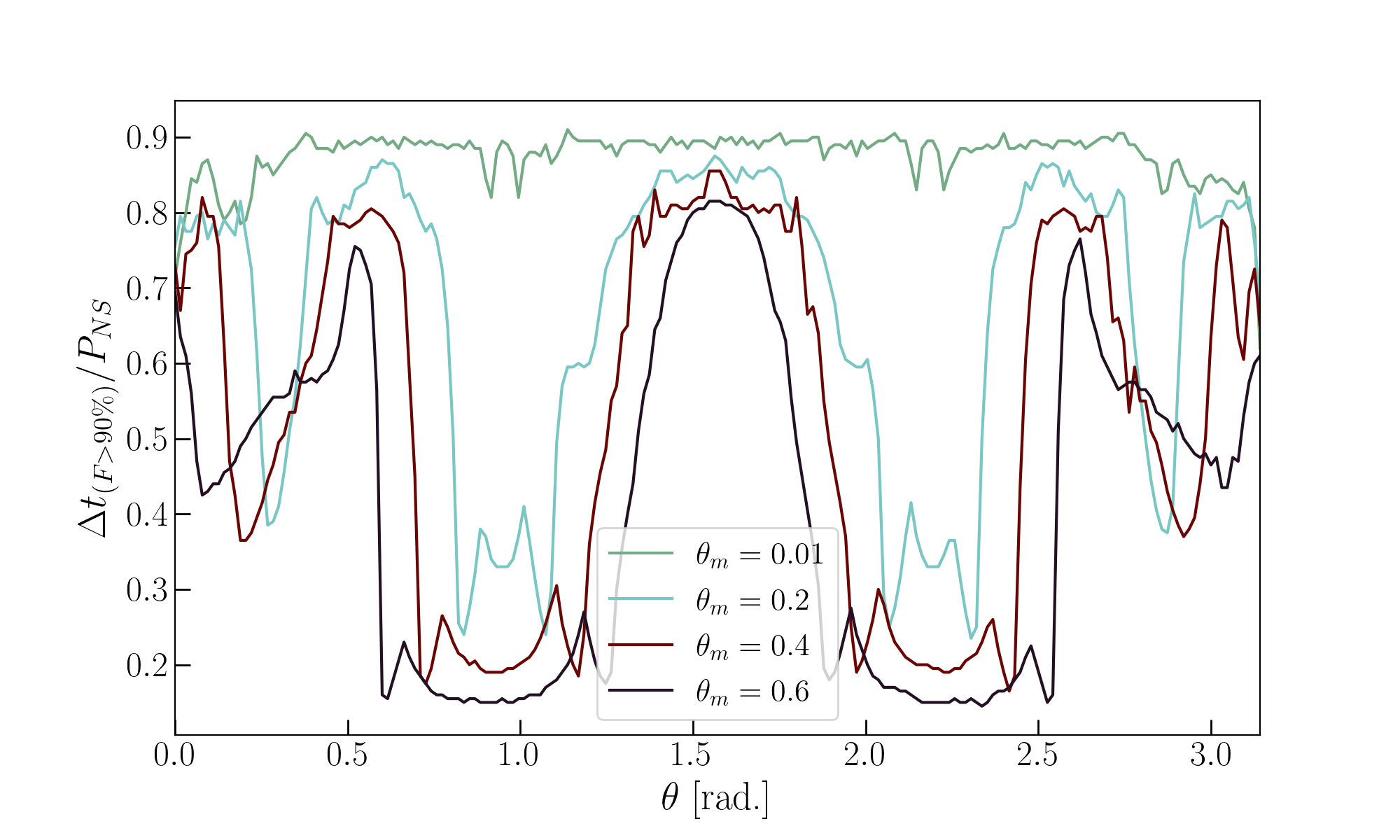}
	\caption{\label{fig:frac_period_TD} Fractional time per rotational period in which $90\%$ of the flux is generated, as a function of viewing angle $\theta$. The four different lines correspond to four choices of $\theta_m$. }
\end{figure}

\section{Time Dependence of Sky Maps}\label{sec:timeD}
An important feature of the signal discussed in this work is the expected time-dependence of the flux --  strong time variations allow for the signal to be more easily disentangled from spurious line signals. As mentioned in the main text, the expected time-dependence of the signal can be read off the flux maps by drawing horizontal lines and tracing the flux over a period of rotation. A single example of this procedure is illustrated in \Fig{fig:sky_map_1}. 
In this section we present two additional illustrations to highlight the time dependence of the flux for a variety of viewing angles and two misalignment angles.

In each case we bin the flux at each value of $\phi$ over a narrow range of viewing angles (defined by angles of constant $\theta$) in order to extract the flux as a function of time. Specifically, we take viewing angles of $\theta = 0.3, 0.6, 1, 1.3,$ and $1.6$ radians (these regions are highlighted in the left panel of \Fig{fig:TimePlt}), and a width in $\theta$ of $0.02$ radians.  We show the projected flux (in arbitrary units) for each slice in the right panels of \Fig{fig:TimePlt}. Depending on the viewing angle, the time variation over a period can span up to a few orders of magnitude, although the likelihood of encountering such strong time variations depends crucially on the misalignment angle.

One way to quantify the strength of the time dependence is to look at the fractional part of the period over which $X\%$ of the flux is generated, with $X$ being an arbitrarily chosen threshold which we set here to be 90. If the quantity is small, it implies that nearly the entirety of the period-averaged flux is generated in a very narrow time window, and thus the time variance must be large. Alternatively, if this fraction is close to 1, the signal must be nearly time-independent. We plot this quantity in \Fig{fig:frac_period_TD} for various misalignment angles as a function of viewing angle. One can see that this fraction typically spans between 20$\%$ and 80$\%$, and is systematically shifted toward smaller values at larger misalignment angles. 

\section{Convergence Checks}\label{sec:converge}
In this section we illustrate the approximate level of convergence achieved in our sky maps by generating many realizations under a fixed number of photon trajectories, and comparing the ratio of the standard deviation to mean $\xi_i \equiv \sigma_i / \mu_i$ in each pixel $i$ across the different realizations. Importantly, convergent sky maps are far more complicated to generate than e.g. a converged estimate of the time-averaged differential power (per unit viewing angle), simply because the latter exploits the azimuthal symmetry. In addition, convergence is also impeded by the size of the conversion surface -- consequently, in order to be conservative we illustrate the convergence below using a mass of $10^{-6}$ eV. Thus it should be understood that all time averaged quantities and larger axion masses have far stronger convergence than presented here.

\begin{figure}
	\includegraphics[width=0.45\textwidth]{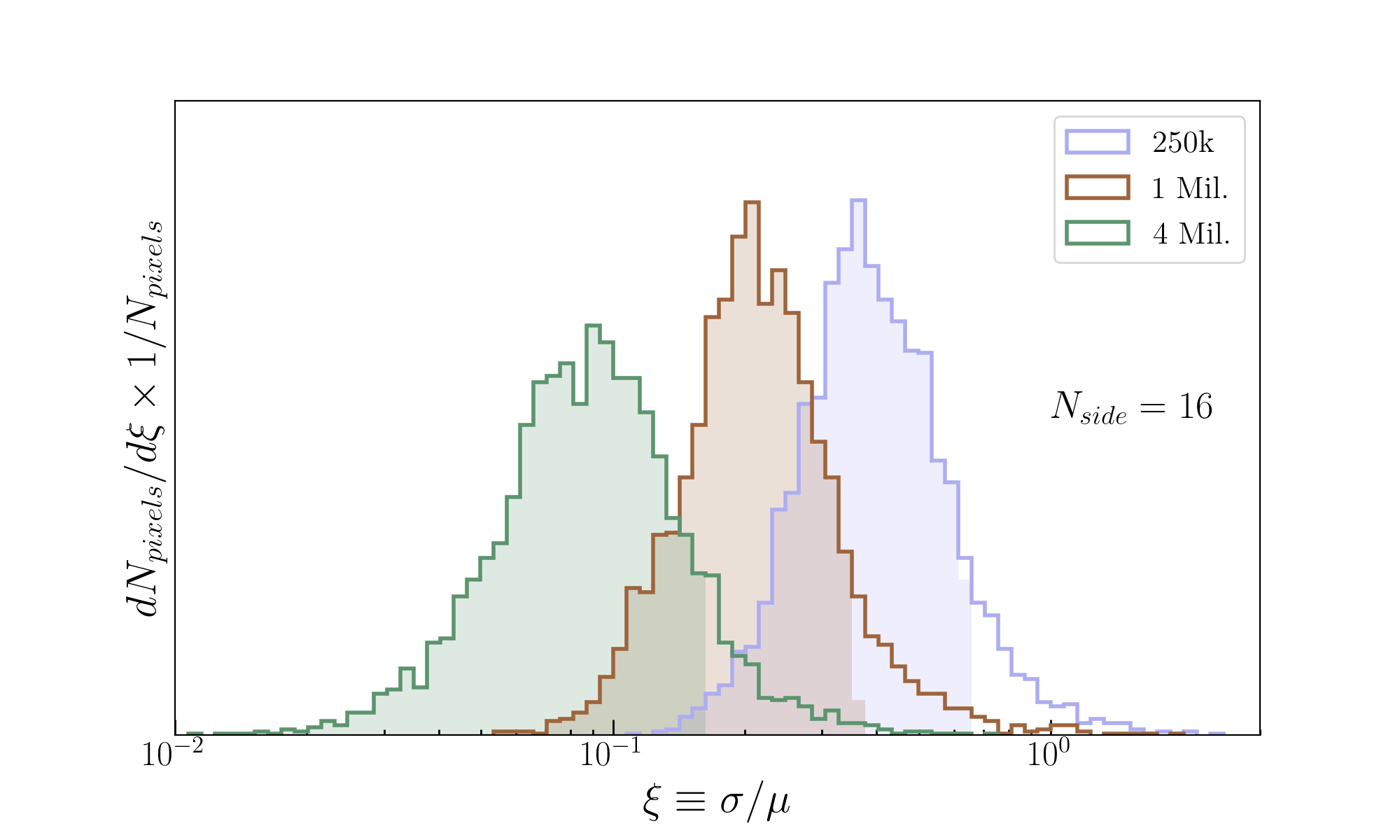}
	\includegraphics[width=0.45\textwidth]{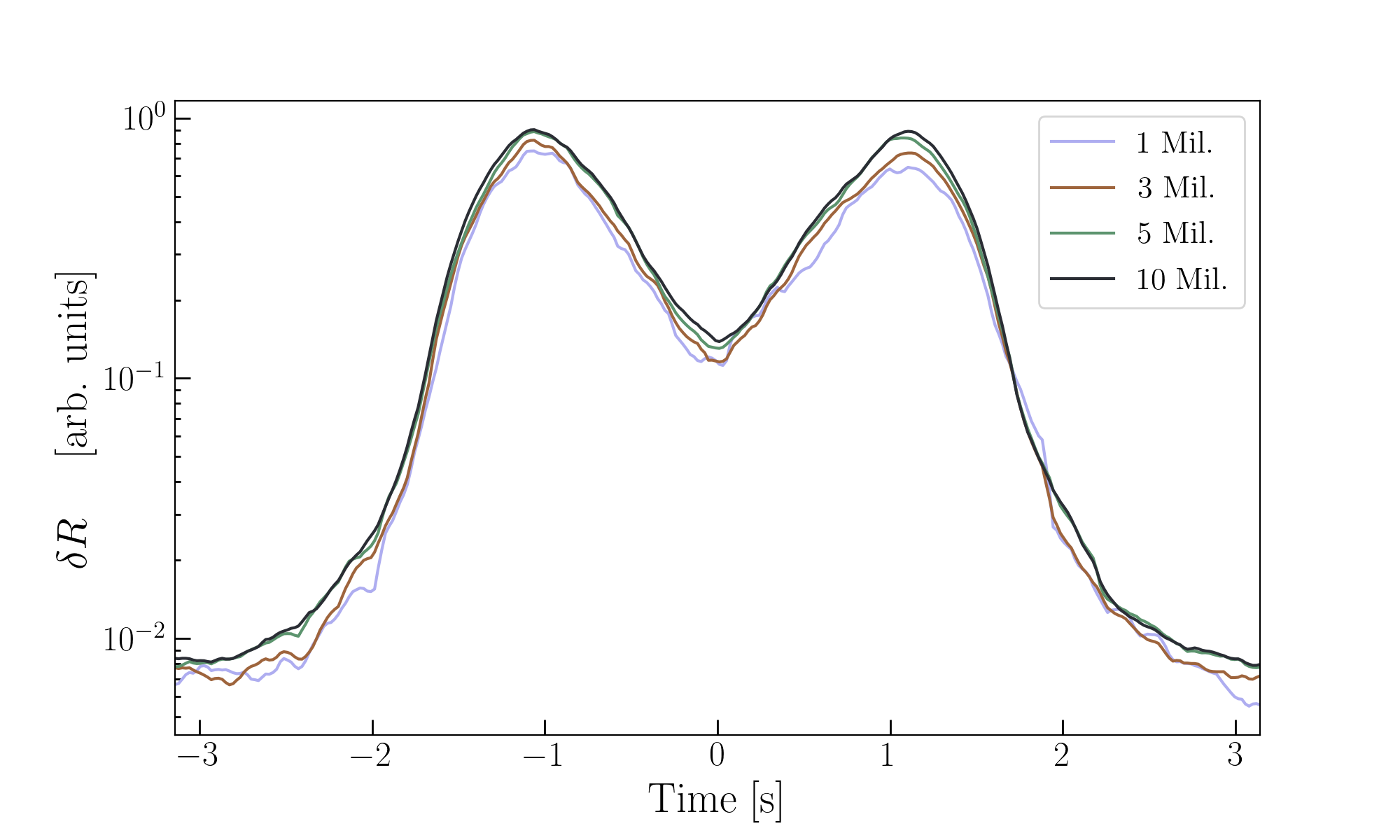}
	\caption{\label{fig:converge} Histogram of $\xi \equiv \sigma / \mu$ in each pixel of an $N_{\rm side}=16$ healpix map, generated using 250 thousand, 1 million, or 4 million trajectories. Shading illustrates 90$\%$ containment for each histogram (left). Time slice shown in right panel of \Fig{fig:sky_map_1}, varying the number of photon trajectories to illustrate the typical level of convergence (right). }
\end{figure}

In the left panel of \Fig{fig:converge} we show the convergence test using an $N_{\rm side} = 16$ healpix map, and for $N_{\rm photons} = 2.5 \times10^5$, $10^6$ and $4 \times 10^6$ trajectories. As an aside, we note that on a single core, generating $10^6$  trajectories in our fiducial model requires approximately between 4 and 20 hours (the time is largely driven by the fractional number of photon trajectories which undergo strong reflections, something which is strongly correlated with e.g. the radial size of the conversion surface, as these trajectories prove to be the most difficult to accurately resolve with high precision) -- this can be straightforwardly parallelized across an arbitrary number of cores. One can see from \Fig{fig:converge} that for maps with order $\mathcal{O}({\rm{few}} \times10^6)$ photons, roughly 90$\%$ of pixels achieve convergence at the $\mathcal{O}(20\%)$ level (in order to aide the reader, we shade the 90$\%$ containment area of the histograms). If we now consider that our time domain analysis is performed using angular patches on the sky roughly half the size than what is generated from an $N_{\rm side} = 16$ pixel, we expect $\mathcal{O}(10^7)$ photons will be enough to generate sufficient convergence in these analyses. In order to verify this estimate we illustrate in the right panel of \Fig{fig:converge} the time evolution of the flux shown in the right panel of \Fig{fig:sky_map_1} using various numbers of trajectories. As can be seen, only minor variations in the time profile appear for $\gtrsim 5\times 10^6$ trajectories, suggesting this number is indeed sufficient. All plots in this work are therefore generated using $10^7$ trajectories.

\end{document}